\documentclass[pmlr]{jmlr_arxiv}
\jmlrworkshop{\href{https://www.mlforhc.org}{Machine Learning for Healthcare, 2021}}
\jmlryear{Last updated: 2021-07-21}
\jmlrvolume{}
\jmlrpages{}

\usepackage[utf8]{inputenc}
\usepackage[T1]{fontenc}

\usepackage{hyperref}
\hypersetup{
    colorlinks,
    citecolor=blue,
    filecolor=blue,
    linkcolor=blue,
    urlcolor=blue,
}

\makeatletter
\def\set@curr@file#1{\def\@curr@file{#1}} 
\makeatother
\makeatletter
\let\Ginclude@graphics\@org@Ginclude@graphics
\makeatother

\setcitestyle{authoryear,open={(},close={)}}

\usepackage{graphicx}
\begingroup
	\expandafter\ifx\csname pdfsuppresswarningpagegroup\endcsname\relax\else\global\pdfsuppresswarningpagegroup=1\relax\fi
\endgroup

\usepackage{multirow}
\usepackage{chngcntr} 

\usepackage{caption}
\usepackage{colortbl}

\title[ABC for an Explicit-Duration HMM of COVID-19 Hospital Trajectories]{Approximate Bayesian Computation for an Explicit-Duration Hidden Markov Model of COVID-19 Hospital Trajectories}
\date{April 2021}

\author{\Name{Gian Marco Visani}$^{1}$
		\Email{\textsc{gian\_marco.visani@tufts.edu}}
\AND
        \Name{Alexandra Hope Lee}$^{1}$
		\Email{\textsc{alexandra.lee@tufts.edu}}
\AND
        \Name{Cuong Nguyen}$^{1}$, M.S.
		\Email{\textsc{cuong.nguyen@tufts.edu}}
\AND
        \Name{David M. Kent}$^{2}$, M.D., C.M., M.Sc.
		\Email{\textsc{dkent1@tuftsmedicalcenter.org}}
\AND
        \Name{John B. Wong}$^{3}$, M.D.
        \Email{\textsc{jwong@tuftsmedicalcenter.org}}
\AND
        \Name{Joshua T. Cohen}$^{4}$, Ph.D.
        \Email{\textsc{jcohen@tuftsmedicalcenter.org}}
\AND        
        \Name{Michael C. Hughes}$^1$, Ph.D.
        \Email{\textsc{michael.hughes@tufts.edu}}
        \\
        \addr $^1$ Dept. of Computer Science, Tufts University, Medford, MA, USA
        \\
        \addr $^2$ Predictive Analytics and Comparative Effectiveness Center, Tufts Medical Center, Boston, MA
        \\
        \addr $^3$ Division of Clinical Decision Making, Tufts Medical Center, Boston, MA
        \\
        \addr $^4$ Center for the Evaluation of Value and Risk in Health, Tufts Medical Center, Boston, MA
}

\renewcommand{\SS}{\mathbf{S}}
\newcommand{\XX}{\mathbf{T}}
\newcommand{\RR}{\mathbf{R}}
\newcommand{\GG}{\mathbf{G}}
\newcommand{\II}{\mathbf{I}}
\newcommand{\VV}{\mathbf{V}}
\newcommand{\p}%
	{\ensuremath{^{\prime}}}
\newcommand{\pp}%
	{\ensuremath{^{\prime \prime}}}

\begin{document}

\setlength{\abovedisplayskip}{2pt plus 3pt}
\setlength{\belowdisplayskip}{2pt plus 3pt}

\maketitle

\vspace{-1.4cm}
\begin{abstract}
We address the problem of modeling constrained hospital resources in the midst of the COVID-19 pandemic in order to inform decision-makers of future demand and assess the societal value of possible interventions.
For broad applicability, we focus on the common yet challenging scenario where patient-level data for a region of interest are \emph{not} available.
Instead, given daily admissions counts, we model daily aggregated counts of observed resource use, such as the number of patients in the general ward, in the intensive care unit, or on a ventilator.
In order to explain how individual patient trajectories produce these counts, we propose an aggregate count explicit-duration hidden Markov model, nicknamed the ACED-HMM, with an interpretable, compact parameterization.
We develop an Approximate Bayesian Computation approach\footnote{Code URL: \url{https://www.github.com/tufts-ml/aced-hmm-hospitalized-patient-trajectory-model}}
 that draws samples from the posterior distribution over the model's transition and duration parameters given aggregate counts from a specific location, thus adapting the model to a region or individual hospital site of interest.
Samples from this posterior can then be used to produce future forecasts of any counts of interest.
Using data from the United States and the United Kingdom, we show our mechanistic approach provides competitive probabilistic forecasts for the future even as the dynamics of the pandemic shift.
Furthermore, we show how our model provides insight about  recovery probabilities or length of stay distributions, and we suggest its potential to answer challenging what-if questions about the societal value of possible interventions.

\end{abstract}

\section{Introduction}
The ongoing COVID-19 pandemic has imposed immense costs to human health, quality of life, and the economy.
There remains a pressing need for forecasting models that can reliably predict demand for scarce hospital resources, such as general-ward beds, ICU beds, and ventilators.
Many previous COVID-19 forecasts have failed~\citep{ioannidisForecastingCOVID19Has2020}.
To be successful, it is especially important that models are likely to extrapolate to the future rather than simply repeat the past, as the pandemic evolves through new waves, new treatments, new disease variants, and the introduction of vaccines.
This need favors \emph{mechanistic models}~\citep{bakerMechanisticModelsMachine2018} that try to capture how patients actually move through the hospital, rather than ``black-box'' predictive models.

In addition to helping hospital administrators or regional authorities plan ahead, mechanistic forecasting models can also help assess the \emph{societal value} of possible interventions.
Throughout this pandemic, many interventions have been considered to improve outcomes for infected and hospitalized individuals, such as screening strategies, pharmaceuticals~\citep{beigelRemdesivirTreatmentCovid192020,chenSARSCoV2NeutralizingAntibody2021}, physical treatments~\citep{koeckerlingAwakePronePositioning2020}, and more~\citep{zhangCurrentStatusPotential2020,ahamadPrimedGlobalCoronavirus2021}.
Estimating the societal value of these interventions remains challenging, as they can impact many different stages of care.

Our motivating problem is to assess which interventions might avoid short-term future demand exceeding hospital capacity, as aggressive shut-downs harm other aspects of health, reduce quality of life, and restrict economic activity~\citep{neumannConsiderationValueBasedPricing2020}.
Policy makers, pharmaceutical companies, and funding agencies can use mechanistic forecasting models to identify interventions to prioritize; such models can also inform health technology assessment and hence questions of reimbursement.
The mechanistic aspect of the model is crucial: a model can shed light on an intervention's value only if it supports  \emph{``what-if''} queries about how an intervention would change patient trajectories.
To make accurate decisions, it is also crucial that the model can be \emph{adapted} to the hospital dynamics in the region of interest.

In this work, we develop and evaluate a \emph{mechanistic} probabilistic model of a COVID-19 patient's trajectory through stages of care in the hospital. This model can be used to make data-driven forecasts of the future daily census counts at different stages of care. That is, we can model the total count of patients in general ward, in the intensive care unit (ICU), or in the ICU on a ventilator. 
Importantly, the parameters driving this forecast are \emph{interpretable} in the sense that they directly correspond to underlying mechanisms, such as transitions between stages of care and dwell-time distributions.
Further, we show how our model supports asking what-if queries, such as ``what if the average stay on the ventilator decreased by 2 days?'', and how it quantitatively answers such questions via downstream forecasts of daily demand for hospital beds or ventilators.
These forecasts could be used at a regional or hospital level to assess if utilization will exceed crucial thresholds.

\paragraph{Desiderata.}
In addition to seeking a \emph{mechanistic} model with interpretable parameters, another key design goal is \emph{portability} to health systems across the globe. We wish to use the model to accurately assess the value of an intervention in a given specific hospital, region, or country.
Healthcare system heterogeneity means that in different regions of interest, different data may be available due to logistical obstacles and legal concerns.
To achieve portability, we assume \emph{no} patient-level data are available (not even individualized length of stay information).
Instead, our models rely only on aggregated daily count data.

Another goal is to explicitly capture our uncertainty given limited data by learning a \emph{posterior distribution} over parameters rather than a point estimate.
We take a Bayesian approach by expressing our prior beliefs, updating those beliefs given some observed data, and then making predictions and decisions by averaging over the remaining uncertainty~\citep{gelmanBayesianDataAnalysis2013}.
Usually, fitting a posterior requires either optimization methods based on variational inference (VI)~\citep{wainwrightGraphicalModelsExponential2008}, or sampling methods based on Markov chain Monte Carlo (MCMC)~\citep{andrieuIntroductionMCMCMachine2003}.
However, for our model and many like it, VI and MCMC have high derivation and implementation costs even for statistical experts.
These costs make VI and MCMC difficult for those who need our models the most: data scientists inside healthcare organizations.
Key discrete variables in our model (difficult to marginalize away) preclude the easy application of recent ``automatic'' VI or MCMC~\citep{hoffmanNoUTurnSamplerAdaptively2014,kucukelbirAutomaticDifferentiationVariational2017}.
Thus, to achieve accessible and extensible posterior estimation, we use simulation methods via Approximate Bayesian Computation (ABC)~\citep{marjoramMarkovChainMonte2003,marinApproximateBayesianComputational2012,cranmerFrontierSimulationbasedInference2020}.
ABC algorithms are easy to implement given a simulation that samples from the model.
Furthermore, ABC allows us to design a custom distance function to assess fit, which we tailor to our application. This would be difficult with an explicit likelihood in VI and MCMC.

\paragraph{Contributions.}
This study makes three key contributions.
\begin{itemize}
\item First, we develop a new model capable of forecasting the daily counts of patients at various stages of care (general ward, intensive care unit (ICU), on ventilator in the ICU, death) when given the daily admissions counts. The model parameters are fully specified by fewer than 20 numbers, each one interpretable by practicing clinicians because they correspond directly to how patients move through typical hospitals.

\item Second, we develop routines for fitting posterior distributions over parameters given only aggregated daily count data from any region of interest. Building on Approximate Bayesian Computation (ABC) extensions of Metropolis-Hastings MCMC~\citep{marjoramMarkovChainMonte2003}, we show how distance metrics and annealing schedules for acceptance thresholds can be carefully designed to generalize well to new target regions. 
We further show how to scale up to \emph{tens of thousands} of hospital stays.
Because our ABC approach is simpler to implement than other possibilities, other researchers can adopt our work or even extend our model without advanced expertise in Bayesian learning.

\item Third, we \emph{validate} our model and posterior estimation procedures on real hospital census counts from both the United States and the United Kingdom. 
We show that the model can provide competitive forecasts of daily counts for both a single hospital and for an entire geographic region, even when the ``training'' period shows rising numbers but the ``testing'' period exhibits falling counts due to underlying shifts in policy and population behavior.
Beyond accuracy, we show that our approach can deliver \emph{insight} into underlying mechanistic parameters
 and can be interrogated with \emph{``what-if'' scenarios} to understand the likely quantitative impact of possible interventions.
\end{itemize}

\subsection*{Generalizable Insights about Machine Learning in the Context of Healthcare}

Many healthcare modeling problems are beset by two challenges.
First, it is difficult to acquire large volumes of patient-level data (due to privacy concerns or logistical obstacles).
However, relevant \emph{aggregated} data with no patient-specific information are often publicly available.
Second, it is usually easy to specify a mechanistic probabilistic model that can \emph{simulate} individual patient trajectories and produce aggregated counts, but difficult to \emph{learn} the posterior distribution over model parameters.
Our work can be seen as a case study of lessons learned in how to apply an under-utilized methodology -- Approximate Bayesian Computation (ABC) -- to address both concerns in a healthcare context.


\section{Related Work}
Projecting the spread of COVID-19 and its impact on hospital utilization has been widely pursued. We briefly survey several threads of related work: regional modeling, local modeling (at either the hospital-level or patient-level), combinations of regional and local models, and efforts to assess societal value.

\paragraph{Regional modeling.}
Many studies have forecasted hospital utilization within a large geographic region~\citep{murrayForecastingImpactFirst2020,jewellPredictiveMathematicalModels2020,reinerModelingCOVID19Scenarios2021}.
Some regional models use Susceptible-Infected-Recovered (SIR) models to predict regional case rates~\citep{zouEpidemicModelGuided2020,liOverviewDELPHIModel2020}, which can then inform bed usage and ventilator usage. 
Others~\citep{murrayForecastingImpactFirst2020} focus on predicting mortality rates given observed deaths, then use an internal simulation to estimate hospital utilization given fatalities.

A common limitation of regional models is their simplified and inflexible characterizations of the care pathways within hospitals.
For example, CHIME\footnote{\url{https://penn-chime.phl.io}}~\citep{weissmanLocallyInformedSimulation2020} allows users to specify the proportions of patients requiring ICU care and mechanical ventilation, as well as the average length of stay in the hospital and in the ICU.
But the model does not accommodate \emph{distributions} for the “dwell times” at each level of care; nor does it allow for different durations among patients who recover and who deteriorate.
The Imperial College model~\citep{flaxmanEstimatingEffectsNonpharmaceutical2020} connects deaths to key epidemiological parameters, but does not capture hospital resources.
The published forecasts of the popular ``IHME model''~\citep{reinerModelingCOVID19Scenarios2021}, maintained by the Institute for Health Metrics and Evaluation (IHME) at the University of Washington, rely on rigid assumptions about how patients progress through the hospital~\citep[Suppl. Sec. 8]{ihmecovid-19forecastingteamSupplementaryInformationModeling2020}. For example, length-of-stay is not probabilistic but fixed for each class of patients (details in App.~\ref{sec:appendix_ihme}).
Users of these forecasts cannot easily modify these internal parameters or \emph{learn} them for a location of interest; instead, the IHME itself must release updates.
While the IHME model has at times performed well at its target task of accurate forecasting of mortality~\citep{friedmanPredictivePerformanceInternational2020}, arguably it has been outperformed by other efforts~\citep{guConcernsIHMEModel2020}.

\paragraph{Hospital-level modeling.}
At the hospital level, \citet{leeForecastingCOVID19Counts2021} develop a non-mechanistic prediction model for univariate counts of COVID-19 patients at a specific hospital site based on auto-regressive statistical models. While this work is portable, we expect our mechanistic model to generalize better (as confirmed in later experiments).

 \citet{epsteinPredictiveModelPatient2020} developed a mechanistic model for making hospital-specific short-term predictions of daily census counts. Given access to length-of-stay (LOS) information at that hospital from previous patients, they can forecast the length-of-stay of current patients at various stages of severity. The LOS data requirement limits portability compared to our approach.

\paragraph{Patient-level modeling.} At the patient level, the multi-state survival model by \citet{roimiDevelopmentValidationMachine2021} predicts how a patient moves day-by-day between clinical states (critical, severe, or moderate) using covariates such as age and sex. 
They use this model to forecast hospital demand in Israel.
Our proposed model is simpler and does not depend on any patient-level data.
Unlike our approach, it would be difficult to transfer \citet{roimiDevelopmentValidationMachine2021}'s model across regions because acquiring the necessary data is challenging.

\paragraph{Multi-level modeling.} ~\citet{qianCPASUKNational2020} present a model for COVID-19 at the national, regional, hospital and individual levels, targeted at the United Kingdom.
Their COVID-19 Capacity Planning and Analysis System (CPAS) makes hospital stay forecasts for current patients via patient-level covariates (e.g. age, sex, comorbidities).
They simulate the hospital stays of future patients via a regional-level trend forecaster to predict admissions.
While admirable, this kind of model is only possible in a nationalized health system with access to patient-level data. 
For many regions of interest, such data would not be available.

\paragraph{Efforts to assess societal value.}
We have identified three health technology assessments of COVID-19 interventions that include hospital utilization projections.    \citet{conglyTreatmentModerateSevere2020} evaluated the administration of either remdesivir or dexamethasone to various groups of patients hospitalized to treat their COVID 19 infections. Assumptions for the model came from the literature. 
\citet{gandjourHowManyIntensive2021} assessed the addition of ICU capacity in German hospitals, using assumptions derived from information released by the German Ministry of Health.
Finally, \citet{joCosteffectivenessRemdesivirDexamethasone2020} examined use of remdesivir and dexamethasone in South Africa.  The assessment relied on clinical trial results to project therapeutic impact on length of stay and clinical outcomes.  But the analysis did not explicitly describe its modeling of patient progression in the hospital setting.
We emphasize that none of these analyses performed any \emph{learning or calibration} of models to ensure consistency with observed data from the region of interest.
Thus, our study advances the state-of-the-art for assessing societal value by applying Approximate Bayesian Computation methods to ensure that models accurately explain data from the target site.

\section{Model}
We now describe our proposed probabilistic model for an individual COVID patient's trajectory through the hospital.
We call our model the Aggregate Counts Explicit Duration Hidden Markov Model, or ``ACED-HMM'' for short.
While we specialize to modeling hospitalized patients with confirmed cases of COVID-19 here, the model could conceptually be used for any disease with appropriate changes to the stages of care considered as well as modified assumptions about transitions and durations.

\begin{figure}[!t]
    \centering
    \includegraphics[width=0.8\textwidth]{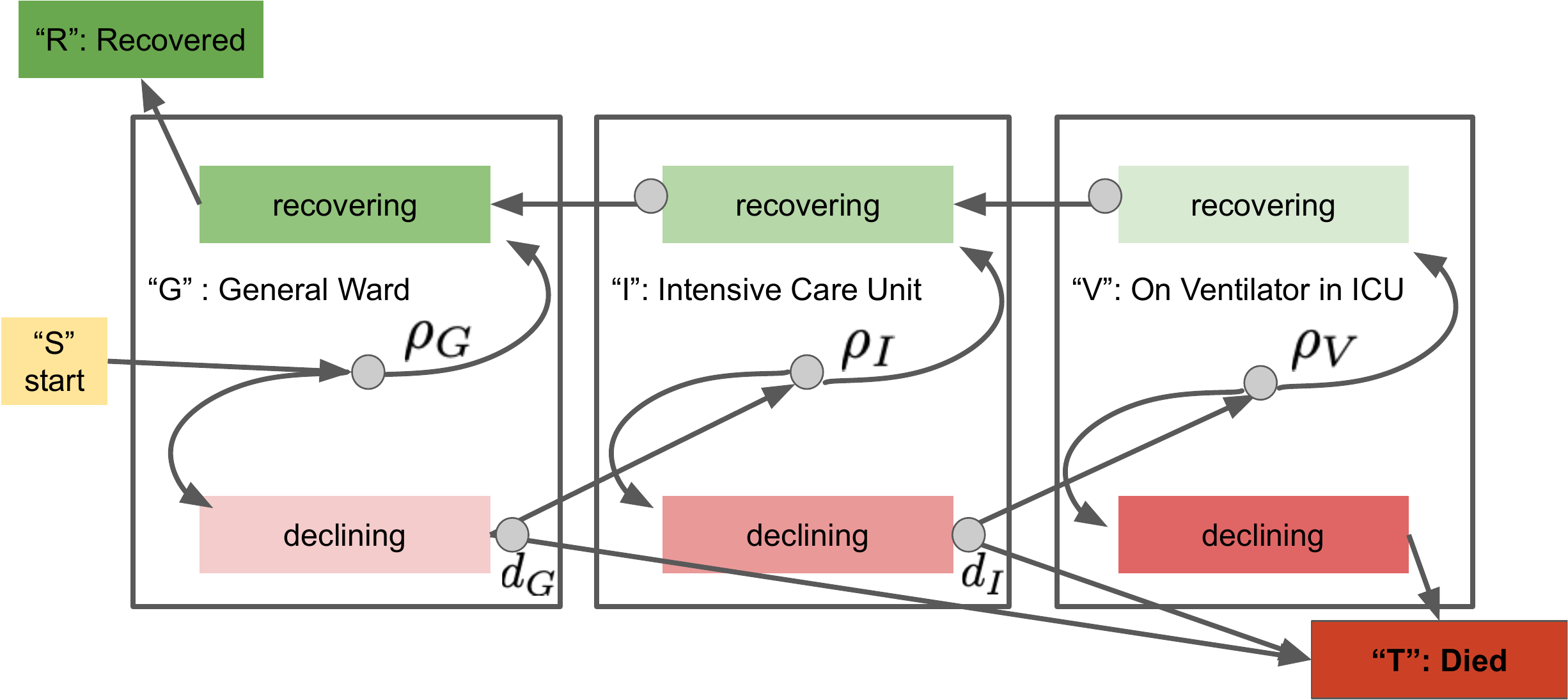}
\caption{
Diagram of the proposed Markovian model for a COVID-19 patient's trajectory through the hospital. 
Rectangles indicate possible states, defined by the patient's current stage of care $s$ (general ward $\GG$, off ventilator in ICU $\II$, on ventilator in ICU $\VV$) as well as health trajectory $h$ (recovering or declining).
Small circles indicate stochastic binary choices, with one path marked by transition probability.
Recovery transition parameters $\rho_k$ give the chance of switching to a recovery path, while death probabilities $d_k$ mark the chance of dying early.
}
\label{fig:model_diagram_transitions}
\end{figure}

Our proposed model represents the patient's current health state and stage of care in the hospital over a period of $T$ days, indexed by $t \in \{1, 2, \ldots T\}$.
There are 5 possible \emph{stages}, each indicated by a unique symbol.
We assume each patient occupies exactly one stage on each day.
Fig.~\ref{fig:model_diagram_transitions} diagrams the possible pathways through these stages and associated transition probabilities.
There are 3 stages in the \emph{intermediate} set $\mathcal{S} = \{\GG, \II, \VV\}$: stage $\GG$ indicates the patient is the general ward, 
stage $\II$ indicates the patient is in the intensive care unit (but not yet on mechanical ventilation), and 
stage $\VV$ indicates the patient is actively receiving mechanical ventilation in intensive care.
There are additionally 2 possible \emph{absorbing} or \emph{terminal} stages: ``terminal death'' $\XX$ or ``fully recovered'' $\RR$.

For each patient, indexed by $n$, our model generates a sequence of care segments indexed by $\ell$, with total length $L_n$ (not all patients will experience the same number of segments).
Each segment is defined by a \emph{stage indicator} $s_{n\ell} \in \{\GG, \II, \VV, \RR, \XX \}$, a \emph{health indicator} $h_{n\ell} \in \{0, 1\}$ corresponding to a ``declining'' or ``recovering'' trajectory within that segment, and a \emph{duration} integer $\Delta_{n\ell} \in \{1, 2, \ldots D\}$ representing the number of \emph{days} spent in that care segment.
Our model generates patient $n$'s sequence of stage indicators $s_{n,1:L_n}$, health indicators $h_{n,1:L_n}$, and durations $\Delta_{n,1:L_n}$ via the generative process defined below.
For simplicity, we drop the patient index $n$ when it is clear we are talking about one patient.

Our model assumes that for each patient, we are given an integer $a_{n} \in \{1, 2, \ldots, T\}$ indicating the day of \emph{admission} to the hospital.
Throughout this paper, we assume all patients begin in the general ward on their first day.
However, if desired it is easy to alternatively allow some incoming patients to directly begin in the ICU ($\II$) or on the ventilator ($\VV$), which might be needed to model transfers to a large facility with specialized care.

\subsection{The ACED-HMM Generative Model}

\textbf{Initialization.}
We begin in the general ward, $s_1 \gets \GG$, and sample the health indicator $h_1 \sim \text{Bern}( \rho_{\GG} )$, where parameter $\rho_{\GG}$ gives the probability of recovery in the general ward.

\paragraph{Generating stage indicators.}
We generate the \emph{stage indicator} $s_\ell \in \{ \GG, \II, \VV, \RR, \XX \}$ for each subsequent segment $\ell \in 2, 3, \ldots $ given the previous health indicator $h_{\ell-1}$ via:
\begin{align}
s_{\ell} &\gets \begin{cases}
 	\XX &~\text{w. prob.} ~ d_{s_{\ell-1}} ~\text{if}~ h_{\ell-1} =0
	\\
	\texttt{next\_stage}(s_{\ell-1}, h_{\ell-1}) &~\text{otherwise}
\end{cases}
\end{align}
where parameter $d_{s}$ defines the chance of early death for each intermediate stage $s \in \mathcal{S}$. The deterministic \texttt{next\_stage} procedure always advances in order $[\GG, \II, \VV, \XX]$ if health is declining ($h_{\ell-1}=0$), and advances in order $[\VV, \II, \GG, \RR]$ if health is recovering ($h_{\ell-1}=1$). 
See Fig.~\ref{fig:model_diagram_transitions} for visual illustration of these stage pathways.
Once we reach either absorbing stage ($\RR$ or $\XX$), we halt all further generation (this event defines the sequence length $L_n$).

\paragraph{Generating health indicators.}
We generate the health indicator $h_{\ell} \in \{0, 1\}$ for each subsequent segment $\ell \in 2, 3, \ldots$  given the previous health $h_{\ell-1}$ and the new stage $s_{\ell}$ via:
\begin{align}
h_{\ell}  &\gets \begin{cases}
	1 ~\text{if}~ h_{\ell-1} = 1
	\\
	u ~\text{if}~ h_{\ell-1} = 0, \quad u \sim \text{Bern}( \rho_{s_{\ell}} ),
\end{cases} \quad \ell \in \{2, 3 \ldots \}.
\end{align}
After each declining segment ($h = 0$), we switch to recovery with a stage-specific recovery probability $\rho_s \in [0,1]$ for each intermediate stage $s \in \mathcal{S}$, but otherwise keep declining.
Once on a path to recovery ($h=1$), we assume the patient always continues to recover.

Assuming recovery continues unabated is a simplifying assumption.
The chance of readmission to the ICU or reintubation on the ventilator is non-zero and has evolved with care practices substantially over the course of the pandemic.
Among all hospitalized patients with COVID-19, only a fraction progress to the ICU and later recover to the general ward.
Among these initially recovered patients, our clinical experts report that for the focus period of our later case studies (Nov. 2020 - Feb. 2021) only a small fraction -- roughly 10\% -- would be later sent back to the ICU. 
Because the overall fraction of patients on pathway to recovery who then decline is small, we suggest that our model, while not capturing all possible pathways, still captures most while benefiting from simplicity in estimation and implementation. Allowing recovering patients to later decline is a  future research opportunity and would require additional parameters that are currently unavailable. Our model and ABC learning could handle this extension: we would simply add a probability of reversal to our Markovian model simulation with an appropriate prior, but would also then need to understand if progression or recovery then differs.

\paragraph{Generating durations.} We generate the duration  $\Delta_\ell$ (in days) for segment $\ell$ as:
\begin{align}
	\Delta_{\ell} | s_\ell, h_\ell \sim \text{Cat}( \pi^{s_\ell, h_\ell}_1, \ldots \pi^{s_\ell, h_\ell}_D ).
\end{align}
where probability vector $\pi^{s,h}$ defines a categorical probability mass function over the first $D$ integers $\{1, 2, \ldots D\}$ representing possible durations spent at intermediate stage $s \in \mathcal{S}$ while in health state $h \in \{0, 1\}$. Notably, this model flexibly allows recovering ($h=1$) and declining ($h=0$) patients to have different duration distributions at each stage.

For tractability we define a maximum number of days $D$ allowed in any single segment. Since there are at most 5 possible segments before a patient reaches a terminal state (see Fig.~\ref{fig:model_diagram_transitions}),  the maximum hospital stay is $5D$ days.
In all experiments, we set $D=22$.
Designating this value as the maximum duration of any segment of care reflects COVID-19 length-of-stay statistics as of September 2020 from the U.S. CDC ~\citep[Table 2]{u.s.cdcCOVID19PandemicPlanning2020}, which only reports 25th, 50th, and 75th percentiles. For patients admitted to the ICU, the 75th percentile estimated for \emph{total hospital length of stay} ranges from 20 days to 25 days across age groups (corresponding to at least 2 segments of our model: general-ward plus ICU). For patients not admitted to the ICU, stays are typically much shorter. By setting $D=22$ in our analyses, the total length of stay could reach up to 66 days for patients admitted to the ICU at any time during hospitalization, and up to 110 days for general ward patients who eventually require mechanical ventilation. Truncation to at most 22 days per segment should thus affect few real patients.

At the suggestion of reviewers, we investigated higher truncation levels (full details in App.~\ref{sec:appendix_durations_ablation}).
While results were almost all indistinguishable, 
we did find that using a larger duration ($D=44$) produced slightly better forecasts for ventilator counts.
Because the computational cost of a larger truncation limit is negligible (except for sampling each duration, runtime and storage scales with the number of segments not $D$), we recommend that future analyses use $D=44$ if possible.
Our public ACED-HMM implementation allows the user to set arbitrary upper bounds for each duration distribution via a plain-text configuration file.


\paragraph{Tractable parameterization of durations.}
To define each duration distribution, we could learn a separate $D-$dimensional probability vector $\pi^{s,h}$ for each state (indexed by stage $s$ and health $h$).
We instead explore a simpler formulation with only two scalar parameters per state: mode location $\lambda^{s,h} > 0$ and temperature $\nu^{s,h} > 0$:
\begin{align}
	[\pi_1^{s,h}, ~\ldots~, \pi_D^{s,h}] \gets 
	\text{softmax} \left(
		\frac{\log \text{PoiPMF}( 1 | \lambda^{s,h} )}{\nu^{s,h}},
		\ldots,
		\frac{\log \text{PoiPMF}( D | \lambda^{s,h} )}{\nu^{s,h}}
	\right).
	\label{eq:tractable_duration_two_parameter_family}
\end{align}
where $\text{PoiPMF}$ denotes the probability mass function of the Poisson distribution.
Fig.~\ref{fig:duration_dist_diagram} shows how possible distributions over $D$ days depend on these two parameters.
Intuitively, $\lambda^{s,h}$ controls the location of the most likely duration (mode).
Large temperatures $\nu^{s,h} \rightarrow \infty$ makes the distribution $\pi^{s,h}$ more uniform, while as $\nu^{s,h} \rightarrow 0$ the distribution becomes peaked at the mode. 
This parameterization encodes two valuable inductive biases: similar stay-lengths have similar probabilities and the overall distribution is either unimodal or flat.
This 2-parameter formulation has been previously used for supervised modeling of ordinal outcomes~\citep{beckhamUnimodalProbabilityDistributions2017}, but not to our knowledge for durations in a Markov model. We compare our formulation to other possibilities (truncated Poisson) in App.~\ref{sec:appendix_durations_ablation}.

\paragraph{Joint Probability Distribution.}
Each patient's entire hospital stay can be captured by a sequence of stage indicators, health indicators, and durations, with joint probability:
\begin{align}
p( s_{1:L_n}, h_{1:L_n}, \Delta_{1:L_n} )
	&= p_{\tau}( s_1, h_1 ) \prod_{\ell=2}^{L_n} p_\tau(s_\ell, h_\ell | s_{\ell-1}, h_{\ell-1} )
	\cdot \prod_{\ell=1}^{L_n} p_{\xi}(\Delta_{\ell} | s_{\ell}, h_{\ell} )
\end{align}
Here we've gathered all transition probabilities into $\tau$, such that $\tau = \{d, \rho\}$ and all duration parameters are denoted $\xi = \{\lambda, \nu\}$ for simplicity.
Notably, this model structure can be mapped to the semi-Markov model known as an explicit-duration (hidden) Markov model ~\citep{fergusonVariableDurationModels1980,mitchellModelingDurationHidden1993,yuHiddenSemiMarkovModels2010}, by mapping the health and stage indicator variable sequences $h_{1:L_n}$, $s_{1:L_n}$ to a discrete state sequence $z_{1:L_n}$
 with 9 possible states: the start state $\SS$, the two absorbing states $\XX, \RR$, and 6 intermediate states $\GG0, \GG1, \II0, \II1, \VV0, \VV1$ where the 0 (1) suffix in the state name represents the declining (recovering) variant of each stage, respectively.
The transition probability parameters for this HMM are specified in Supplementary Table~\ref{tab:transition_probas}.
Throughout the rest of the paper, we use the compact notation $\mathbf{z}_{n} = z_{n,1:L_n}$ below to represent the state sequence for patient $n$, which can be easily mapped back to the stage sequence $s_{n,1:L_n}$ and health sequence $h_{n,1:L_n}$.

\subsection{Prior beliefs about transition and duration parameters}

\textbf{Priors on transition parameters.}
	We can use published statistics to inform our prior beliefs about transition probability parameters $\tau = \{\rho, d\}$.
	Published probability statistics from the U.S. Centers for Disease Control (CDC) (see Table 2 of \citet{u.s.cdcCOVID19PandemicPlanning2020}) indicate the fraction of hospitalized COVID-19 patients who are admitted to ICU, placed on the ventilator, and eventually die, for each of three age groups (18-49, 50-64, and 65+).
	We translate this information into Beta priors for each $\rho_k$, so that each prior distribution has a mean matching these statistics while covering the overall range across age groups with non-trivial mass.
	See Appendix~\ref{sec:app_prior} for details.

	We next set the early death probabilities. Based on conversations with clinical experts, we expect the chance of early death in the general ward is very low ($\mathbb{E}[d_{\GG}] \approx 1\%)$ and the chance of early death in the ICU is only slightly larger ($\mathbb{E}[d_{\II}] \approx 2\%)$.
	We again set Beta priors to match these means with little uncertainty around them (such that the 95th percentile is below 0.05) to reflect our confidence that early death is possible but rare.
	
\paragraph{Priors on duration parameters.}
To determine priors over the durations spent in each state (indexed by $s,h$), we draw relevant parameters independently from a common prior:
\begin{align}
\lambda^{s,h} \sim \text{TruncNormal}( \mu^{s,h}, ( \sigma^{s,h} )^2, \text{lower}=0, \text{higher}=D),
\quad
\log_{10} \nu^{s,h} \sim \mathcal{N}(0.5, 0.5^2).
\label{eq:prior_on_durations}
\end{align}
Visuals of our prior's imposed distribution on durations in each state can be seen in Fig.~\ref{fig:posterior_visualization_MA}.

Our truncated normal prior for the mode location $\lambda^{s,h}$ of each state reflects that these quantities are difficult to know without reliable data about previous patient experience in the region of interest.
We select a mean of $\mu^{s,h} = 8$ days with a standard deviation $\sigma^{s,h} = 3$.
Under this Normal prior, the mode of each segment's duration distribution could plausibly range from 1 to 17 days (99\% of the  probability mass lies within the interval $\mu^{s,h} \pm 3\sigma^{s,h}$) .

\begin{figure}[!t]
    \centering
    \includegraphics[width=0.6\textwidth]{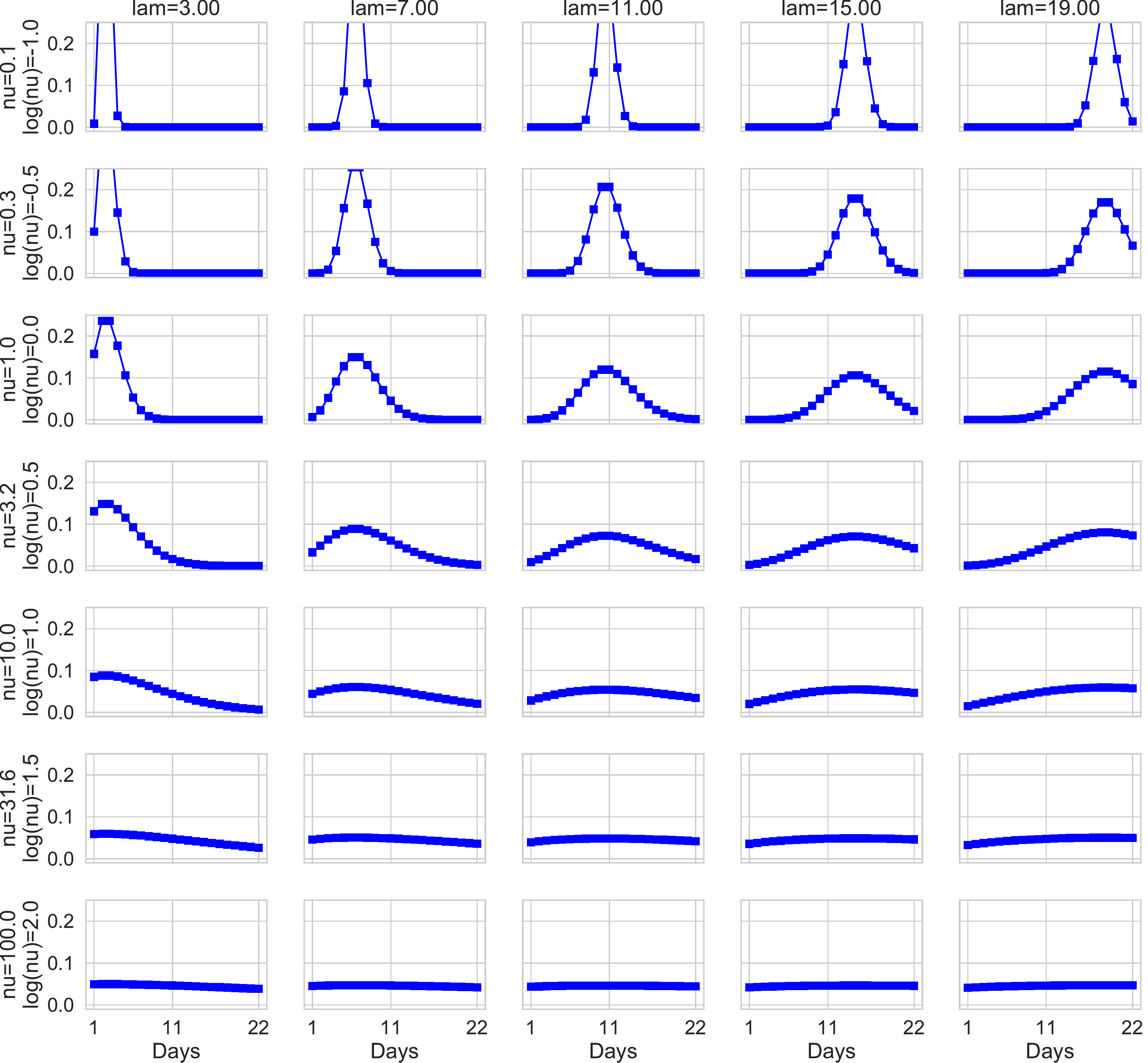}
\caption{
    Visualization of possible duration distributions under our tractable parameterization (Eq.~\ref{eq:tractable_duration_two_parameter_family}).
    Parameter $\lambda > 0$ (increasing across columns) controls the location of the mode, while temperature $\nu > 0$ (increasing top-to-bottom across rows) controls entropy around the mode.
}
    \label{fig:duration_dist_diagram}
\end{figure}

Our prior for each state's temperature $\nu^{s,h}$ assumes the \emph{log} of this value will be normally distributed. By using a prior for $\log_{10} \nu^{s, h}$ with mean 0.5 and standard deviation 0.5, we allow plausible temperatures to span both high and low entropy distributions (see Fig.~\ref{fig:duration_dist_diagram}).

\section{Estimating the Posterior Distribution over Parameters using ABC}
\label{sec:learning_abc}
In order to train our model so its parameters reasonably explain hospital trajectories in a specific region or site of interest, 
we assume we have access to observed daily counts $\mathbf{y}_{1:T} = \{y^k_{1:T} \}_{k \in \mathcal{K}}$ during the training period of interest.
Here the set of observed stages $\mathcal{K}$ could include any of the intermediate or terminal stages; typically $\mathcal{K} = \{\GG, \II, \VV, \XX\}$.
We also assume knowledge of the admission counts $a_{1:T}$ for the training period, which represent a total of $N$ patients who enter the hospital system, with  $N = \sum_{t=1}^T a_t $.

Given observed counts $\mathbf{y}_{1:T}, a_{1:T}$, we wish to be able to draw samples from the posterior $p( \tau, \xi, \mathbf{z}_{1:N}, \mathbf{\Delta}_{1:N} | \mathbf{y}_{1:T}, a_{1:T} )$.
Samples from this distribution allow us to properly understand the range of possible values for the model's transition parameters $\tau$ and duration parameters $\xi = \{\lambda, \nu\}$.
Samples of parameters can further allow us to produce \emph{forecasts}, as we can draw samples of counts $\mathbf{y}_{T+1:T+F}$ for a ``testing'' period of $F$ days -- $T+1, \ldots T+F$ -- by taking some assumed admissions $a_{T+1:T+F}$ for this period and sampling forward from the ACED-HMM.

For our model, it is easy to \emph{simulate} patient care sequences and to produce aggregate counts from these sequences (simply counting the total number of patients simulated in each stage at each day).
However, it is cumbersome to define an explicit probability mass function for both the individual patient care sequences $\mathbf{z}_n, \mathbf{\Delta}_n$ and the observed aggregate counts $\mathbf{y}_{1:T}$.
This would only become more difficult if we explore more complex extensions of our model (e.g. add age-dependent transition or duration probabilities).
We would like to simply define a \emph{simulation} that produces samples of aggregate counts, and then estimate a \emph{posterior distribution} over the parameters of this simulation that fits observed counts. 

\subsection{Approximate Bayesian Computation}

To fit posterior distributions, we turn to Approximate Bayesian Computation (ABC) ~\citep{marjoramMarkovChainMonte2003,marinApproximateBayesianComputational2012,cranmerFrontierSimulationbasedInference2020}, a method for fitting models that are naturally expressed as simulations. ABC has found applications in ecology~\citep{beaumontApproximateBayesianComputation2010}, population genetics~\citep{beaumontApproximateBayesianComputation2002}, and epidemiology~\citep{blumHIVContactTracing2010}.
To use standard Markov chain Monte Carlo methods in our application would require an explicit ability to compute the likelihood of observed counts arising from our model's simulations. 
ABC offers a simpler approach: we need only the ability to simulate counts and to decide if simulated counts are ``close enough'' to observed data.

Our approach is based on the ABC MCMC algorithm presented by~\citet{marjoramMarkovChainMonte2003}.
We iteratively propose new candidate values for parameters $\tau, \xi$, and then decide to accept or reject these based on an acceptance criterion designed to ensure the overall sequence of sampled parameters converges to the target posterior distribution. 
This looks superficially similar to conventional Metropolis-Hastings MCMC~\citep{metropolisEquationStateCalculations1953,andrieuIntroductionMCMCMachine2003}.
However, the test for acceptance uses an evaluation of a distance function $d$ between observed data $\mathbf{y}_{1:T}$ and our simulation's sampled counts $\mathbf{\tilde{y}}_{1:T}$.
This distance threshold check replaces the evaluation of an explicit likelihood (which would require specification of a probability mass function and the ability to evaluate it efficiently). 

An advantage of our ABC approach is that we do not need to compute or even define an explicit probability distribution for the likelihood $p(\mathbf{y}_{1:T} | \mathbf{z}_{1:N}, \mathbf{\Delta}_{1:N})$.
This allows us to avoid a potentially detrimental \emph{misspecified} parametric form chosen for computational convenience.
Another advantage of ABC is that we can  customize the distance function to meet application-specific needs; specifying a likelihood that achieves the same goals and is also a valid probability mass function might be more challenging.
For example, our chosen distance below emphasizes fitting the most recent daily counts in the training period (so our forecasts will generalize to the future well) and fitting later stages of the model well (since those stages are harder to estimate).


\paragraph{Distance computation.}
Let $\mathcal{K}$ be the set of observed stages, and $K$ the number of stages ($K = |\mathcal{K}|$). 
Let $y^{k}_{t}$ denote the observed count at time $t$ in stage $k$, and $\tilde{y}^{k}_{t}$ denote the simulation's sampled count.
We define the distance as a weighted mean absolute error:
\begin{align}
    d( \mathbf{y}_{1:T}, \mathbf{\tilde{y}}_{1:T} ) 
    &= \frac{1}{K \cdot T}
    	\sum_{t=1}^{T} \sum_{k=1}^{K} w_{tk} \frac{| y^k_{t} - \tilde{y}^k_{t}|}{\max(y^k_{t}, \tilde{y}^k_{t})}
    	\label{eq:weighted_distance_abc}
\end{align}

where $w_{tk}$ is a scalar weight that determines the relative ``importance'' of matching the counts at time $t$ and stage $k$.
We define $w_{tk}$ as a product of a time-specific weight $v_t$ and a stage-specific weight $u_k$.
For timestep weight $v_t$, we apply a recency bias: we linearly interpolate between $v_1 = 0.5$ at day 1 and $v_T = 1.5$ at day $T$, so that our model is better at making forecasts after day $T$ (and the average weight is 1).
We use stage-specific weights $u_k$ because preliminary experiments suggested the \textit{later} stages of the model are harder to fit (while all $N$ patients enter the general ward, only a smaller fraction will reach the ICU or the ventilator).
Thus, we set $u_k$ values so later stages are worth more.
See App.~\ref{sec:appendix_abc} for further details and justification of our chosen distance, and App.~\ref{sec:appendix_distance_ablation} for ablation studies.

 
\paragraph{Proposal distributions $q$.}
Our ABC algorithm requires a procedure to draw a new plausible parameter value given a current value.
For each type of parameter (transition probabilities, duration modes, duration temperatures), we define a standard ``random walk'' proposal distribution $q$ with the mean at the previous value and a variance that encourages modest exploration. See Appendix~\ref{sec:appendix_abc} for parameter-specific details.

\paragraph{ABC algorithm.}
Initial values for the transition parameters $\tau$ and duration parameters $\xi$ are samples from our chosen priors $p(\tau)$ and $p(\xi)$ defined above.
Until convergence, the algorithm visits each individual parameter in turn, either a specific transition probability parameter $\rho_s$ or $d_s$ within $\tau$, or a state-specific duration mode $\lambda^{s,h}$ or temperature $\nu^{s,h}$ within $\xi$.
At each visit, the algorithm draws a candidate value from the proposal distribution $q$ defined above.
Then, we decide to ``accept'' or ``reject'' the candidate.
Denote the current parameters as $\tau, \xi$, and new candidate parameters (with one entry updated) as $\tau^*, \xi^*$.

There are two stages of acceptance.
First, counts are simulated from our ACED-HMM progression model for the entire training period using the proposed parameters $\tau^*, \xi^*$, generating $\mathbf{\tilde{y}}_{1:T}$.
Then, we compute the distance $d$ between the samples $\tilde{\mathbf{y
}}_{1:T}$ and true counts $\mathbf{y}_{1:T}$.
If the distance $d$ is below a chosen tolerance threshold $\varepsilon$, the algorithm moves to the second stage of acceptance; otherwise the proposal is rejected.
In stage two, the proposal is accepted if a random uniform draw between 0 and 1 is less than the acceptance ratio  $\alpha = \frac{p(\tau^*, \xi^*)q(\tau^*, \xi^* \rightarrow \tau, \xi)}{p(\tau, \xi)q(\tau, \xi \rightarrow \tau^*, \xi^*)}$.
The first stage of acceptance checks the fit to the data (replacing the likelihood used in conventional MCMC), whereas the second stage further filters the acceptance via information from the prior and from the proposal, ensuring the Markov chain will converge to the target posterior distribution~\citep{marjoramMarkovChainMonte2003}.

\paragraph{Scheduling the distance tolerance threshold.}
Because random initial parameters are unlikely to explain observed data well, we find it useful to start with a lenient value of the data-agreement threshold $\varepsilon$, i.e. one that accepts all proposals. We then decay $\varepsilon$ across iterations toward a smaller value that better ensures fit to observed data.
In App.~\ref{sec:appendix_abc}, we describe two phases for our ABC algorithm: 
a \textit{burn-in} phase and a \textit{sampling} phase.
During \textit{burn-in}, we gradually decay the value of $\varepsilon$ over many iterations to reach a good explanation of the data.
Then, during \textit{sampling} we hold $\varepsilon$ fixed and collect parameter samples, which are treated as representative of the target posterior distribution.
During burn-in, we find it helpful to periodically \emph{increase} $\varepsilon$ sharply and then decay gradually to escape local optima.
See Fig.~\ref{fig:distances_plot_south_tees} for a visual representation of escaping a local optima using this idea.

\paragraph{Ensembling.}
To obtain robust estimates of the posterior, we combine samples from multiple independent runs of our ABC algorithm into an ensemble.
For any given dataset, we expect there might be several different parameter configurations that explain the training counts well (due to the flexibility of the model and possible identifiability issues).
Ensembling allows us to capture more of these configurations than a single run of ABC might recover.
By aggregating, we gain statistical strength in the forecasts and a deeper understanding of the uncertainty in learned parameter values.

\paragraph{Synthetic validation.} We conducted several experiments on synthetic ``toy'' datasets (generated by our model) to verify that our ABC procedures can (approximately) recover plausible posterior distributions. See Appendix~\ref{sec:synthetic_data} for a thorough description.

\section{Data for Case Studies}
We fit our proposed model to pursue two case study applications over a range of national contexts and geographic scales.
First, as a primary use case we pursued \emph{regional} forecasts for several states in the United States.
State-level modeling allows us to make forecasts that interest policy makers and public health officials within that state.
We also show how such regional forecasts might be used to assess societal value.
Second, we performed \emph{site-level} forecasts for specific hospitals in the United Kingdom (UK).
These experiments allow us to assess potential utility to hospital administrators for planning purposes.
  

Below, we describe the data acquired for training and evaluating our forecasting model in both cases.
Each dataset is assumed to provide census counts for the training period, denoted as days $1, 2, \ldots T$, for some subset or additive combination of the stages of our model (general ward $\GG$, off-ventilator in ICU $\II$, on-ventilator in ICU $\VV$, terminal death $\XX$, recovered $\RR$).
To evaluate forecasts against a ``ground truth'', we further require the same counts for the future period of interest, starting on day $T+1$ and ending on $T+F$.

In order to assess our model, during both training and forecasting we assume access to true daily \emph{admissions} counts to the general ward.
For a real forecast after day $T$, these counts will not be known.
We leave this complex scenario to future work. We could use hand-designed projections representing plausible future admission scenarios from clinical experts, or develop a forecasting model to project admissions (as in \citet{qianCPASUKNational2020}).


\subsection{U.S. state hospital data for regional forecasts}

We obtain hospital utilization counts for four US states: Massachusetts (MA), Utah (UT), South Dakota (SD), and California (CA).
We use counts from November 11th, 2020 to January 11th, 2021 as training data, and forecast one month into the future.
We selected these states to represent diverse geographic regions as well as diverse hospital utilization dynamics and policy responses to the pandemic during the selected period.
Hospitalizations in SD peaked in late November, in MA in early January, and in UT in late January.

For all states, we use data from the COVID Tracking Project~\citep{theatlanticmonthlygroupCOVIDTrackingProject2021}\footnote{\tiny \url{https://covidtracking.com/data}} and the \citet{u.s.dept.ofhealthandhumanservicesCOVID19ReportedPatient2021}\footnote{\tiny \url{https://healthdata.gov/Hospital/COVID-19-Reported-Patient-Impact-and-Hospital-Capa/g62h-syeh}}. For each state we have access to a daily time series of General Ward ($\GG$) counts, counts in the ICU off the ventilator ($\II$), counts on the ventilator ($\VV$) and terminal death counts ($\XX$).
For Utah and California in particular, we have access only to aggregate ICU counts ($\II +\VV$)  that do not distinguish between on and off the ventilator.
For all states, we found that \emph{smoothing} the terminal counts produced more sensible data (in the raw data, some weekends and holidays report implausibly low counts for deaths that reflect imperfect reporting processes).
 See Appendix~\ref{sec:app_datasets} for data collection and preprocessing details.

\subsection{United Kingdom hospital data for site-level forecasts}

We obtain daily occupancy counts at different care stages from two U.K. hospitals: South Tees and Oxford University, made available by the UK National Health Service\footnote{\tiny \url{https://www.england.nhs.uk/statistics/statistical-work-areas/covid-19-hospital-activity/}}.
We use counts from November 3rd to January 3rd as training data, and forecast one month into the future (counts past Feb. 3rd were unavailable when we ran the experiments). 
We selected these two hospitals because they represented large daily volumes with fewer data irregularities than other alternatives.
After data collection (see details in Appendix~\ref{sec:app_datasets}), for each site we obtained daily counts for total beds used ($\GG+\II+\VV$), ventilators used ($\VV$), and recovered patients discharged from the hospital (stage $\RR$ of our model).

\section{Results of Case Studies}
\label{sec:results}
We now describe several experimental results that assess our model's ability to forecast accurately as well as its scalability, sensitivity to available data, and ability to answer what-if questions. 

\subsection{Forecast assessment: Can this model capture qualitative trends?}

\begin{figure}[t!]
    \centering
    \includegraphics[width=0.8\textwidth]{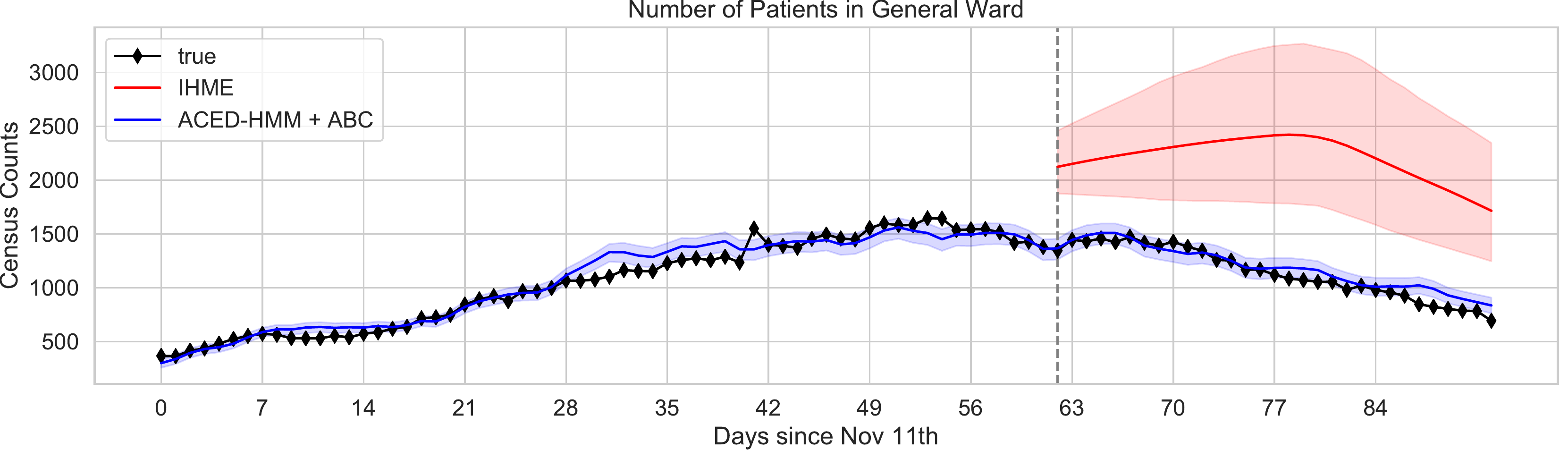}
    \includegraphics[width=0.8\textwidth]{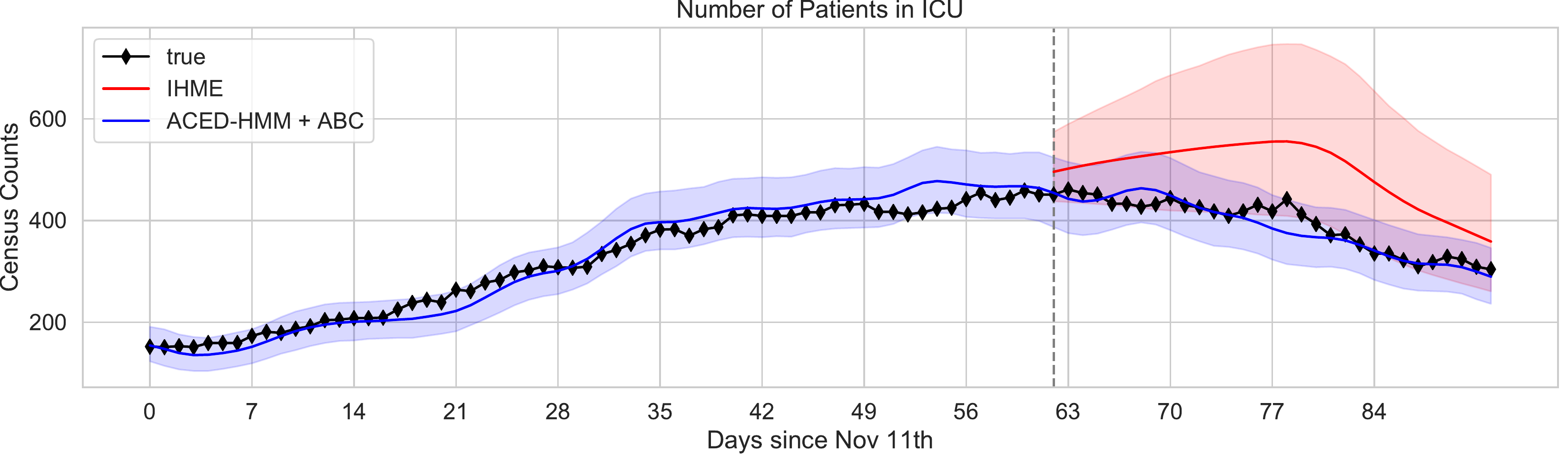}
    \caption{\textbf{Fit and forecasts on General Ward ($\GG$) and Total ICU ($\II + \VV)$ counts for Massachusetts.}
   Using ABC, we estimate a posterior over model parameters from 2 months of training counts (Nov. 11 2020 - Jan. 11, 2021, left of the dashed line).
    We stress that our ABC method is provided true daily admission counts for all days (0 - 92).
    We also include forecasts from IHME released on Jan. 15 2021. See App.~\ref{sec:appendix_ihme} for details about comparing IHME's forecasts to ours.
    For our method, shaded intervals show the 2.5th and 97.5th percentiles of 2000 posterior samples.
    For IHME, shared intervals show their published ``upper'' and ``lower'' forecasts.
    }
\label{fig:forecast_MA}
\end{figure}

\begin{figure}[h!]
    \centering
    \includegraphics[width=0.8\textwidth]{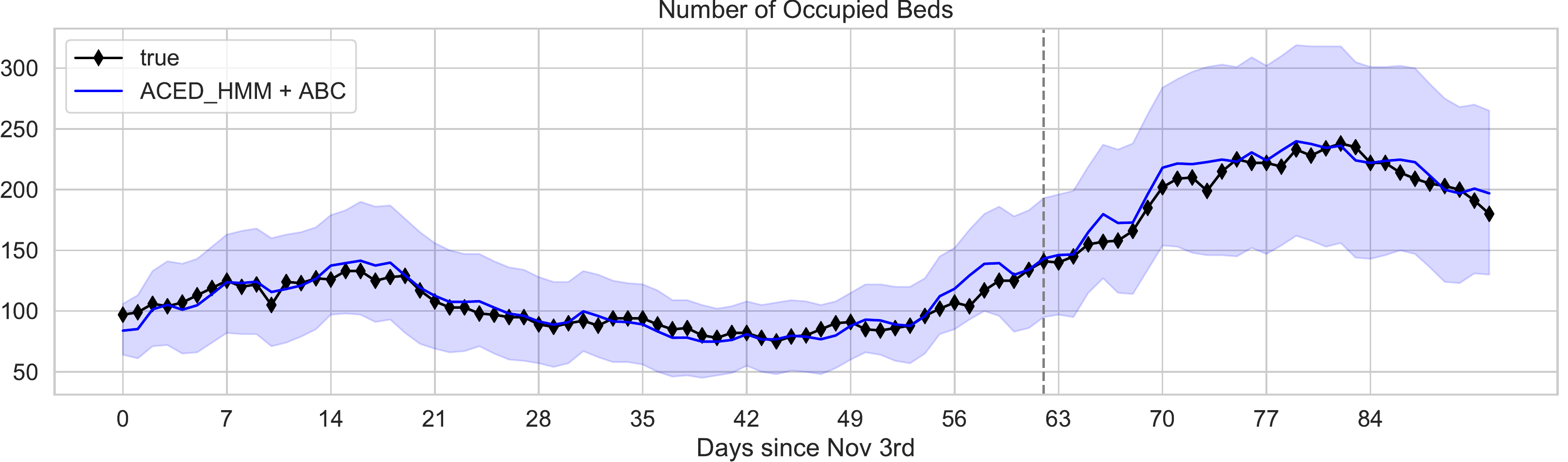}
    \includegraphics[width=0.8\textwidth]{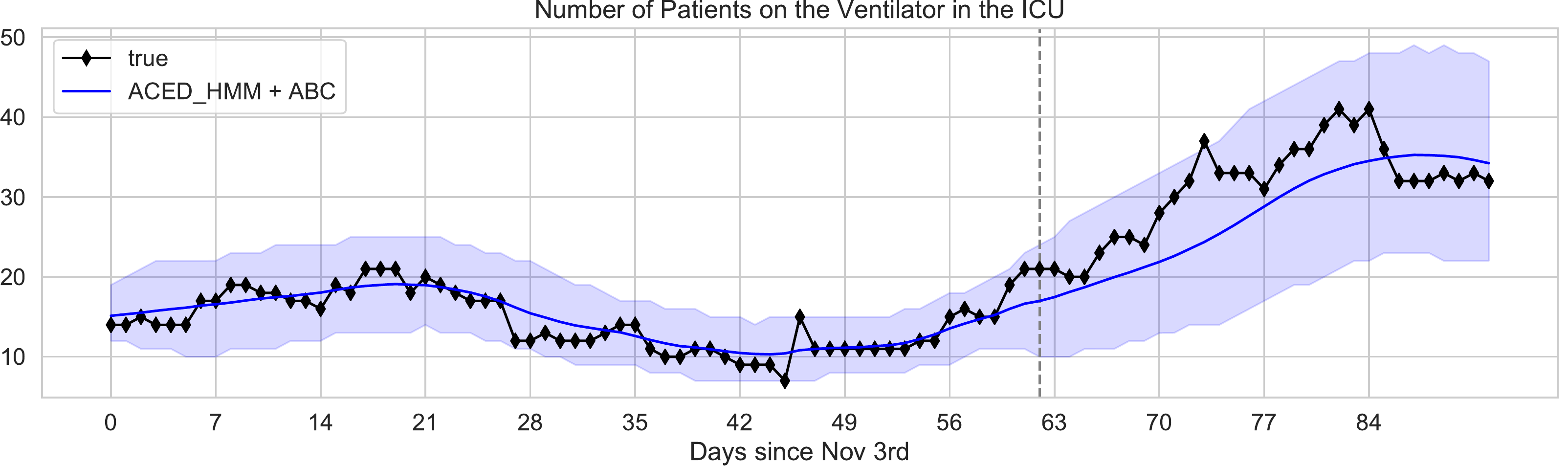}
    \caption{\textbf{Fit and forecasts on Occupied Beds ($\GG + \II + \VV$) and Ventilator ($\VV$) counts for South Tees hospital in the UK.} Parameters were selected via ABC to fit 2 months of counts (Nov. 3 '20 - Jan. 3, '21, left of the dashed line). Shaded intervals show the 2.5th and 97.5th percentiles of 2000 posterior samples.}
\label{fig:forecast_SouthTees}
\end{figure}

We show our method's forecasted distribution over daily counts for Massachusetts in Fig.~\ref{fig:forecast_MA}.
Importantly, our ACED-HMM mechanistic model (blue lines), trained during a period of \emph{rising} counts, can reasonably forecast the \emph{downward} trend in hospital counts seen in MA after mid-January 2021. 
The figure also shows the published forecasts for MA from IHME~\citep{reinerModelingCOVID19Scenarios2021} made on Jan. 15th. The IHME model is primarily trained to match death counts from the training period and to our knowledge is not informed by state-specific hospital census counts. Consequently, we see large systematic overestimates.
We do stress that our model incorporates actual admissions data (which IHME does not apparently use), so this is not quite a fair comparison (see App.~\ref{sec:appendix_ihme} for thorough discussion), but it is still a compelling demonstration of benefits of our mechanistic model.



We further examine a forecast for a specific hospital site in the UK (``South Tees'') in Fig.~\ref{fig:forecast_SouthTees}.
Here, we see how our method can capture \emph{rising} trends during the evaluation period for both overall occupied beds (the sum of several stages: $\GG + \II + \VV$) as well as the number of ventilators in use (stage $\VV$), despite the fact that during the training period the counts were mostly flat and had only recently begun to rise.

Together, these qualitative results indicate that our model's learned transition and duration distributions can successfully \emph{translate} several weeks into the future in some cases.
Again, we stress that our model incorporates the true admission counts for the testing period.
However, the accurate ventilator demand prediction shown in Fig.~\ref{fig:forecast_MA}-\ref{fig:forecast_SouthTees}  suggests that our ACED-HMM's transition and duration modeling is reasonably robust.

\subsection{Quantitative assessment: How accurate are forecasts?}
\newcommand{\intervalname}{2.5-97.5\% interval on MAE}
\newcommand{\ihmeintervalname}{MAE of lower - MAE of upper}
\begin{table}[!t]
\centering
\resizebox{\textwidth}{!}{%
\texttt{%
\begin{tabular}{c|r|r r|r r|r r|r r| }
  & 
    & $\GG$~ & \textnormal{\bf InGeneralWard}
    & $\II + \VV$ & \textnormal{\bf InICU}
    & $\VV$~ & \textnormal{\bf OnVentInICU}
    & $\XX$~ & \textnormal{\bf Death (smooth)}
    \\ 
  & \textnormal{\bf Method}
    & MAE~ & lower - upper~
    & MAE~ & lower - upper
    & MAE~ & lower - upper
    & MAE~ & lower - upper
 \\ \hline \hline
 \multirow{8}{*}{MA}
  & Median of train set
                & 211.2~ & NA~~~~~~~~~ 
                & 73.5~ & NA~~~~~~~~
                & 80.1~ & NA~~~~~~~~
                & 17.0~  &  NA~~~~~~~~ \\   
  & \cellcolor{gray!15} Bayes LR (day)
                & \cellcolor{gray!15} 931.3~ & \cellcolor{gray!15} 927.2 - ~935.3
                & \cellcolor{gray!15} 187.2~~& \cellcolor{gray!15} 186.5 - 188.0  
                & \cellcolor{gray!15} 103.0~~& \cellcolor{gray!15} 102.7 - 103.3
                & \cellcolor{gray!15} 26.9~ &  \cellcolor{gray!15} 26.7 - 27.2~~\\   
  & Bayes LR (day + adm.)
                & 277.8~ & 275.6 - ~280.5  
                & 24.3~~& 23.8 - ~24.8  
                & 24.3~~& 23.9 - ~24.8
                &  8.0~ &  7.8 - ~8.2~~\\ 
  & \cellcolor{orange!25} ACED-HMM + ABC
                &  \cellcolor{orange!25}  65.0~ &  \cellcolor{orange!25} ~61.5 - ~~68.9  
                &   \cellcolor{orange!25}  15.7~~&  \cellcolor{orange!25}  ~14.5 - ~16.9 
                &  \cellcolor{orange!25}  34.0~~&  \cellcolor{orange!25} ~32.6 - ~35.7
                &  \cellcolor{orange!25} ~8.0~ &    \cellcolor{orange!25} 7.8 - ~8.3~~\\
  & IHME
                &1066.1~ & *543.4 - 1754.9 
                & 104.0~~& *15.6 - 257.9  
                &  25.7~~&  *41.4 - 106.5
                & 21.6~ &  *8.7 - 49.7~~\\                
   & \cellcolor{gray!15} AR-Poisson
                & \cellcolor{gray!15}498.0~ & \cellcolor{gray!15}438.3 - ~553.6  
                & \cellcolor{gray!15}162.2~~&\cellcolor{gray!15}148.6 - 173.5  
                & \cellcolor{gray!15}128.3~~&\cellcolor{gray!15} 118.5 - 138.6
                &  \cellcolor{gray!15}34.8~ &\cellcolor{gray!15}  28.4 - 42.0~~\\ 
  & ACED-HMM + Prior
                & 868.6~ & 779.6 - ~945.8  
                & 293.6~~&272.6 - 313.3  
                & 119.6~~& 113.8 - 124.8
                &  40.7~ &  39.1 - 42.2~~\\
  \cline{2-10}
  & Mean test-set $y$
  				& 1141.6~ &                
  				& 392.5~~&               
  				&  249.1~ &              
  				& 66.5~ & \\ 
 \cline{2-10}
 \hline \hline
 \multirow{6}{*}{SD}
  & Median of train set
                & 234.0~ & NA~~~~~~~~~ 
                & 55.1~ & NA~~~~~~~~ 
                & 23.9~ & NA~~~~~~~~ 
                & ~8.2~  &  NA~~~~~~~~ \\   
  & \cellcolor{gray!15} Bayes LR (day)
                & \cellcolor{gray!15}55.0~ &\cellcolor{gray!15} 54.3 - ~~55.8
                & \cellcolor{gray!15}5.8~ &\cellcolor{gray!15} ~5.5 - ~~6.1  
                & \cellcolor{gray!15}7.0~~&\cellcolor{gray!15} 6.7 - ~~7.2
                & \cellcolor{gray!15}2.6~ &\cellcolor{gray!15}  2.4 - ~2.8~~\\   
  & Bayes LR (day + adm.)
                & 53.1~ & 52.6 - ~~53.9
                & 4.3~ & 4.2 - ~~4.5
                & 5.1~ & 4.9 - ~~5.3
                & 2.8~ &  2.7 - ~2.9~~\\ 
  & \cellcolor{orange!25} ACED-HMM + ABC
                &  \cellcolor{orange!25} 9.2~ &\cellcolor{orange!25}   8.5 - ~~10.1  
                &  \cellcolor{orange!25} 4.5~ &\cellcolor{orange!25} ~4.3 - ~~4.8  
                &  \cellcolor{orange!25}4.8~ &\cellcolor{orange!25}  4.5 - ~~5.1 
                & \cellcolor{orange!25}~2.8~ & \cellcolor{orange!25} 2.7 -~~2.9~ \\ 
  & IHME 
                &  71.8~ &   *6.6 - ~166.4 
                &  17.7~ & *4.2 - ~46.0  
                &  5.9~ & *6.3 - ~20.6 
                & ~2.5~ & *3.6 -~~5.0~ \\ 
 \cline{2-10}
  & Mean test-set $y$
  				& 102.5~ & 				  
  				& 34.9~  & 			   
  				& 23.6~ &  		      
  				& 7.8~  & \\
 \cline{2-10}
 \hline \hline
 \multirow{6}{*}{UT}
  & Median of train set
                & 75.9~ & NA~~~~~~~~~ 
                & 42.1~ & NA~~~~~~~~ 
                & NA~ & 
                & ~2.7~  &  NA~~~~~~~~ \\   
  & \cellcolor{gray!15} Bayes LR (day)
                & \cellcolor{gray!15} 68.7~ &\cellcolor{gray!15} 67.9 - ~~69.5
                & \cellcolor{gray!15} 33.6~ & \cellcolor{gray!15} ~33.1 - ~34.2  
                & \cellcolor{gray!15} NA~ & \cellcolor{gray!15}
                & \cellcolor{gray!15} 2.7~ & \cellcolor{gray!15}  2.7 - ~2.8~~\\   
  & Bayes LR (day + adm.)
                & 63.4~ & 62.8 - ~~63.9
                & 32.6~ & 32.1 - ~33.1
                & NA~ & 
                & 2.6~ &  2.5 - ~2.7~~\\ 
  & \cellcolor{orange!25} ACED-HMM + ABC
                &  \cellcolor{orange!25} 20.8~ & \cellcolor{orange!25}  20.2 - ~~21.3 
                & \cellcolor{orange!25} 18.2~ & \cellcolor{orange!25} 17.4 - ~19.2  
                & \cellcolor{orange!25} NA~   & \cellcolor{orange!25}            
                & \cellcolor{orange!25} ~2.4~ & \cellcolor{orange!25}  2.3 -~~2.5~ \\
  & IHME
                & 332.9~ &*142.6 - ~578.4 
                & 105.0~ &*35.7 - 213.9  
                &  NA~   &             
                & ~7.4~ & *2.7 -~15.7~ \\            
  \cline{2-10}
  & Mean test-set $y$ 
  				& 272.9~ &               
  				& 164.3~ &               
  				& NA~   &             
  				& 11.7~ &
  \\ \hline
\end{tabular}
}
}
\caption{\textbf{Quantitative error assessment on 3 U.S. states.} We measure error with respect to the observed counts during the test period (Jan. 12 - Feb. 11, 2021) at each possible care stage: general ward $\GG$, in ICU (including on and off ventilator) $\II + \VV$, on ventilator in ICU $\VV$, and \emph{smoothed} death counts $\XX$).
For all states and stages, we report the mean count $y$ across the testing period to give a sense of scale.
Cells marked NA were not available for that state (see discussion in App.~\ref{sec:app_datasets}).
AR-Poisson baseline is described in App.~\ref{sec:appendix_gar}, IHME baseline in App.~\ref{sec:appendix_ihme}.
We communicate uncertainty by reporting lower and upper estimates of the MAE.
To obtain these intervals for our ACED-HMM and other probabilistic baselines, we repeat all MAE computations across 100 separate batches of posterior samples and then report the 2.5th and 97.5th percentiles across these batches.
For the IHME baseline, a distribution of forecasted samples is not published.
Instead, we evaluate and report the MAE of the provided ``lower'' and ``upper'' forecasts.
This second kind of interval (marked with *) should not be directly compared to the first, as they capture different aspects of uncertainty.
}
\label{tab:error_metrics_US}
\end{table}

We quantitatively evaluate our forecasts for 3 US states and 2 UK hospitals.
For all U.S. states, we compare to forecasts from the IHME model~\citep{reinerModelingCOVID19Scenarios2021}.
We further compare to non-mechanistic Bayesian linear regression models that treat observed counts in each stage as linear functions of the day index or the day plus the last 3 weeks of admissions counts.
Finally, we tried a non-mechanistic autoregressive model with Poisson likelihood designed for single hospital forecasting~\citep{leeForecastingCOVID19Counts2021}, and an ACED-HMM using our prior (not a posterior fit to the data). Because these last two models performed poorly on two sites, we did not apply them to others.

First, we calculate the Mean Absolute Error (MAE) between each method's mean prediction and the observed counts during the testing period.
See MAE results for the US in Table~\ref{tab:error_metrics_US}.
For Massachusetts, our ACED-HMM achieves a noticeably better MAE of 65 for general ward (stage $\GG$), while all other methods have MAE values above 211.
In addition to this relative gain, we find that on a typical day, our prediction is within 4-10\% of the actual general ward count (on average during the test period MA hospitals had 1141.6 patients in the general ward (range 694-1474).
Thus, our general ward predictions are sufficiently accurate to be operationally useful for planning.
Similar MAE improvements for our method can be found for ICU beds in MA; for ventilator and death counts our method is competitive but not superior to others.
For the states of SD and UT, each much smaller than MA, our mean absolute error (MAE) across all stages remained below 21, while the MAE values for other methods are considerably worse (e.g. the IHME baseline exceeds 100 MAE for the ICU in UT and exceeds 71 MAE for the general ward in SD).
These results illustrate that our method can produce accurate forecasts across several diverse scenarios. Further results for the UK showing similar advantages for our ACED-HMM can be found in App.~\ref{sec:appendix_mae}.

Second, we assess the \emph{coverage} of the learned posterior distribution across the 50\%, 80\% and 95\% centered intervals of the posterior.
For an ideal, well-calibrated posterior, we should observe that X\% of the actual counts in the testing period fall within an interval chosen to cover X\% of the distribution's mass.
See results for the US and the UK in Appendix~\ref{sec:appendix_coverage}.
We find the coverage of all methods could be improved substantially.
For example for MA the interval our ACED-HMM predicts will cover 95\% of data for stages $\GG$ and $\II$ actually covers only 68\% and 65\% of observed counts.
Improving coverage is challenging and an important area for future work.



\subsection{Interpretation of learned posteriors over parameters}

\newcommand{\PWW}{0.24} 
\setlength{\tabcolsep}{0.1cm}
\begin{figure}[t!]
\centering
\begin{tabular}{c c | c c}
	\includegraphics[width=\PWW\textwidth]{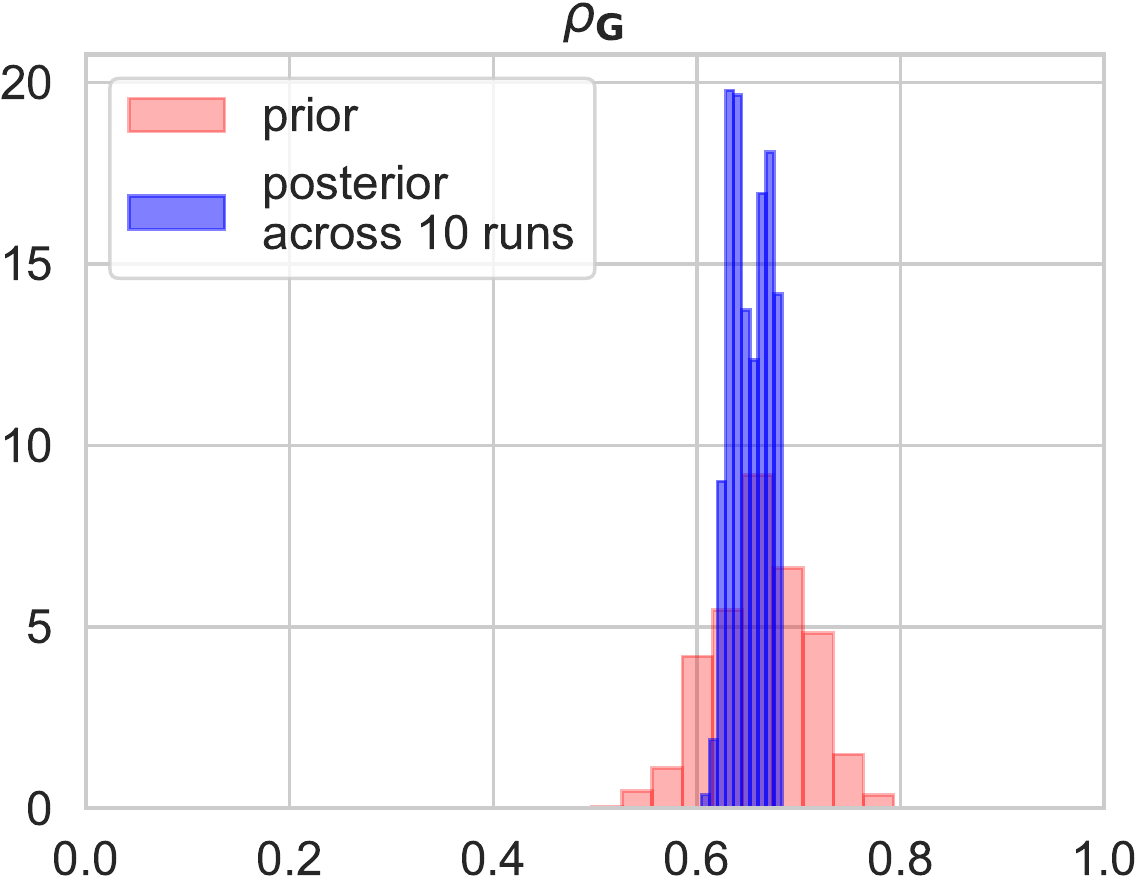}
	& 
	\includegraphics[width=\PWW\textwidth]{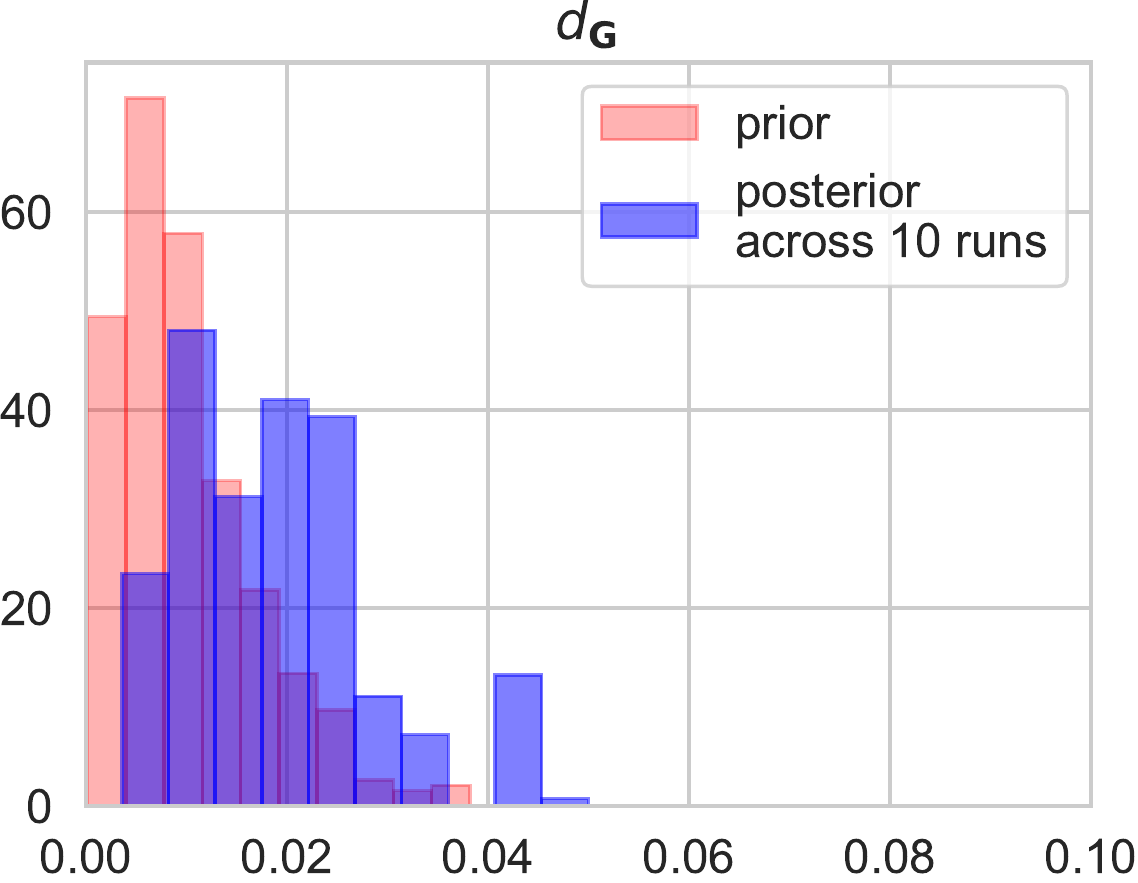}
	&
	\includegraphics[width=\PWW\textwidth]{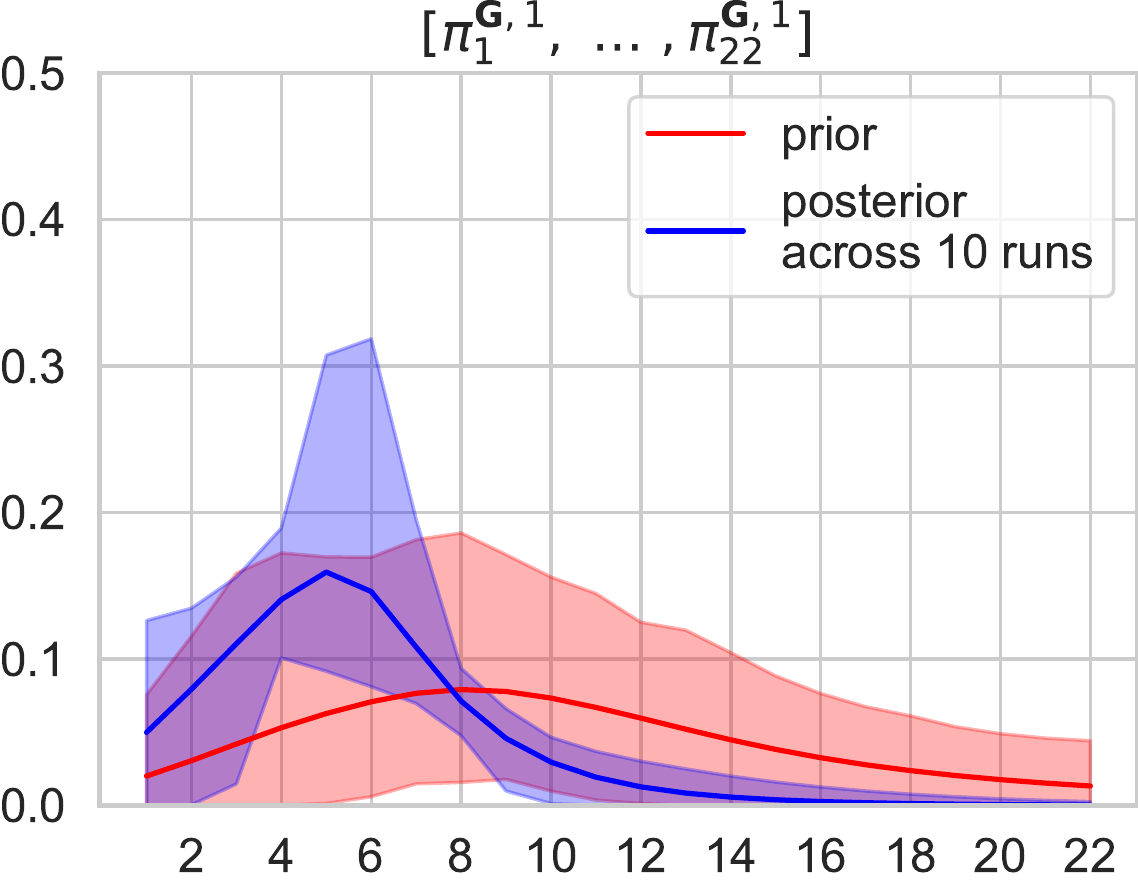}
	&
		\includegraphics[width=\PWW\textwidth]{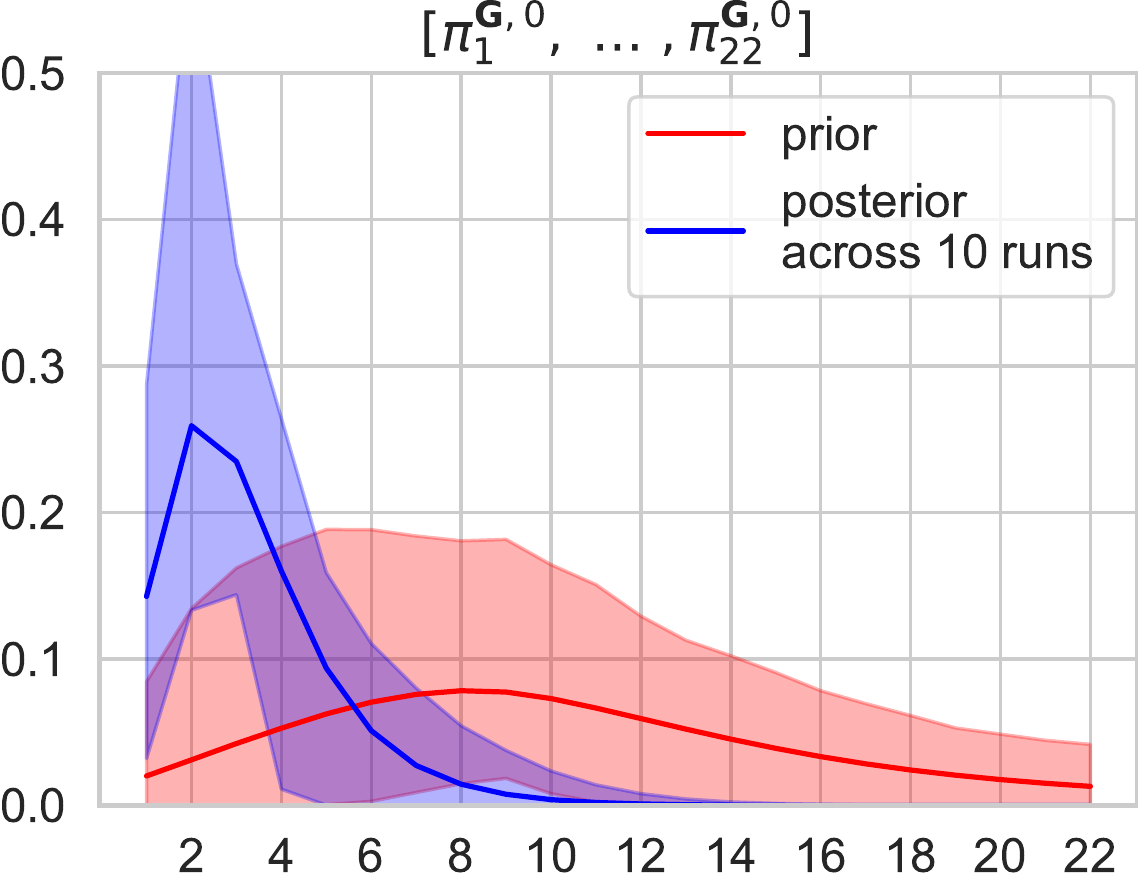}
\\
	\includegraphics[width=\PWW\textwidth]{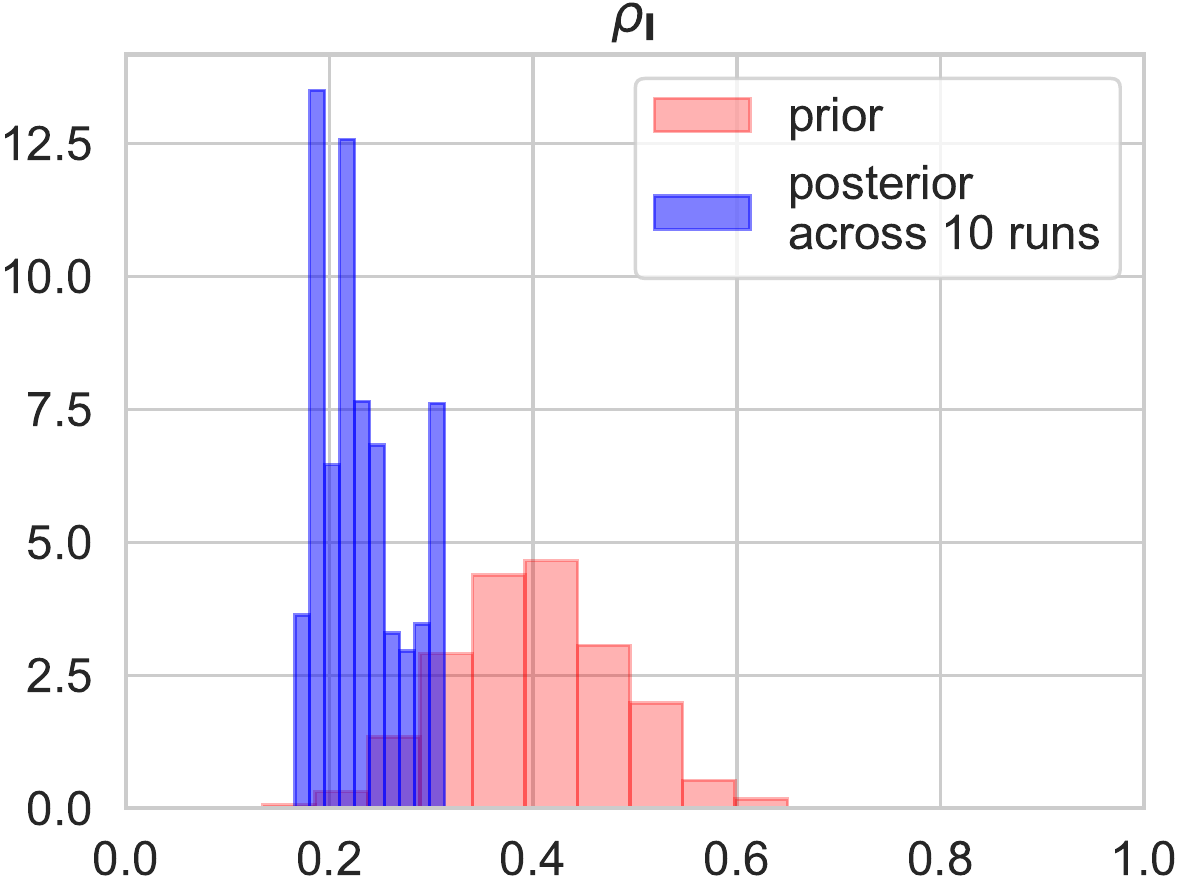}
	& 
	\includegraphics[width=\PWW\textwidth]{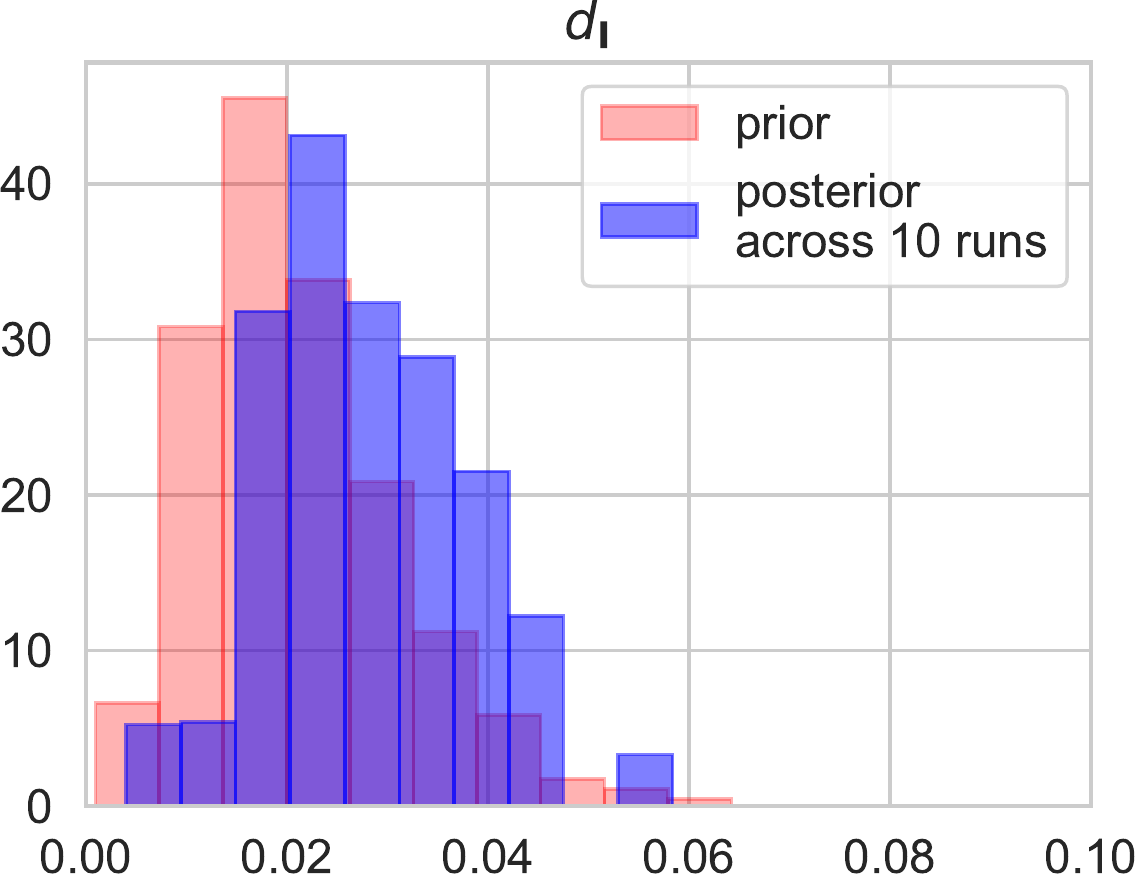}
	&
	\includegraphics[width=\PWW\textwidth]{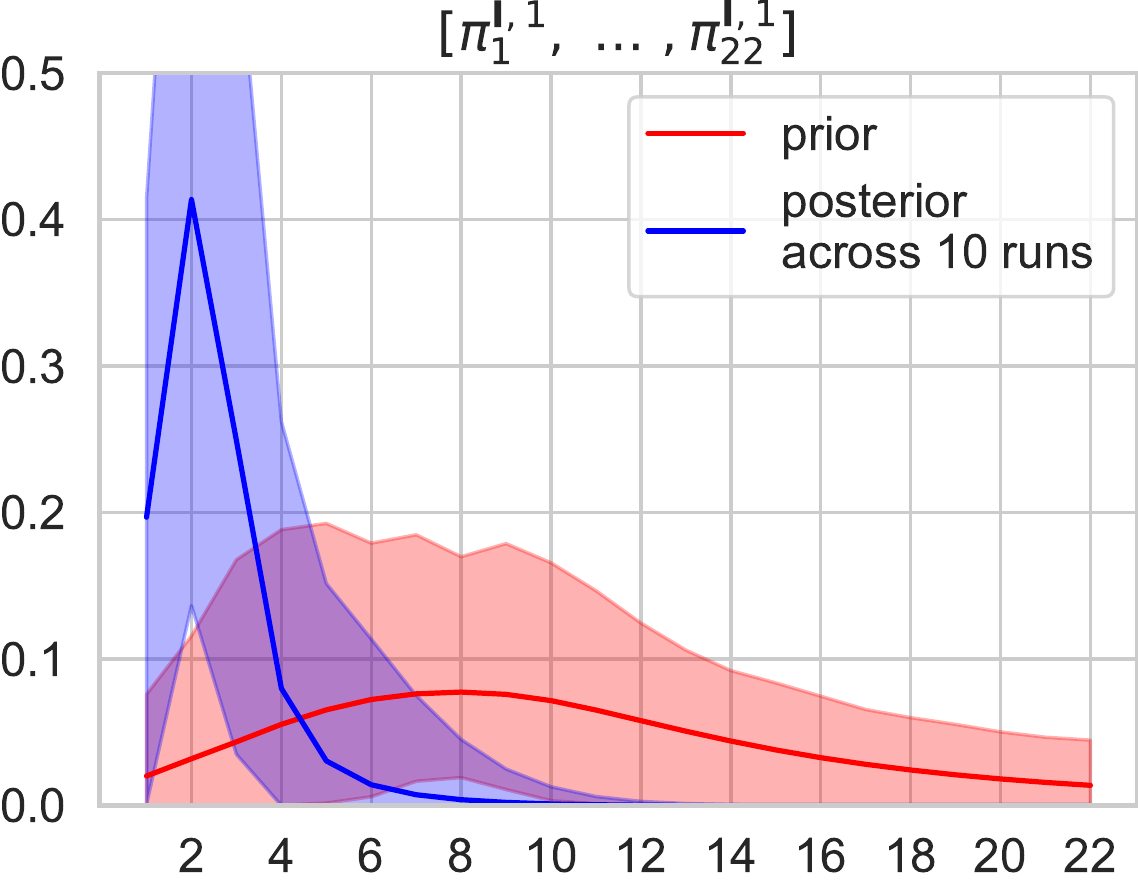}
	&
	\includegraphics[width=\PWW\textwidth]{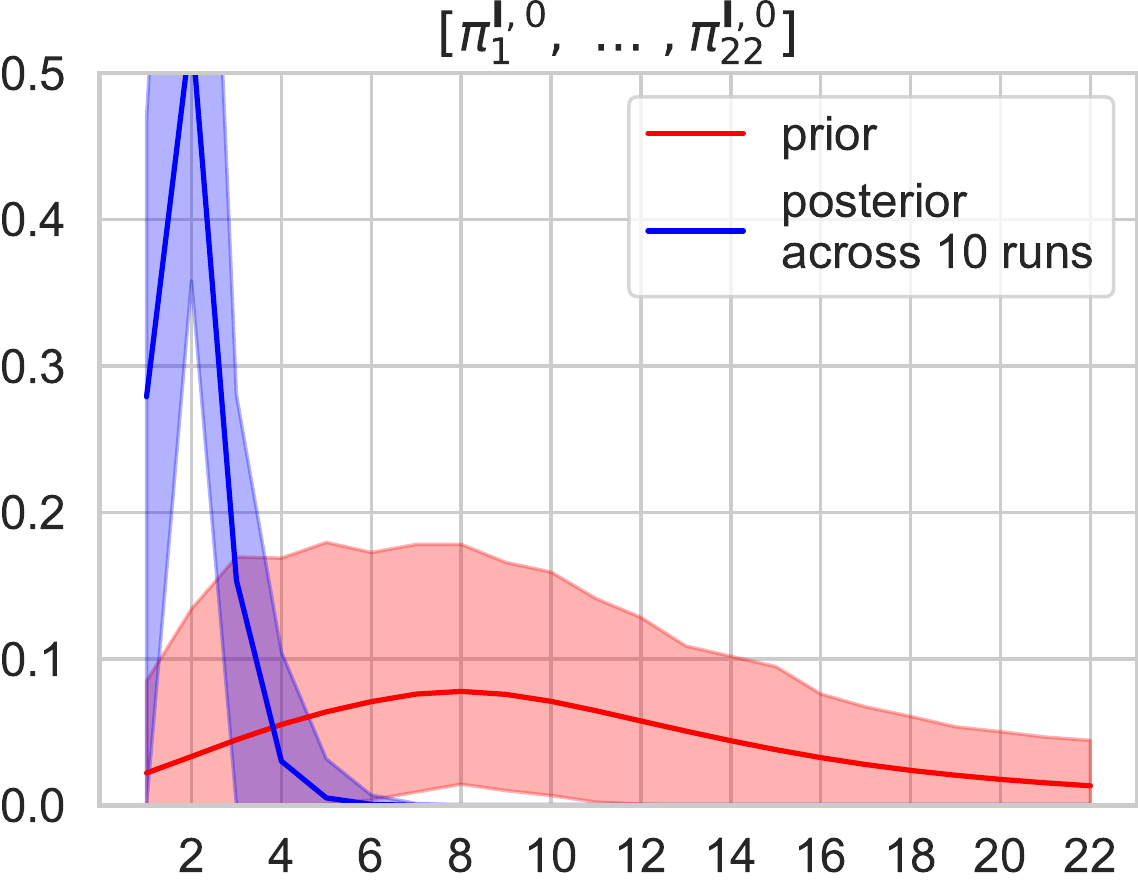}
\\
	\includegraphics[width=\PWW\textwidth]{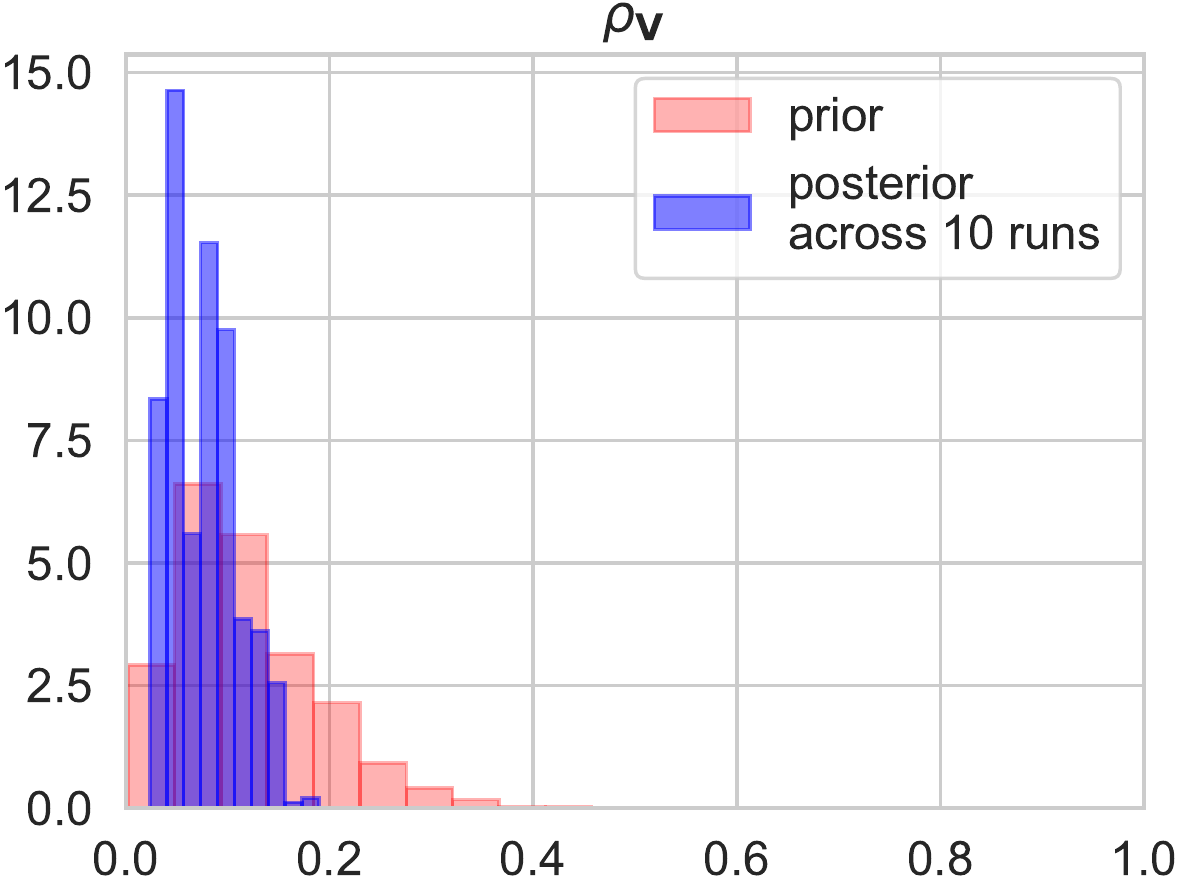}
	& 
	&
	\includegraphics[width=\PWW\textwidth]{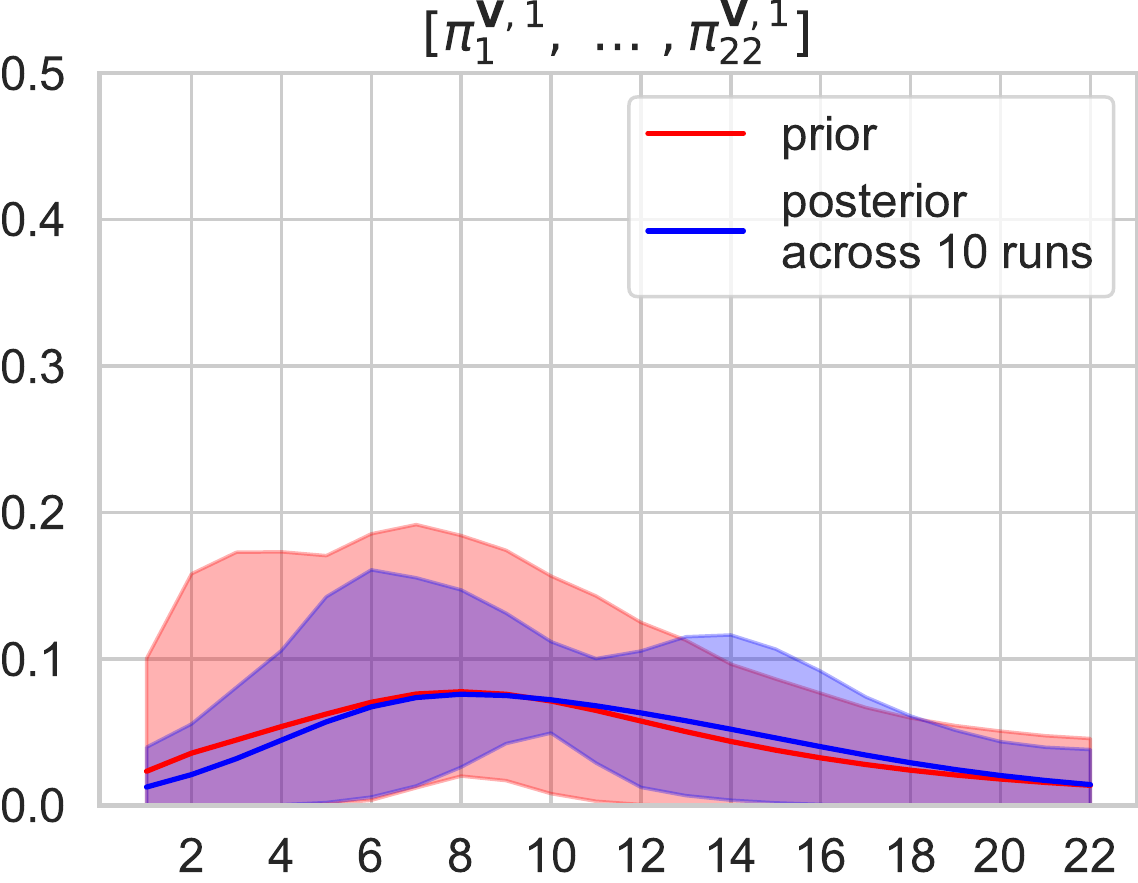}
	&
	\includegraphics[width=\PWW\textwidth]{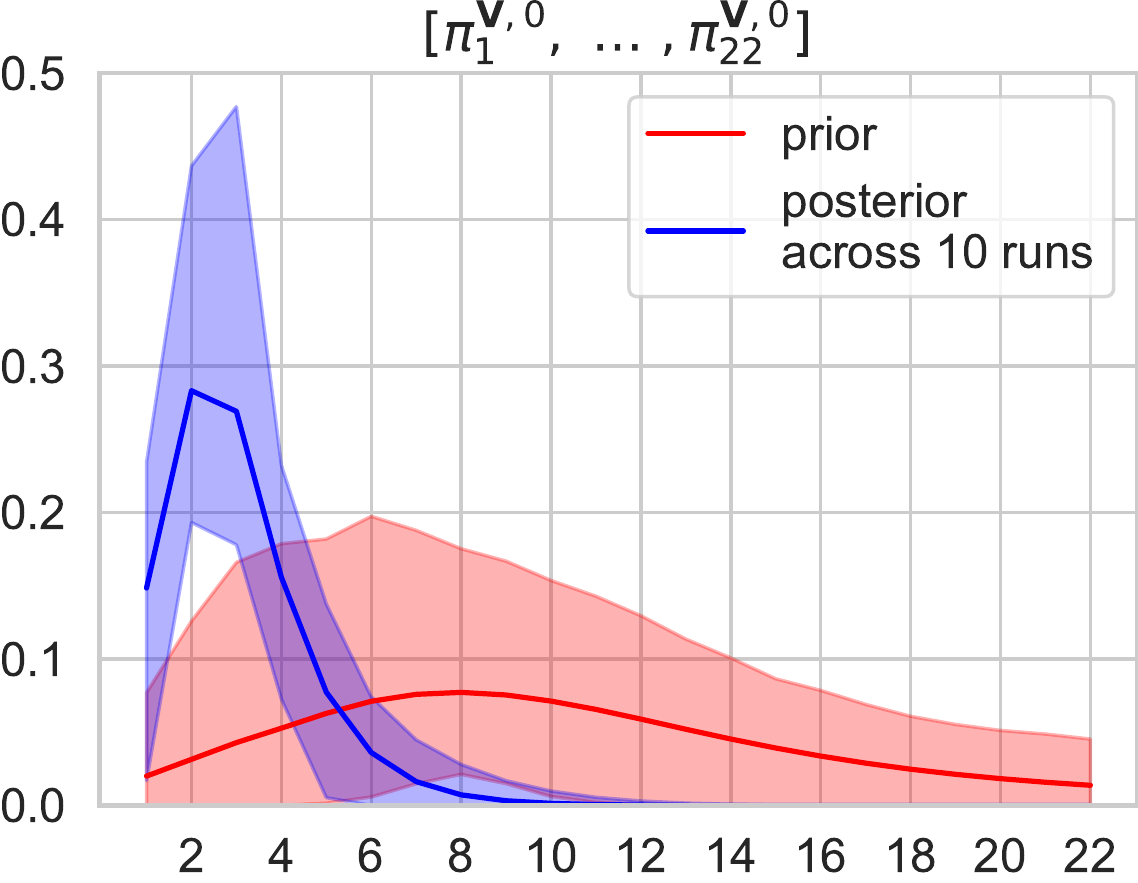}
\end{tabular}
\caption{
\textbf{Posterior distributions over parameters for Massachusetts (trained on Nov. 11 - Jan. 11).}
We show transition parameters (left) and duration parameters (right) after fitting on 2 months of counts, where each day we used available census counts for $\GG, \II, \VV$, and $\XX$. The colored interval of duration plots shows the 2.5 - 97.5th percentile intervals of 2000 samples (10 runs, each with 200 samples).
The prior is also shown for comparison. 
}
\label{fig:posterior_visualization_MA}
\end{figure}

In Fig.~\ref{fig:posterior_visualization_MA}, we show the learned posterior distributions over \emph{all} model parameters after fitting to the MA data, revealing several insights about hospital dynamics during the training period.
For instance, the learned recovery probability in $\II$ is lower than that indicated by our CDC-informed prior, indicating a potential deviation in hospital dynamics or in the patient age distribution in MA compared to the US as a whole. The learned posterior indicates, with high confidence, that the average length of stay in $\II$ (ICU \emph{not} on the ventilator) is very low, both for recovering ($h=1$) and especially for declining ($h=0$) patients, indicating that a patient who is on a declining path in the ICU is very quickly moved to the ventilator.
Last, patients on the ventilator seem to stay there for shorter periods of time when \emph{declining} rather than when \emph{recovering}.
However, we note that the duration posterior for recovering on the ventilator closely matches the prior and is bimodal, which may suggest some chains are producing poor posterior estimates.
This finding is not surprising: only a few patients recover from the ventilator, and thus with little data it is a difficult parameter to recover.

In Fig.~\ref{fig:posterior_visualization_SouthTees} in the Appendix, we further illustrate the learned posteriors for the South Tees hospital site in the UK.
As expected, we observe much higher variance in these learned posteriors for a single hospital than the posteriors for the (much larger) Massachusetts dataset above.
However, we notice some potentially valuable differences compared to MA. Duration distributions appear to be longer, and patients appear more likely to recover both in the General Ward and in the ICU.


\subsection{Data granularity ablation: Are results sensitive to missing some stages?}

Public data sources may differ in terms of the granularity of available counts. For example, for U.S. states we can usually gather separate counts of for each of the $\GG, \II,$ and $\VV$ intermediate stages as well as the terminal stage $\XX$. In contrast, for the U.K. we only have counts of all beds ($\GG + \II + \VV$), patients on ventilation ($\VV$), and recovered ($\RR$). Other sites may yet offer other different subsets of counts $\{ \GG, \II, \VV, \XX, \RR \}$.

To assess sensitivity to data granularity, we trained the ACED-HMM on MA data after an intentional preprocessing that aggregated all intermediate stages. 
We performed a fit to this reduced data -- only counts of all beds ($\GG+\II+\VV$) and death ($\XX$) -- and to the original data.
App.~\ref{sec:app_data_granularity} describes detailed results. 
As expected, forecasts for model stages where training data was not provided are generally of lesser quality, unless closely-related counts informed training (e.g. $\VV$ forecasts were reasonable given $\XX$ data).

\subsection{Scalability assessment: Can our ABC scale to large populations?}
\label{sec:scalability}

A key design requirement for our ABC procedure is the ability to fit posteriors to data from high-population regions or even entire countries.
We propose a simple but effective way to \emph{scale up}: given a scaling factor $r > 1$, we simulate only a fraction $\frac{1}{r}$ of all admissions (at a fraction of the computational cost).
Then we rescale the resulting aggregate counts produced by the simulation appropriately: $r \tilde{y}^k_{t}$.
In expectation, this approach should produce an unbiased estimate of simulated counts from the original admitted population.
In Fig.~\ref{fig:exact_vs_5x_scaling_MA}, we show that our proposed scaling procedure is effective.
Simulating only 20\% of admissions in MA and rescaling, we see a qualitatively similar forecast with only modestly increased uncertainty.
With exact simulation of Massachusetts already feasible, by applying 10-x or 20-x scaling we can easily forecast for any U.S. state of interest.

\subsection{What-if scenario assessment: Can we assess the impact of interventions?}
\label{sec:whatif}

We now consider using our model to assess the societal value of possible \emph{interventions}.
Our mechanistic ACED-HMM model can consider changes to its parameters or assumptions about admissions. We stress this analysis is exploratory, not a proper causal model.

Below, we study two possible scenarios for the state of California (CA), which chose to enact significant \emph{restrictions} in late November 2020 in light of rising hospital utilization and especially rising ICU utilization.
We imagine each intervention is rolled out on Nov. 19th (when CA's Limited Stay at Home Order was announced).
We used estimated posteriors trained on CA count data from November 11th to January 11th (using 30-x scaling). 
Because the training period overlaps with the intervention period, this analysis should be interpreted as a \emph{retrospective} characterization of what might have happened.
Our method can forecast into the future if desired, but we have not done so here for simplicity.

\begin{figure}[!h]
\centering
\begin{tabular}{c}\includegraphics[width=0.8\textwidth]{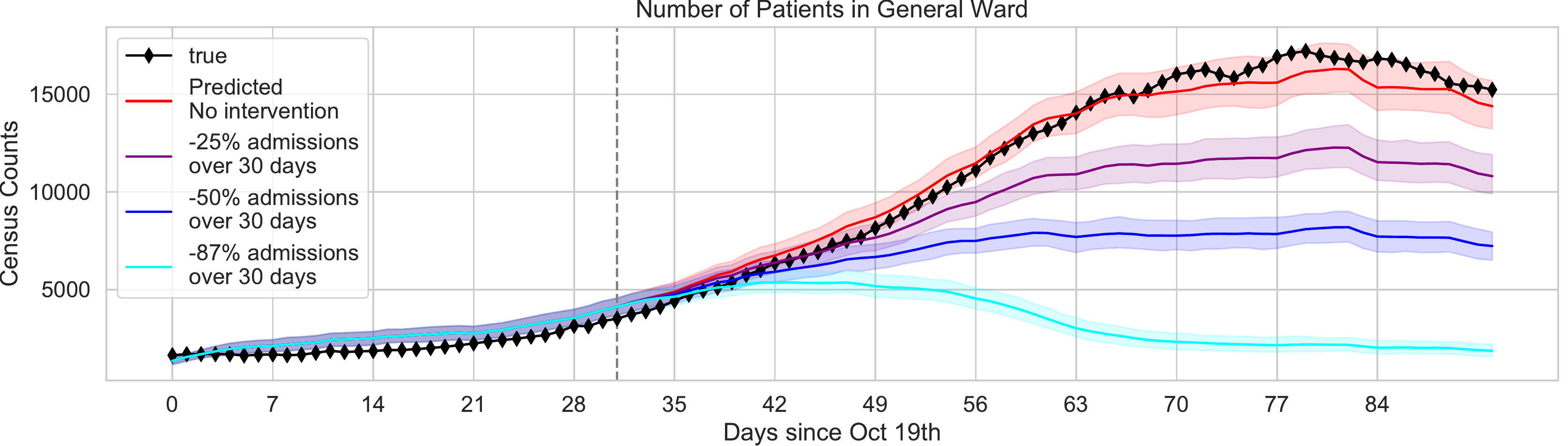}
	\\
	\includegraphics[width=0.8\textwidth]{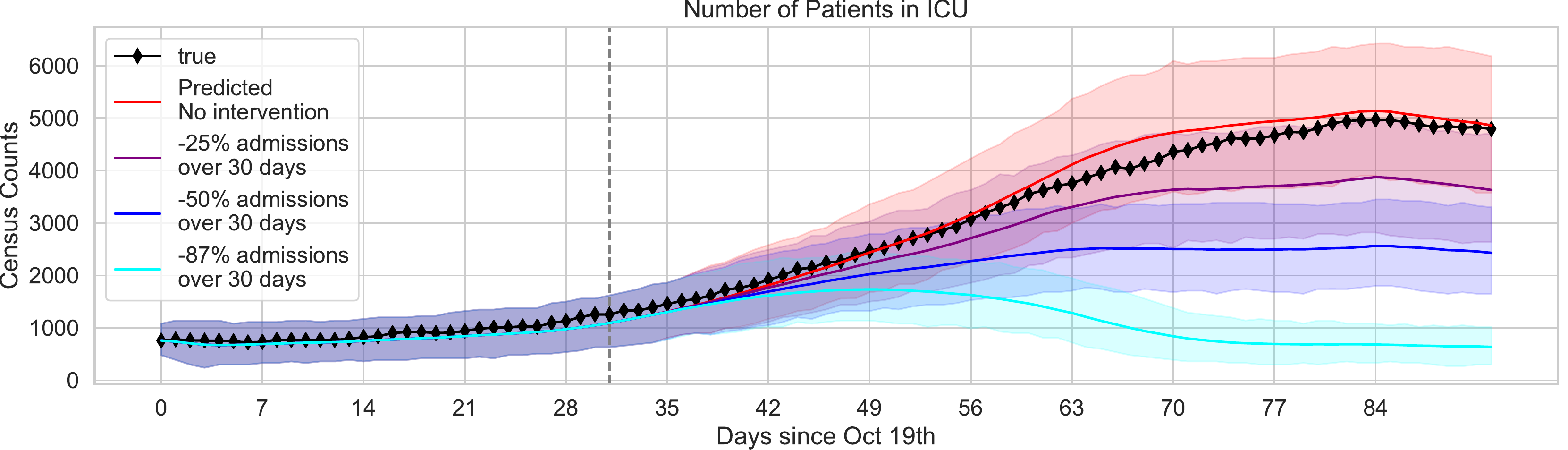}
\end{tabular}
\caption{
    \textbf{Forecasted demand for General Ward (top) and ICU (bottom) beds in California under Scenario 1 (Sec.~\ref{sec:whatif}).}
}
\label{fig:forecast_scenario1_CA}
\end{figure}

\paragraph{What-if Scenario 1: Intervention that lowers hospitalization rates.}
A recent Phase 3 trial suggests that two monoclonal antibody drugs (bamlanivimab and etesevimab) together reduced hospitalizations and death by 87\% when given to severely ill COVID-19 patients before they needed the hospital~\citep{elilillyandcompanyLillyBamlanivimabEtesevimab2021}.
Suppose that policy makers could have aggressively deployed this treatment across the state of California after Nov. 19th, 2020.
In Fig.~\ref{fig:forecast_scenario1_CA}, we show our model's forecasts for several scenarios: what actually happened (no reduction in admissions), an ideal retrospective rollout (a linear ramp up to a full 87\% reduction in hospital admissions after 30 days), and two more realistic scenarios that assume a 50\% and 25\% reduction in hospital admissions, respectively.
Given Fig.~\ref{fig:forecast_scenario1_CA} plus knowledge of capacity limits, policy makers could use such quantitative forecasts to decide which rollout strategies to prioritize. 

\paragraph{What-if Scenario 2: Pharmaceutical that lowers recovery times in the hospital.}

We next consider an intervention that changes the \emph{number of days} spent when recovering in the hospital. 
Our Scenario 2 supposes that a new treatment reduces in-hospital recovery time by 25\%.
This is a conservative translation of a recent clinical trial described by~\citet{beigelRemdesivirTreatmentCovid192020}, suggesting that the drug remdesivir reduced median time to recovery from 15 days (placebo) to 10 days (treatment) for hospitalized patients with COVID-19.

To operationalize this scenario, for each segment $\ell$ of patient $n$'s trajectory we impose a 25\% reduction in the sampled duration $\Delta_{n\ell}$ when the patient is recovering (when $h_{n\ell} = 1$). 
Durations when health is declining are not modified.
When we reduce durations, we round to an integer number of days in a way that preserves expected value (e.g. if the down-scaled duration is 9.75 days, we return a value of 10 days 75\% of the time and 9 days 25\% of the time).
We also enforce that each segment's duration cannot be reduced below 1 day.

\begin{figure}[!h]
\centering
\begin{tabular}{c c}
	\includegraphics[width=0.5\textwidth]{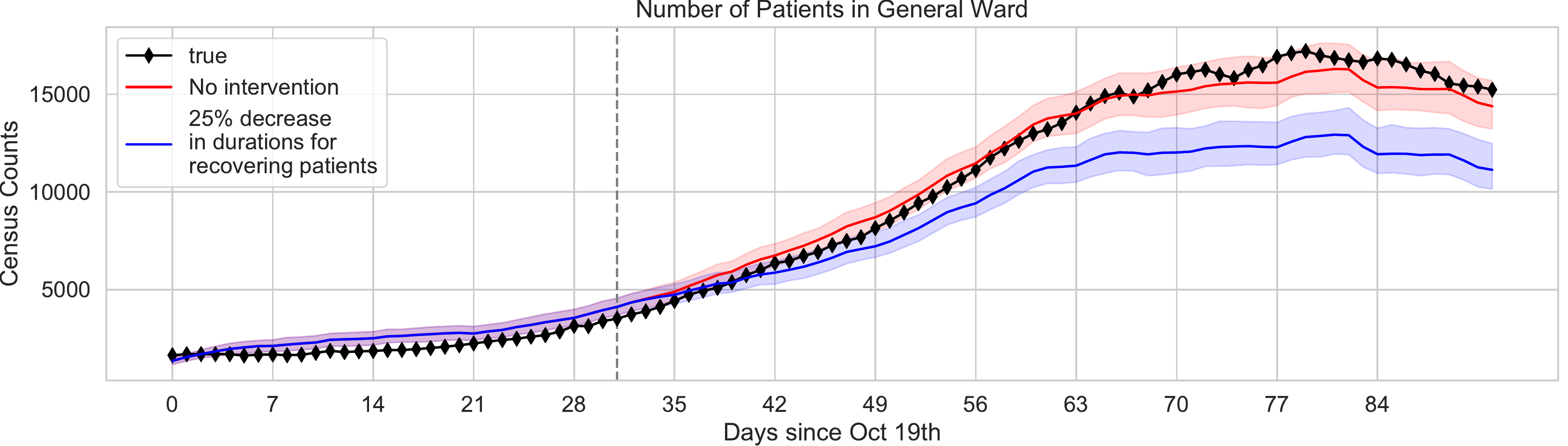}
	&
	\includegraphics[width=0.5\textwidth]{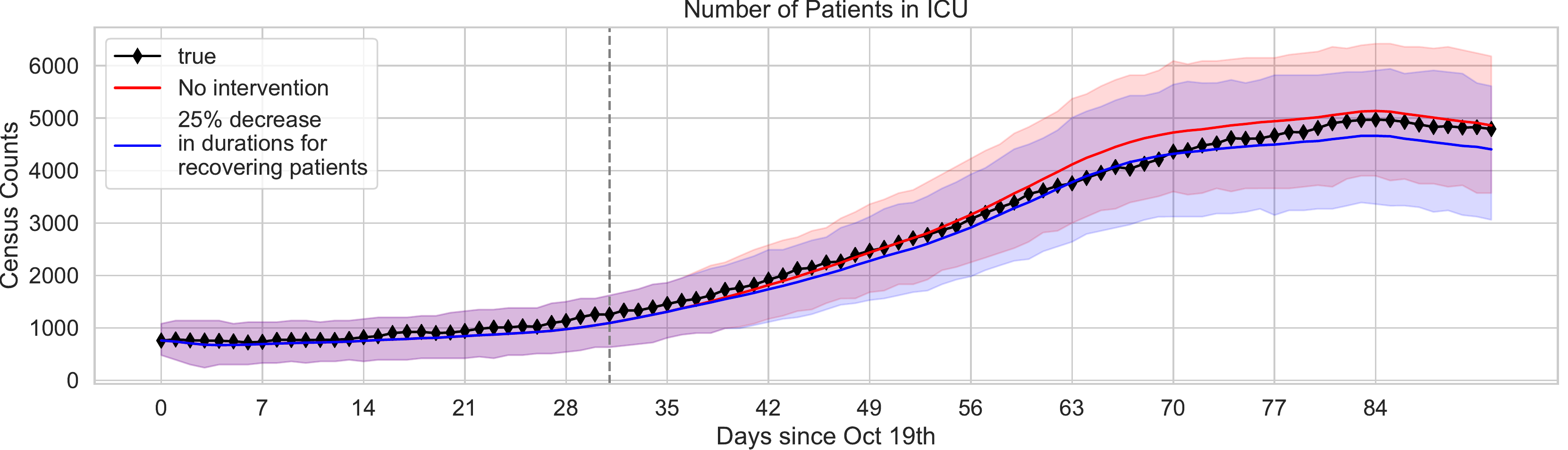}
	\\
	\includegraphics[width=0.5\textwidth]{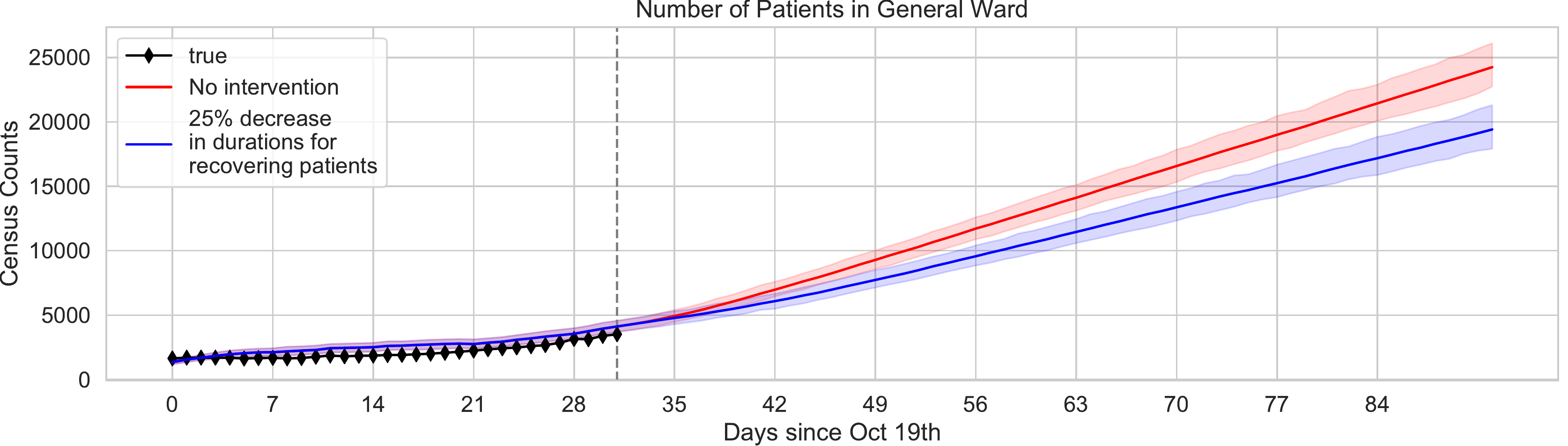}
	&
	\includegraphics[width=0.5\textwidth]{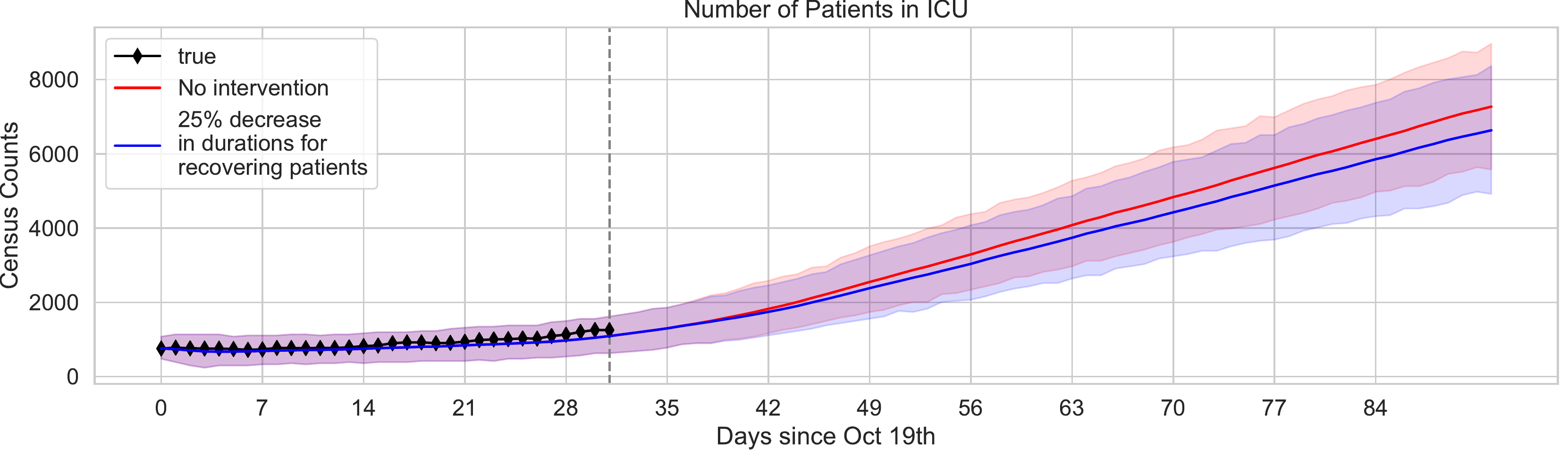}
\end{tabular}
\caption{
    \textbf{Forecasted demand for General Ward beds (left) and ICU beds (right) using actual admission counts (top) and a worst-case projection of admission counts (bottom) in California under Scenario 2 (Sec.~\ref{sec:whatif}).}
}
\label{fig:forecast_scenario2_CA}
\end{figure}

Fig.~\ref{fig:forecast_scenario2_CA} shows the difference in utilization between the standard and 25\% reduction in recovery durations.
As expected, we see noticeable differences in the populations in the general ward. However, the model suggests the ICU counts would not change much, probably because the ICU is unfortunately dominated by \emph{declining} rather than \emph{recovering} patients.

\section{Discussion}
We have contributed a new probabilistic model, developed a scalable ABC algorithm to draw samples from the (approximate) posterior distribution over model parameters, and validated these contributions on several datasets relevant to hospital demand modeling.

\paragraph{Limitations.}
Our proposed ACED-HMM model has several limitations. 
A major limitation is the assumed availability of admission counts when making forecasts, when of course for a real forecast into the future these counts are unknown. 
We hope in future work to explore integrating models for forecasting future admissions, such as the model by~\citet{qianCPASUKNational2020}.
Another major limitation is that we assume that the values of most parameters are static over time.
In reality, we know that many factors change over time, including properties of the disease (e.g. variants with different infection dynamics), characteristics of the infected population (e.g. age distributions and vaccination status), and hospital care practices. In the language of statistics, we know the real progression of patients through the hospital is neither \emph{homogenous} nor \emph{Markov}, so we should not expect our forecasts to be accurate far into the future.
We could allow the model to \emph{learn} how recovery probabilities or duration distributions might change over time.
Finally, our model would likely be improved by capturing age as a key prognostic covariate that informs both the recovery probabilities and the duration probabilities.
Naturally this would require more information, such as estimates of the age distribution on each day in the target region and the effect of mandated non-pharmaceutical interventions or vaccination on the evolving age distribution.
However, if available this information could increase the model's power as an explanatory tool.

Further limitations can be ascribed to our  ABC methodology. 
Though our ABC procedure is simple and scalable, it is inefficient in the way it proposes samples via a random walk, with rather high rejection rates.
Future work may consider Hamiltonian ABC~\citep{meedsHamiltonianABC2015}, using \emph{gradient} information to improve proposals and accelerate posterior exploration.

\paragraph{Advantages.} Towards our twin goals of being able to provide reasonable forecasts and assess societal value, the \emph{extreme portability} of our approach is a key advantage. 
The model requires only daily count data that are available from public health agencies throughout the world.
There is no requirement for patient-level data. We further emphasize that the model and learning procedures are easily extended. 
Changing the model (such as adding age-dependent duration probabilities or allowing recovering patients to later decline) would simply require editing the simulation code that samples each patient's trajectory and specifying relevant prior and proposal distributions for any additional parameters.\\
\\
Overall, we hope this study and our open-source code release leads to an improved ability to make hospital demand forecasts that are portable (by requiring only aggregate count data), interpretable and extrapolate-able (by using a mechanistic model), accurate (by fitting model parameters to local conditions via our ABC algorithm), and appropriately uncertain (by producing posterior distributions rather than point estimates). Although further work is needed to properly assess the societal value of possible interventions with this model, we believe our work represents a promising first step.

\section*{Acknowledgements}

Authors GMV, CN, JTC, and MCH gratefully acknowledge funding support from a consortium of industry partners for this project -- AstraZeneca, Bristol-Myers Squibb, Eli Lilly, Gilead Sciences, Janssen, Merck, Pfizer, and Regeneron -- via an agreement with the Center for the Evaluation of Value and Risk in Health (CEVR) at Tufts Medical Center.

Authors MCH and DMK thank the Office of the Vice Provost for Research at Tufts University for support for this project under a ``Tufts Springboard'' award.

We thank several graduate students in the Department of Computer Science at Tufts University who assisted with various early aspects of this project: Preetish Rath, Zhe Huang, and Panos Lymperopoulos (early model prototyping efforts) as well as Gabriel Appleby, Diana Eastman, and Ab Mosca (visualization efforts). 
We thank Jinny Park (RA at Tufts Medical Center) for assistance with literature surveys and project management.

\bibliography{c19.bib}

\appendix
\counterwithin{table}{section}
\setcounter{table}{0}
\counterwithin{figure}{section}
\setcounter{figure}{0}
\counterwithin{algocf}{section} 
\setcounter{algocf}{0}

\clearpage
\section{Datasets}
\label{sec:app_datasets}

In our public code release, we include CSV files of the datasets we used for both the US and UK. Below, we describe how we obtained and preprocessed this data from original sources.

\subsection{U.S. data.}
Data representing state-level counts came from the \citet{u.s.dept.ofhealthandhumanservicesCOVID19ReportedPatient2021} and the Covid Tracking Project (CTP, \citet{theatlanticmonthlygroupCOVIDTrackingProject2021})\footnote{\url{https://api.covidtracking.com/v1/states/daily.csv}}.
We accessed the data in February 2021.
These sources provide data for all 50 states, we selected a subset of states of interest: California (CA), Massachusetts (MA), South Dakota (SD), and Utah (UT).
HHS data provided daily counts of hospital admissions and general ward occupancy counts (stage $\GG$), while CTP data provided ICU counts ($\II$ and $\VV$), as well as deaths (stage $\XX$).

As much as possible, we validated the consistency of these 2 public data sources by assuring that these counts matched with available corresponding data provided directly by state government public health agencies. For example, we verified the MA data matched counts released by the Massachusetts Department of Public Health\footnote{\url{https://www.mass.gov/info-details/archive-of-covid-19-cases-in-massachusetts}}.

In summary, our US predictive models all use standardized counts representing adult hospitalizations for \emph{confirmed} COVID-19 cases.
We are certain that our data for the general ward ($\GG$) accounts for the adult population while excluding pediatric cases.
Through data-consistency checks across the 2 data sources, we convinced ourselves that CTP's $\II, \VV$, and $\XX$ data can be assumed as a good representation of adult populations (even though CTP reports uncertainty about whether their numbers include pediatric cases).

\paragraph{HHS fields used.} HHS data descriptions are found on the clickable columns of the data web page\footnote{\url{https://healthdata.gov/Hospital/COVID-19-Reported-Patient-Impact-and-Hospital-Capa/g62h-syeh/data}}.  We used the following data fields, with quoted definitions extracted from the data source.
\begin{itemize}
    \setlength\itemsep{0em}
    \item \textbf{previous day admission adult covid confirmed} : ``Number of patients who were admitted to an adult inpatient bed on the previous calendar day who had confirmed COVID-19 at the time of admission in this state'' (We considered previous day admissions as a feasible current day admission because admissions rates were not abruptly changing in orders of magnitudes between days. We recommend considering making the 1 day adjustment in future use of this data field.)
    \item \textbf{total adult patients hospitalized confirmed covid} : ``Reported patients currently hospitalized in an adult inpatient bed who have laboratory-confirmed COVID-19. This include those in observation beds''
    \item \textbf{staffed icu adult patients confirmed covid} : ``Reported patients currently hospitalized in an adult ICU bed who have confirmed COVID-19 in this state''
\end{itemize}

\paragraph{CTP fields used.} CTP data descriptions are found in the 'historic values for all states' section of the data API webpage page\footnote{\url{https://covidtracking.com/data/api}}. We used the following data fields, with quoted definitions extracted from the data source. 

\begin{itemize}
    \setlength\itemsep{0em}
    \item \textbf{inIcuCurrently} : ``Individuals who are currently hospitalized in the Intensive Care Unit with COVID-19. Definitions vary by state / territory, and it is not always clear whether pediatric patients are included in this metric. Where possible, we report patients in the ICU with confirmed or suspected COVID-19 cases''
    \item \textbf{onVentilatorCurrently} : ``Individuals who are currently hospitalized under advanced ventilation with COVID-19. Definitions vary by state / territory, and it is not always clear whether pediatric patients are included in this metric. Where possible, we report patients on ventilation with confirmed or suspected COVID-19 cases.''
    \item \textbf{deathIncrease} : ''Daily increase in death, calculated from the previous day’s value.''
\end{itemize}

We note that as of 2021-04-14, the CTP website relates the following message: ''\textit{The COVID Tracking Project has ended all data collection as of March 7, 2021. These files are still available, but will only include data up to March 7, 2021. We’ll be publishing research through May, and then we will—fully and accessibly—archive our work and be done.}''

\paragraph{Obtaining stage-specific counts from raw data fields.}

\begin{itemize}
    \item Counts for daily admissions to the general ward were obtained from field \textbf{previous day admission adult covid confirmed}. 
    \item $\GG$ : Daily counts for general ward occupancy were obtained from field \textbf{total adult patients hospitalized confirmed covid} subtracting field \textbf{staffed icu adult patients confirmed covid}
    \item $\II$ : Daily counts for ICU-off-ventilator were obtained from field \textbf{inIcuCurrently} subtracting field \textbf{onVentilatorCurrently}
    \item $\VV$ : Daily counts for ICU-on-ventilator were obtained from field  \textbf{onVentilatorCurrently}
    \item $\XX$ : Daily counts obtained from field \textbf{deathIncrease}
\end{itemize}

\paragraph{Variable data granularity across states.}
Not all states report counts from all stages.
For example, UT and CA only provide aggregate ICU counts, and thus do not distinguish between on and off the ventilator. 
We thus take the provided count as the sum of the two ICU stages $\II, \VV$ of our model, and adapt our ABC accordingly to evaluate a distance on this combined stage.

\paragraph{Smoothing terminal counts.}
For all states, we find the daily counts for the terminal death stage $\XX$  to be both noisy and unreliable due to human factors in the reporting process.
In particular, the reported counts often drop to unrealistic values at or close to zero during holidays and weekends (e.g. in some states we observed zero deaths on Thanksgiving and Christmas).
Thus, we \emph{smooth} these counts by replacing each day's raw value with the centered moving average across 5 days, being careful to not leak test-counts information into the moving average of training counts.

\subsection{UK data.}

We selected two hospital sites in England, United Kingdom with consistent data availability and a large volume of cases. We used public data sourced from the UK National Health Service (NHS)\footnote{\url{https://www.england.nhs.uk/statistics/statistical-work-areas/covid-19-hospital-activity/}}, which provided the count of beds occupied by COVID-19 patients on each day at each hospital. We used the following data fields:
\begin{itemize}
    \setlength\itemsep{0em}
    \item mechanical ventilators used for covid patients ($\VV$)
    \item total number of beds used for covid patients ($\GG + \II + \VV$)
    \item patients discharged from covid hospitalization ($\RR$)
    \item patients admitted with covid
    \item patients diagnosed in-hospital with covid but who were not admitted as covid patients
\end{itemize}
In our experiments, we aggregated daily covid admissions and daily in-hospital covid diagnoses to form the daily admissions in $\GG$ that are then fed into ACED-HMM. The two hospital sites we selected had the same available counts. Unlike for US data, we did not find the need to smooth any count.
\section{Additional Results}


We offer extended commentary on the experimental protocols and results of Sec.~\ref{sec:results} here.


\subsection{Quantitative Error Assessment: UK Results}
\label{sec:appendix_mae}

Here, in Table~\ref{tab:error_metrics_UK} we provide mean absolute error assessments for our method and various baselines for two hospital sites in the U.K.
This matches the main paper's Table~\ref{tab:error_metrics_US}, which focused on U.S. state data. 

Table~\ref{tab:error_metrics_UK} shows consistent gains for our proposed ACED-HMM especially in the ``all beds'' and ``on ventilator'' stages. 
For the ``recovered'' stage, our method is competitive for South Tees, and mildly competitive for the Oxford site (Bayes LR seems to offer slightly better MAE on that site: 10.5 vs. our 13.4).

\begin{table}[!t]
\centering
\resizebox{\textwidth}{!}{%
\texttt{
\begin{tabular}{c|r |r r|r r|r r| }
  & 
    & $\GG+\II + \VV$ & \textnormal{\bf All Beds} 
    & $\VV$ & \textnormal{\bf OnVentInICU }
    & $\RR$ & \textnormal{\bf Recovered}  \\ 
  & \textnormal{\bf Method}
    & MAE~ & lower - upper
    & MAE & lower - upper~
    & MAE & lower - upper
\\  
 \hline
 \hline
 \multirow{7}{*}{South Tees}
  & ACED-HMM + ABC
                &  8.7~ &  ~~7.7 - ~10.2   
                &  4.6  & ~~3.9 - ~~~5.1 
                &  5.7  & ~~5.5 - ~~5.8  \\
  & ACED-HMM + Prior
                & 118.4~ & 109.8 - 125.9  
                &  10.7  & ~10.0 - ~~11.4  
                &  5.9   & ~~5.7 - ~~6.2  \\
  & AR-Poisson
                &  49.2~ &  ~40.3 - ~60.5  
                & 417.3  &  ~15.8 - 2209.5 
                & 49.2   &  ~~9.7 - 518.7 \\
  & Median predictor
                & 102.0~ &  NA~~~~~~~~
                & 17.1  &  NA~~~~~~~~~
                & 9.9   &  NA~~~~~~~~ \\
  & Bayes LR (day)
                &  114.0~ &  ~113.4 - 114.6
                & 20.5  &  ~20.4 - ~~20.6 
                & 13.6   &  ~~13.4 - ~13.7 \\
  & Bayes LR (day+adm.)
                &  81.7~ &  ~81.3 - ~82.1
                & 13.9  &  ~13.8 - ~~14.0 
                & 12.4   &  ~~12.3 - ~12.6 \\
  \cline{2-8}
  & Mean Test $y$
                &  199.5~ &                
                &   31.1 &                
                & 20.8  &              \\
 \hline
 \hline
 \multirow{5}{*}{Oxford}
  & ACED-HMM + ABC
                & 18.0~ &  ~~15.4 - ~22.4 
                & ~~6.3 &  5.5 - ~~~6.9 
                & 13.4 &  ~13.2 - ~13.6 \\
  & Median predictor
                & 216.7~ &  NA~~~~~~~~
                & 34.7  &  NA~~~~~~~~~
                & 25.1   &  NA~~~~~~~~ \\
  & Bayes LR (day)
                & 71.5~ &  ~70.5 - ~72.5
                & 16.2  &  ~16.0 - ~~16.3
                & 12.5   &  ~~12.3 - ~12.7 \\
  & Bayes LR (day+adm.)
                & 21.8~ &  ~21.4 - ~22.3
                & 10.5  &  ~10.4 - ~~10.7 
                & 10.5   &  ~~10.3 - ~10.7 \\
  \cline{2-8}
  & Mean Test $y$
                & 274.7~ &  ~~ 
                & 37.7  & ~~   
                & 33.1  &  ~~   
\end{tabular}}}
\caption{Quantitative error assessment for predictions for two UK hospital sites, evaluated over the test period Jan. 4 - Feb. 3, 2021. 
We assess mean absolute error (MAE) at each stage where we have observed data: total hospital beds $\GG + \II + \VV$, on ventilator in ICU $\VV$, and recovered $\RR$.
For all sites and stages, we report the mean count in the test period to indicate scale.
We report lower and upper estimates on the MAE, via repeating all MAE computations across 100 separate batches of posterior samples.
We report the 2.5th and 97.5th percentiles across these batches.
}
\label{tab:error_metrics_UK}
\end{table}

\subsection{Quantitative Error Assessment: Baseline Details}
\label{sec:appendix_baselines}

In Tables \ref{tab:error_metrics_US} and \ref{tab:error_metrics_UK}, we compared our proposed ACED-HMM to several reasonable baselines. (We did not try all methods on all datasets due to limited computational resources; poor performing baselines such as ACED-HMM+Prior and AR-Poisson were cut from further experiments to save time and energy).
Further details about these baselines are below.

\paragraph{ACED-HMM + Prior.}
First, we consider ``ACED-HMM + Prior'', which makes forecasts using the \emph{prior} distribution over our model's parameters, rather than the posterior fit via ABC.
This could indicate how reliable our model would be if we did not adapt parameters  to the specific region of interest but instead chose values for transition and duration probabilities from a reasonable literature survey (as our prior is chosen).
This could also indicate how important other factors (such as the provided admission counts in testing period) are to the forecast.

\paragraph{AR-Poisson.}
Second, we compare to ``AR-Poisson'', a non-mechanistic probabilistic autoregressive model with a Poisson likelihood that forecasts each univariate count time-series individually, inspired by the COVID-19 single site hospital forecasts of \citet{leeForecastingCOVID19Counts2021}.
This is a custom univariate forecasting model fit to each stage separately.
It does not use admission counts at all.
The AR-Poisson uses modern Hamiltonian MCMC to fit posteriors (rather than our less-efficient-to-explore ABC).
See Appendix~\ref{sec:appendix_gar} for complete details.
As documented in~\citet{leeForecastingCOVID19Counts2021}, we believe this model represents a reasonable baseline when forecasts need a roughly linear function of the training data. However, when test period looks different than training in trends (as in Fig.~\ref{fig:forecast_MA}, we should expect (and indeed see in practice) that performance suffers.

\paragraph{Median of train set.} This baseline simply predicts the median count observed in the training set of each stage. We selected median (not mean) because it is the simple summary statistic that provably optimizes mean absolute error (MAE) as a performance metric.

\paragraph{Bayes LR.} We used two versions of a Bayesian linear regression model (as implemented in \texttt{sklearn} with default parameters). The first version (``Bayes LR (day)'') used as its only feature the absolute time in days (day 0 is the start of the training set, then the days increase through the entire training period and into the testing period).
The second version (``Bayes LR (day+adm.)'') took as input a feature vector containing absolute time in days, as well as the admission count from each of the last 21 days (we set the admissions values to $0$ for days before day 0 in the training set). This model has the key property of using recent admissions as its covariates, which are also used by ACED-HMM to make forecasts. 

\paragraph{IHME.}
For all U.S. states we compare against the January 15th public release of IHME model forecasts~\citep{reinerModelingCOVID19Scenarios2021}. These forecasts were produced by the IHME team by training on data available before January 12th, thus matching the start of our testing period almost exactly.
This model does not make use of admission counts from the testing period.
See Appendix~\ref{sec:appendix_ihme} for details.

\paragraph{Uncertainty estimation details.}
For all probabilistic models (both versions of ACED-HMM, the AR-Poisson model, and the two Bayesian Ridge regression baselines), we compute the MAE and its confidence interval in Tab.~\ref{tab:error_metrics_US} as follows.
We draw 100 samples from the posterior (prior for ACED-HMM + Prior) distribution over the model parameters, we produce individual forecasts for each sample, and compute the MAE using the mean of these forecasts.
We repeat the process 100 times, and we report the average MAE with the 2.5th - 97.5th percentile range across these batches as confidence interval.
The public release of forecasts from IHME only provides three daily forecast values - a mean estimate, a lower bound and an upper bound - thus making it impossible to compute a confidence interval analogous to that which we compute for the other three models.
Thus, we simply report the MAE computed using each of the three estimates.
The IHME uncertainty intervals (marked with *) are given for completeness, but should be recognized as different from the other intervals and not directly compared.

\subsection{Quantitative Assessment: Posterior Coverage}
\label{sec:appendix_coverage}

In Tab.~\ref{tab:coverage_all} below, we evaluate the posterior coverage of our estimated posteriors on heldout data at target fractions of 50\%, 80\% and 95\%, for both U.S. and U.K. data.
Ideal probabilistic models that are well-calibrated would have an empirical fraction of observations \emph{exactly equal to} the target fraction.

\begin{table}[!b]
\begin{center}
{\footnotesize 
\texttt{%
\begin{tabular}{c c}
U.S. & U.K.
\\
\resizebox{0.5\textwidth}{!}{%
\begin{tabular}{ |c|r|r|r|r|r|r|r| } 
 \hline
  & Coverage (\%) & \textbf{G} & \textbf{I} & \textbf{V} & \textbf{I+V} & \textbf{T} & \textbf{T sm.} \\ 
 \hline
 \hline
 \multirow{3}{*}{SD} & ACED-HMM + ABC \textit{95} & 87 & 77 & 81 & 100 & 45 & 77 \\ 
  &  \textit{80} & 52 & 48 & 71 & 97 & 23 & 68 \\
  &  \textit{50} & 42 & 26 & 32 & 55 & 6 & 32 \\
 \hline
 \hline
 \multirow{3}{*}{UT} 
 & ACED-HMM + ABC \textit{95} & 71 & NA & NA & 100 & 55 & 100 \\ 
  & ~~ \textit{80} & 39 & NA & NA & 100 & 32 & 87 \\ 
  & ~~ \textit{50} & 19 & NA & NA & 61 & 23 & 39 \\
 \hline
 \hline
 \multirow{9}{*}{MA} 
 & AR-Poisson \textit{95} & 81 & 100 & 45 & 61 & 100 & 100 \\ 
  & ~~ \textit{80} & 45 & 90 & 26 & 35 & 68 & 87 \\ 
  & ~~ \textit{50} & 29 & 35 & 13 & 13 & 32 & 10 \\
  \cline{2-8}
  & ACED-HMM + Prior \textit{95} & 52 & 39 & 61 & 71 & 42 & 23 \\ 
  & ~~ \textit{80} & 6 & 35 & 42 & 39 & 26 & 23 \\ 
  & ~~ \textit{50} & 6 & 19 & 19 & 23 & 23 & 19 \\
  \cline{2-8}
  & ACED-HMM + ABC \textit{95} & 68 & 65 & 42 & 97 & 55 & 84 \\ 
  & ~~ \textit{80} & 48 & 45 & 16 & 94 & 39 & 74 \\ 
  & ~~ \textit{50} & 32 & 16 & 3 & 74 & 16 & 58 \\
 \hline
\end{tabular}
}
&
\resizebox{0.45\textwidth}{!}{%
\begin{tabular}{ |c|r|r|r|r| } 
 \hline
  & Coverage (\%) & \textbf{R} & \textbf{G+I+V} & \textbf{V} \\ 
 \hline
 \hline
 \multirow{9}{*}{S. Tees} 
 & AR-Poisson \textit{95} & 90 & 100 & 100 \\ 
  & \textit{80} & 52 & 100 & 100 \\
  & \textit{50} & 3 & 58 & 74 \\
  \cline{2-5}
  & ACED-HMM + Prior \textit{95} & 77 & 100 & 84 \\ 
  &  \textit{80} & 58 & 13 & 42 \\
  &  \textit{50} & 29 & 10 & 26 \\
  \cline{2-5}
  & ACED-HMM + ABC \textit{95} & 77 & 100 & 97 \\ 
  &  \textit{80} & 48 & 100 & 87 \\
  &  \textit{50} & 26 & 97 & 39 \\
 \hline
 \hline
 \multirow{3}{*}{Oxford} 
 & ACED-HMM + ABC 95 & 55 & 100 & 100 \\ 
  &  \textit{80} & 19 & 100 & 77 \\ 
  &  \textit{50} & 10 & 87 & 35 \\
 \hline
\end{tabular}
}
\end{tabular}
}
}
\end{center}
\vspace{-5mm}
\caption{
Coverage assessment for estimated posteriors fit to U.S. data (left) and U.K. data (right).
This metric assesses how reliably calibrated the forecasted \emph{distributions} each method produces may be on future unseen data.
Given many samples from a predicted distribution, we obtain a centered interval of specified coverage percentage by computing relevant percentiles from posterior samples (e.g. for the 80\% percentage our interval is defined by the 10th and 90th percentiles).
We report the observed percentage of daily count data within the testing period at each stage that fall within that predicted interval. Ideally, the observed fraction should be close to the intended coverage fraction.
}
\label{tab:coverage_all}
\end{table}

Across Tab.~\ref{tab:coverage_all}, we see that all models could be better calibrated. 
Our proposed ACED-HMM appears ``too confident'': it primarily underestimates its posterior mass intervals. For example for MA the interval it predicts will cover 95\% of data for $\GG$ and $\II$ stages actually covers only 68\% and 65\% of observed counts.
One remedy that may help is the scalability approximation that we propose (Sec.~\ref{sec:app_scalability}) provides more conservative uncertainty estimates.
ABC may be an issue too here: using a proper likelihood (rather than a distance function) may lead to improved calibraton.

Our ACED-HMM trained with ABC does have a reasonable fit to the $\GG+\II+\VV$ counts and $\VV$ counts on both UK hospitals, as well as a slight over-estimate of the interval for the total ICU bed count ($\II+\VV$) on all US states. These bed counts have been commonly used by US policy makers to determine the imposition of social-distancing measures.

We wish to emphasize that coverages need to be put in the context of their MAEs as well. For instance, while ACED-HMM has worse 95\% coverage than AR-Poisson on $\GG$ counts, that is because AR-Poisson has much worse MAE with a very high uncertainty interval, effectively making AR-Poissons's predictions less useful than those of ACED-HMM.

\subsection{Interpretation of Learned Posteriors: UK Results}
\label{sec:appendix_posteriors_uk}

\begin{figure}[!b]
\centering
\begin{tabular}{c c | c c}
	\includegraphics[width=\PWW\textwidth]{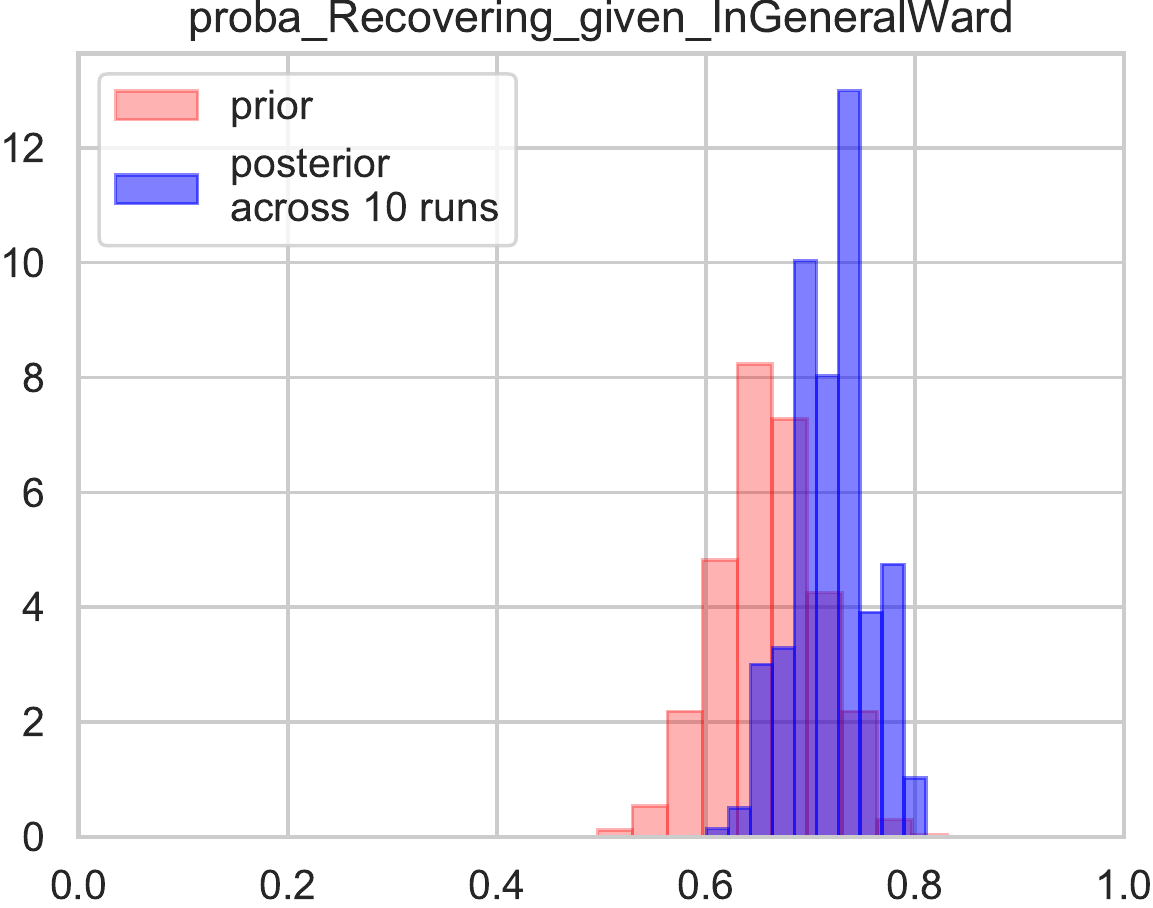}
	& 
	\includegraphics[width=\PWW\textwidth]{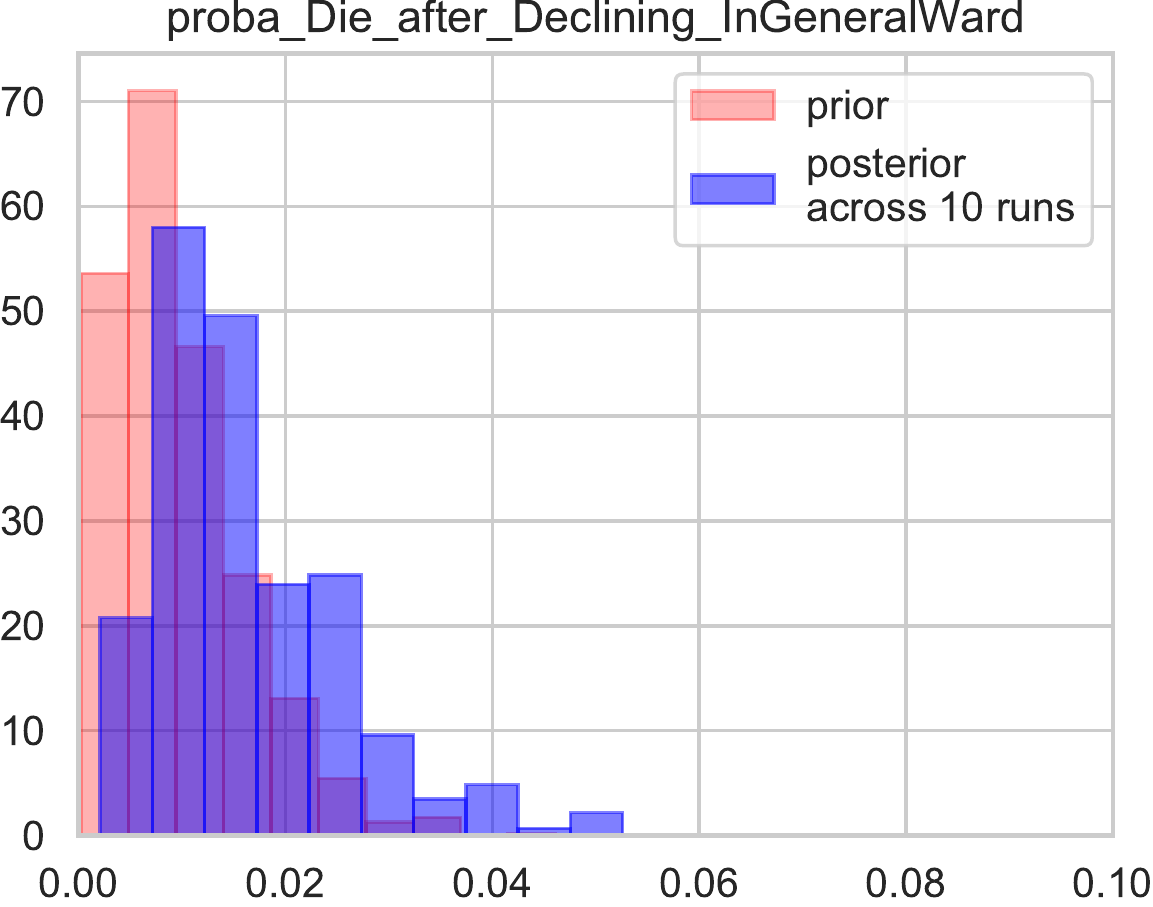}
	&
	\includegraphics[width=\PWW\textwidth]{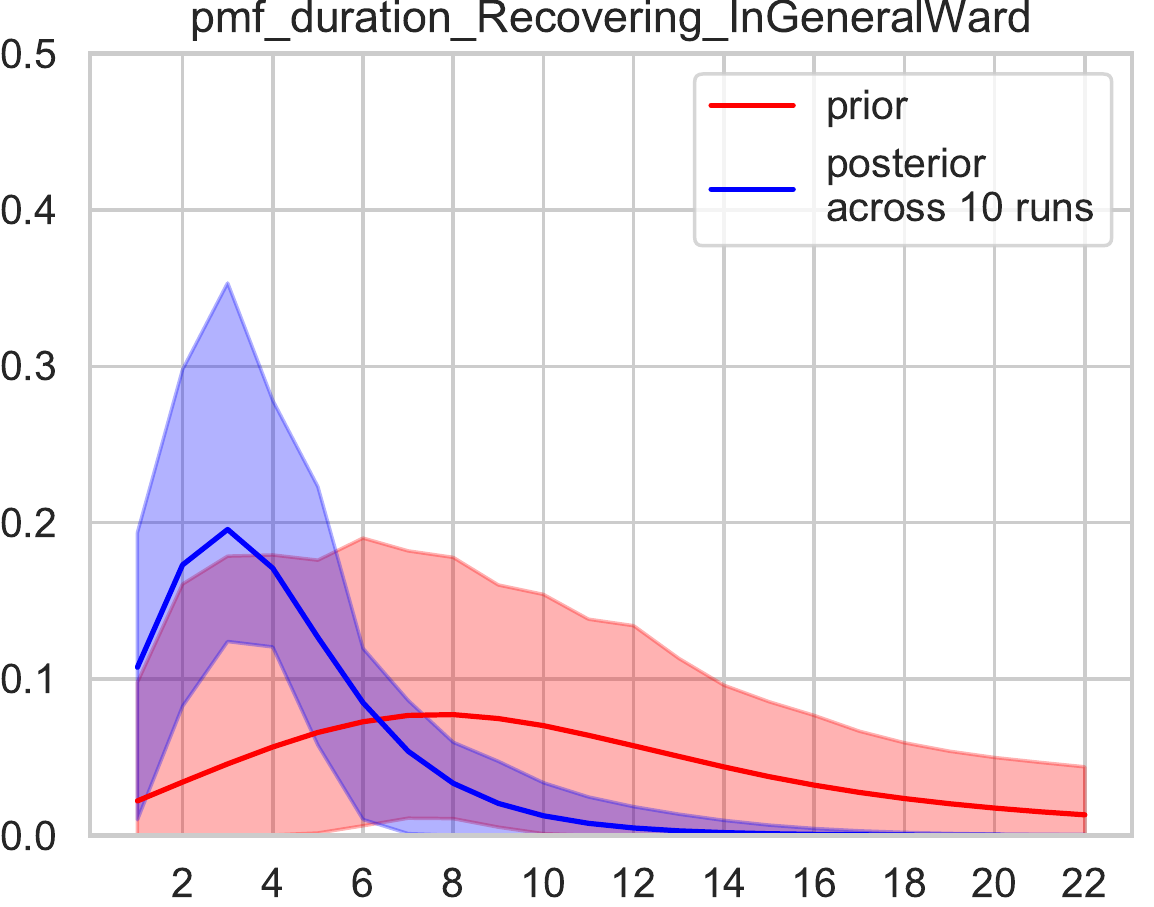}
	&
		\includegraphics[width=\PWW\textwidth]{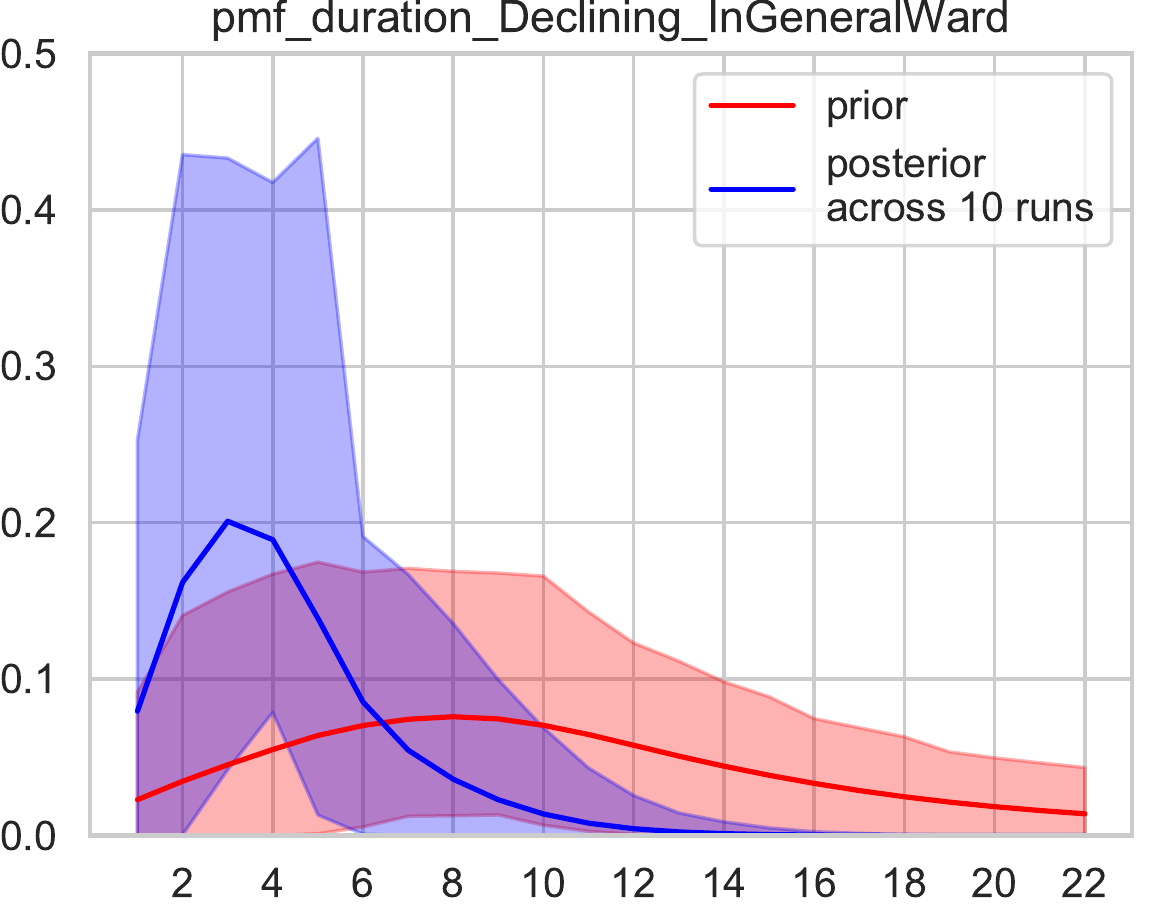}
\\
	\includegraphics[width=\PWW\textwidth]{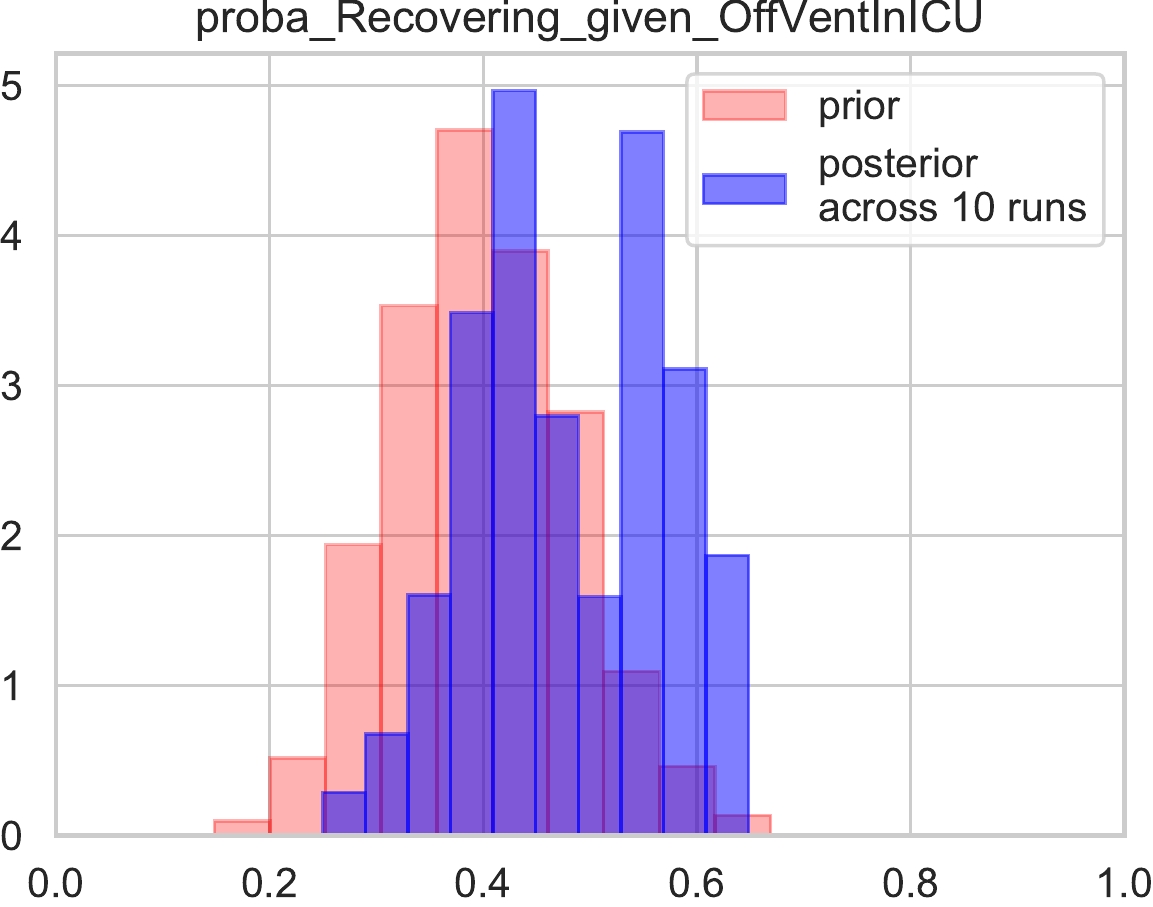}
	& 
	\includegraphics[width=\PWW\textwidth]{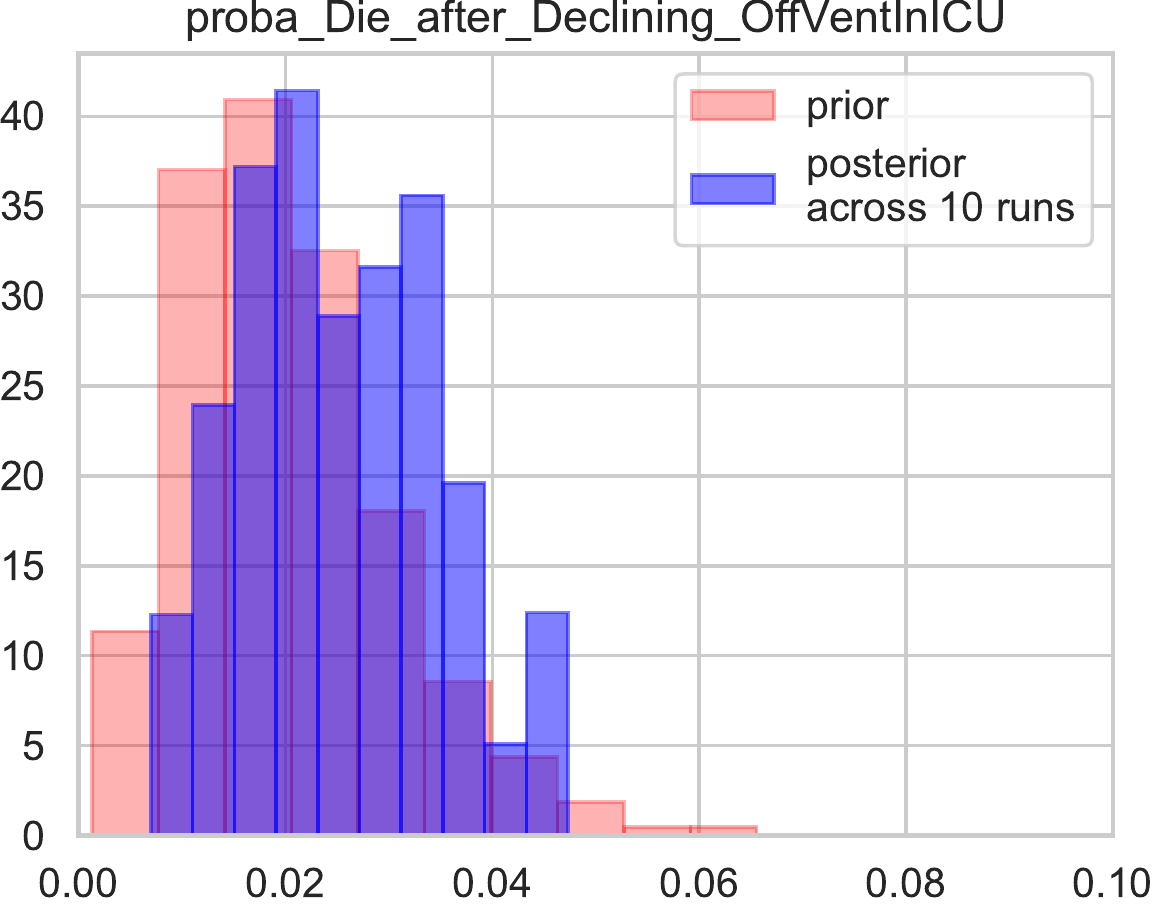}
	&
	\includegraphics[width=\PWW\textwidth]{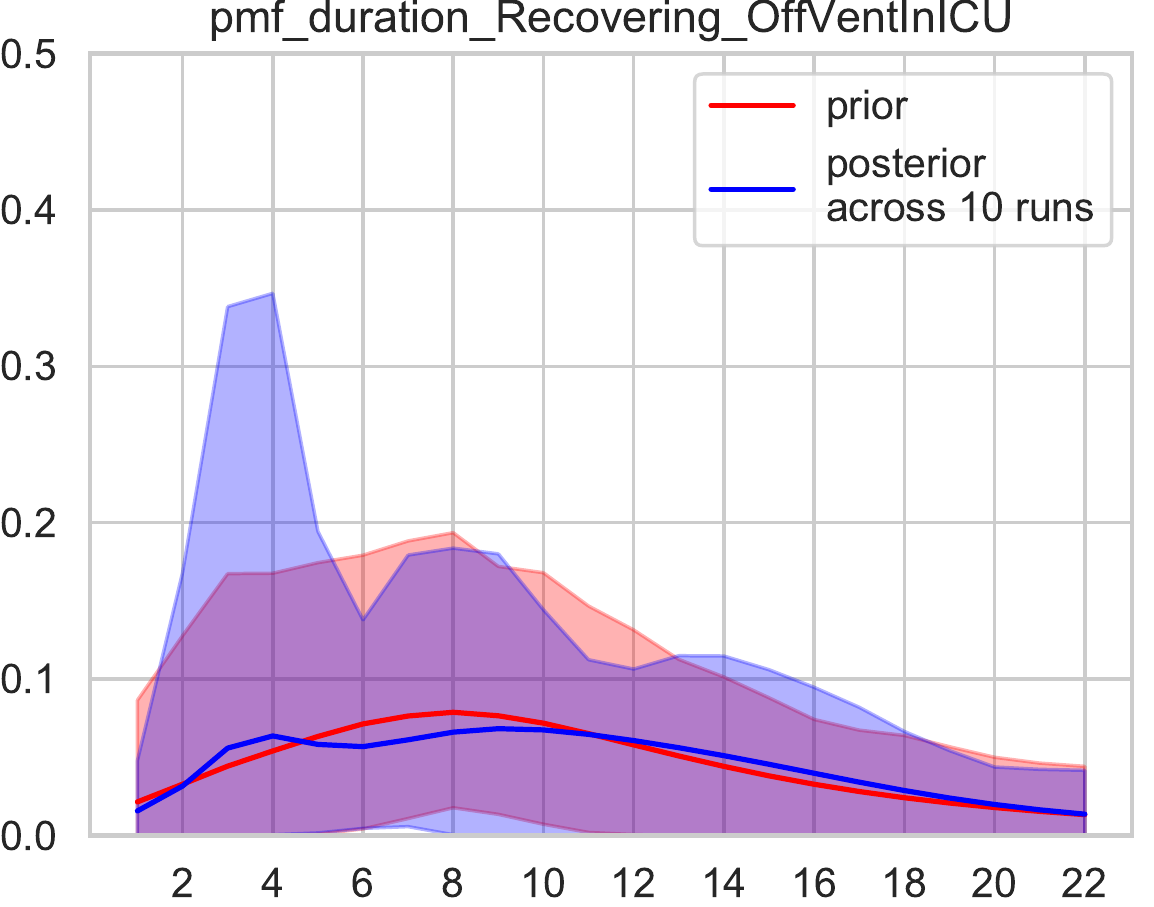}
	&
	\includegraphics[width=\PWW\textwidth]{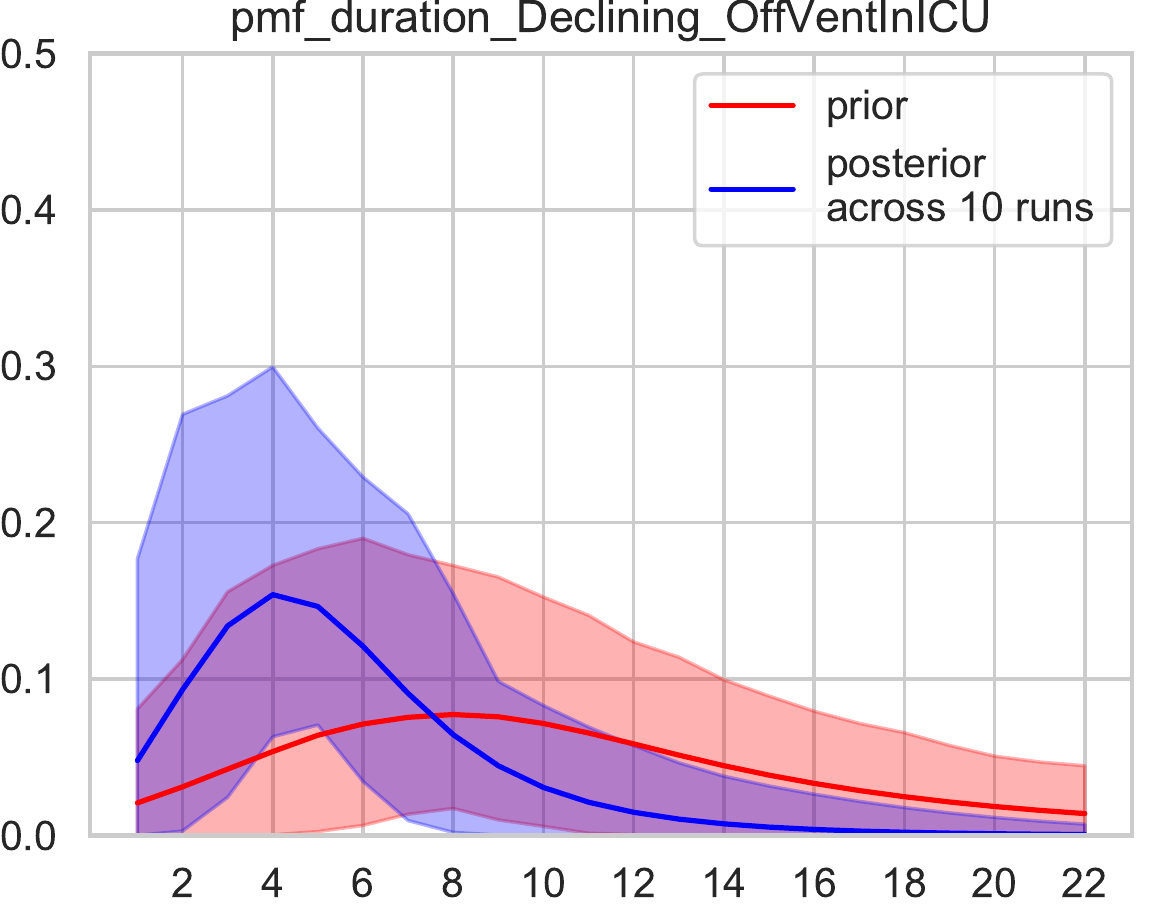}
\\
	\includegraphics[width=\PWW\textwidth]{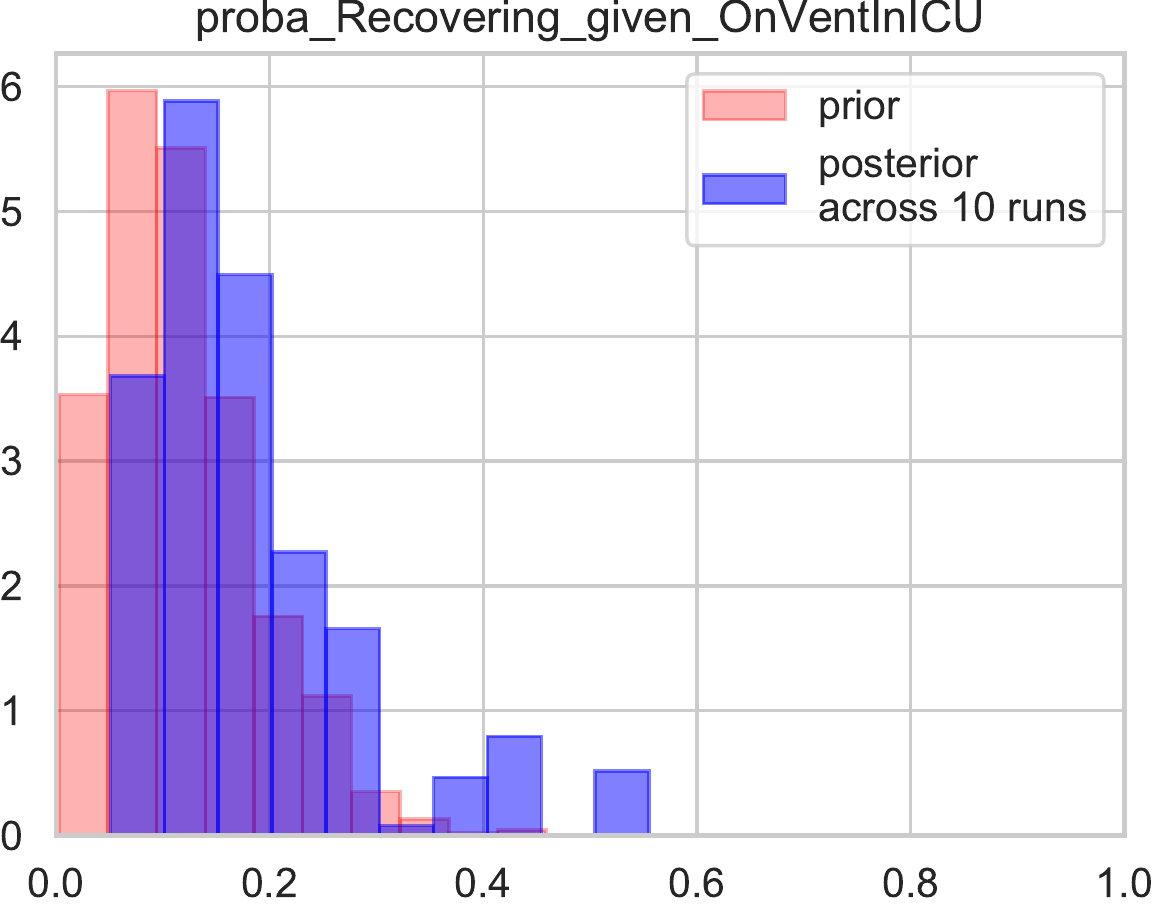}
	& 
	&
	\includegraphics[width=\PWW\textwidth]{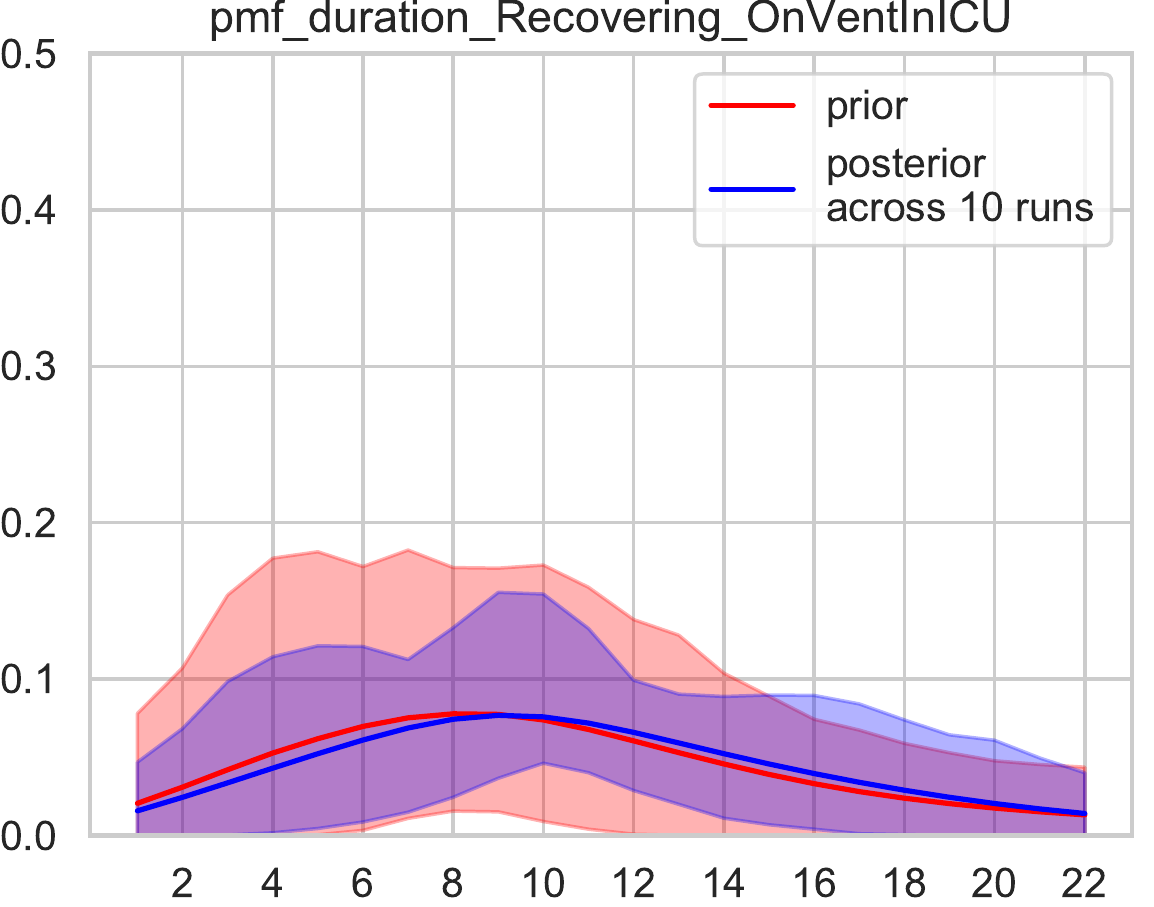}
	&
	\includegraphics[width=\PWW\textwidth]{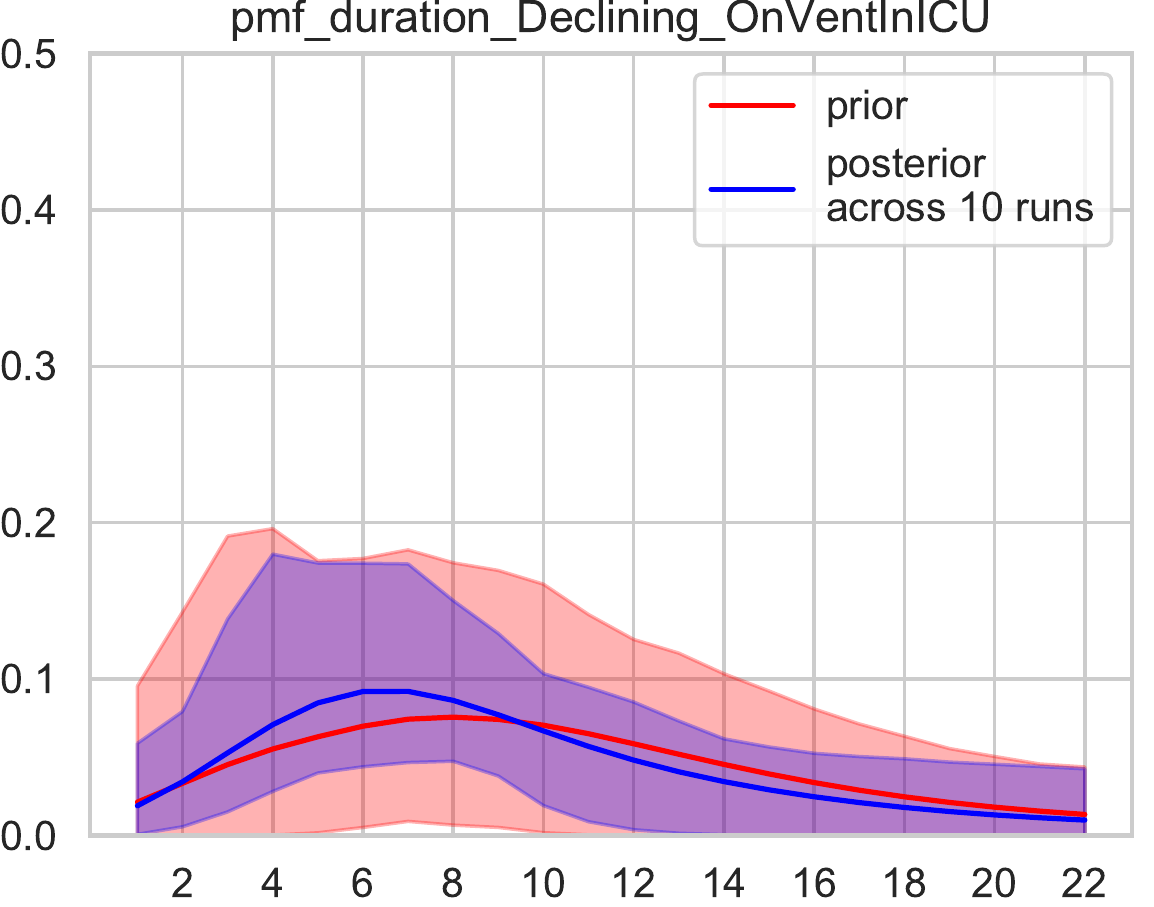}
\end{tabular}
    \caption{
    \textbf{Posterior distributions over parameters for South Tees hospital in the UK (trained on Nov. 3rd - Jan. 3rd).}
    We show transition parameters (left) and duration parameters (right) after fitting on 2 months of counts, where each day we used available census counts for $\RR, \GG+\II+\VV$, and $\VV$.
The colored interval of duration plots shows the 2.5 - 97.5th percentile intervals of 2000 samples (10 runs, each with 200 samples).
The prior is also shown for comparison.
}
\label{fig:posterior_visualization_SouthTees}
\end{figure}

We visualize the learned distributions for all parameters of our ACED-HMM in Fig.~\ref{fig:posterior_visualization_SouthTees}.

As expected, these posteriors (derived from data for a single hospital) show a higher variance (uncertainty) than the posteriors derived from the state of Massachusetts in the main paper's Fig.~\ref{fig:posterior_visualization_MA}.
The distributions shown for this U.K. hospital appear to favor longer durations. Patients appear more likely to recover both in the General Ward and in the ICU than in MA.

\subsection{Data granularity ablation study}
\label{sec:app_data_granularity}

\begin{figure}[!t]
\begin{tabular}{c c}
	\includegraphics[width=0.47\textwidth]{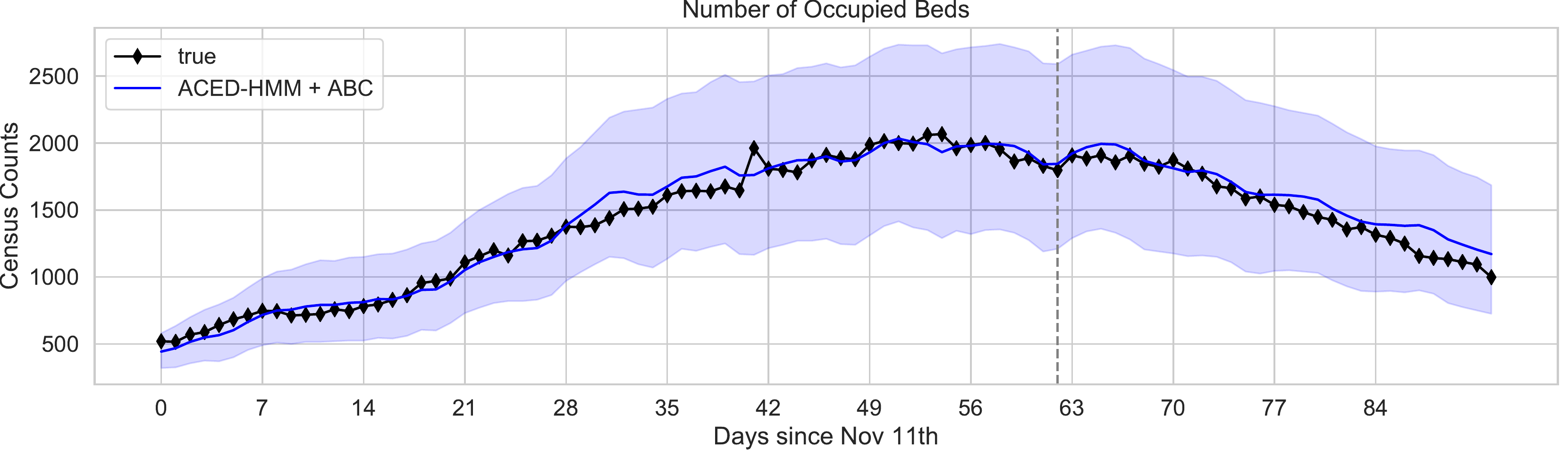}
	&
	\includegraphics[width=0.47\textwidth]{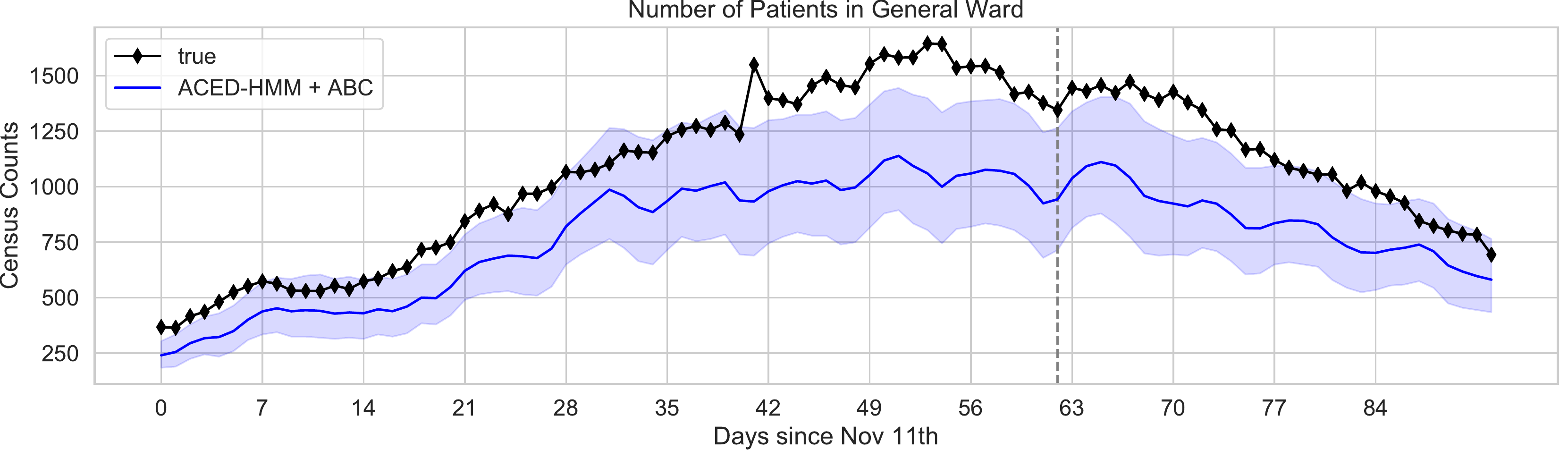}
	\\	
	\includegraphics[width=0.47\textwidth]{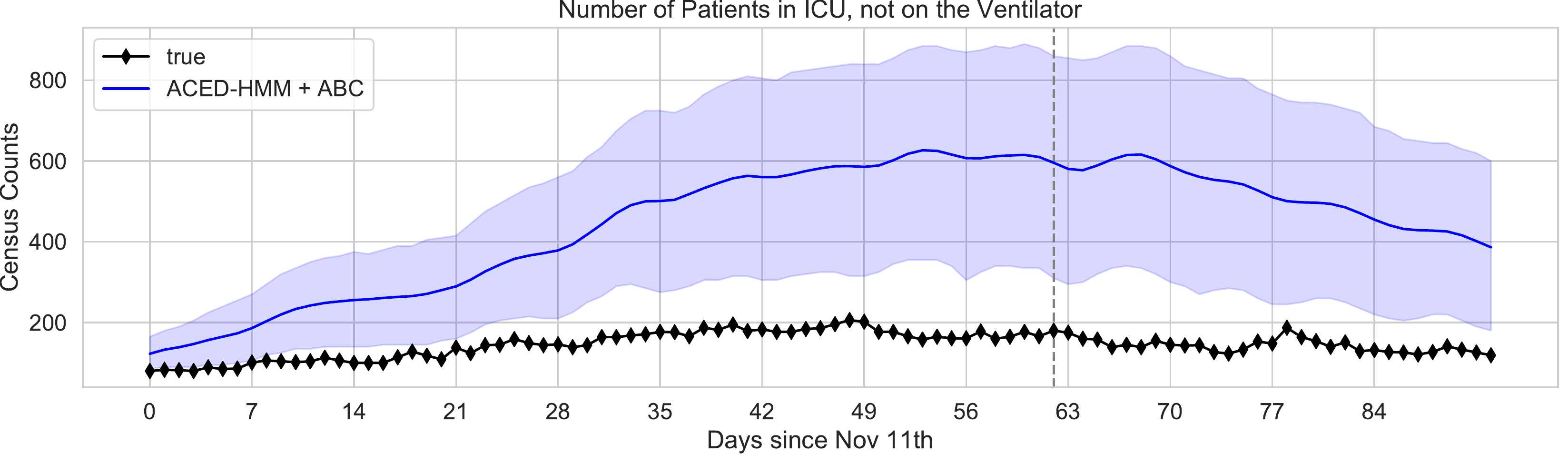}
	&
	\includegraphics[width=0.47\textwidth]{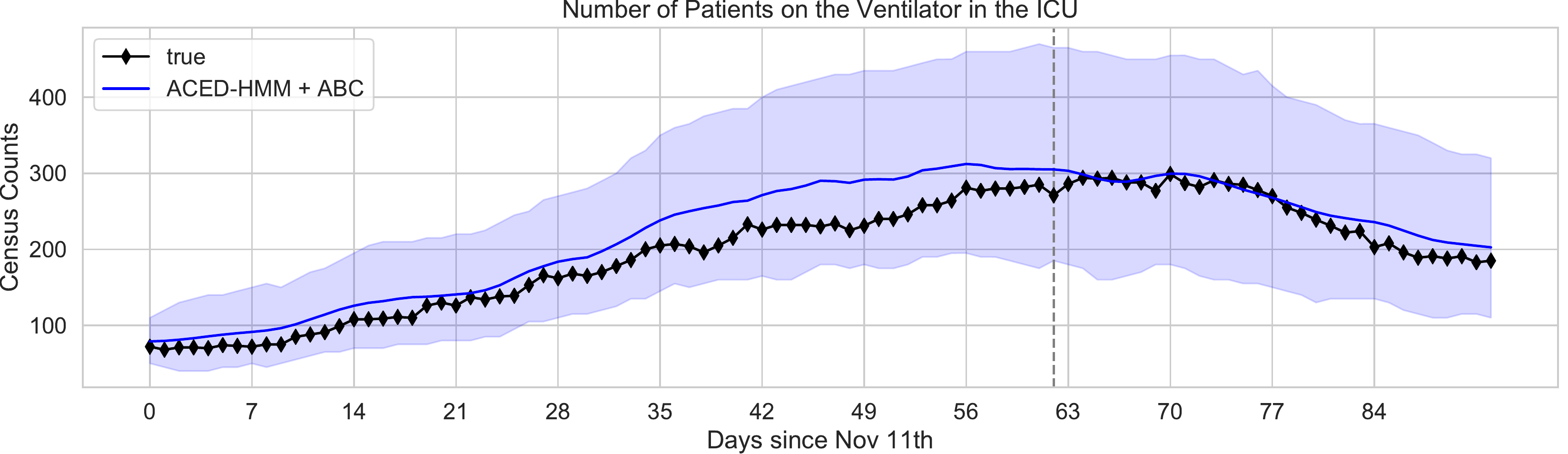}
	\\
\end{tabular}
\caption{\textbf{Fit and forecasts on MA data for our data granularity ablation study.} Here, the ACED-HMM was trained to match the total number of hospital beds ($\GG+\II+\VV$) as well as terminal counts ($\XX$) for MA state-level data. (Compare to Fig.~\ref{fig:forecast_MA}, which fit on counts for individual stages $\GG$, $\II$, $\VV$, and $\XX$).
The model is able to fit and forecast the total bed counts accurately (upper-left), but did not recover the counts at some individual compartments ($\GG$ : lower left, $\II$ upper right) well. The ventilator counts ($\VV$ : lower right) are near-optimally recovered as they have a very direct influence over the observed $\XX$ counts.}
\label{fig:data_ablation_MA}
\end{figure}

To assess the model's sensitivity to data granularity, we trained the ACED-HMM on the Massachusetts data we used in our other experiments, but with a key manually imposed difference in available data: instead of fitting to the finest granularity of available counts -- $\GG$, $\II$, $\VV$ and $\XX$ separately -- we instead aggregate all intermediate stages and fit the parameters to daily counts of all beds ($\GG+\II+\VV$) and death ($\XX$) only.
See Fig.~\ref{fig:data_ablation_MA} for fits and forecasts to $\GG$, $\II$, $\VV$ and $\GG+\II+\VV$ counts, and Fig.~\ref{fig:posterior_visualization_MA_ablation} for the learned model parameters.

We find that, while the model trained with less-granular data can fit the aggregate ``all beds'' counts well in both training and test periods, the model could not recover individual stage counts from the $\GG$ and $\II$ stages well.
In particular, the algorithm converged to parameters that undershot $\GG$ counts and overshot $\II$ counts. The differing signs of these errors roughly cancel out and thus aggregate counts are predicted well.
Interestingly, the model managed better predictions of $\VV$. We believe this is due to the presence of $\XX$ counts at training time, as these values provide a more direct signal for what the $\VV$ counts should be. 

This experiment shows that, as expected, forecasts on fine-grained counts that the model was not trained on are generally not reliable, unless some closely-related count is provided (e.g. $\XX$ is provided as a signal for $\VV$ or $\RR$ is provided as a signal for $\GG$).
Stronger, more informative priors about the intermediate stages may also improve performance when fine-grained signals are not available.

\renewcommand{\PWW}{0.2}
\begin{figure}[!t]
\centering
\begin{tabular}{c c | c c}
	\includegraphics[width=\PWW\textwidth]{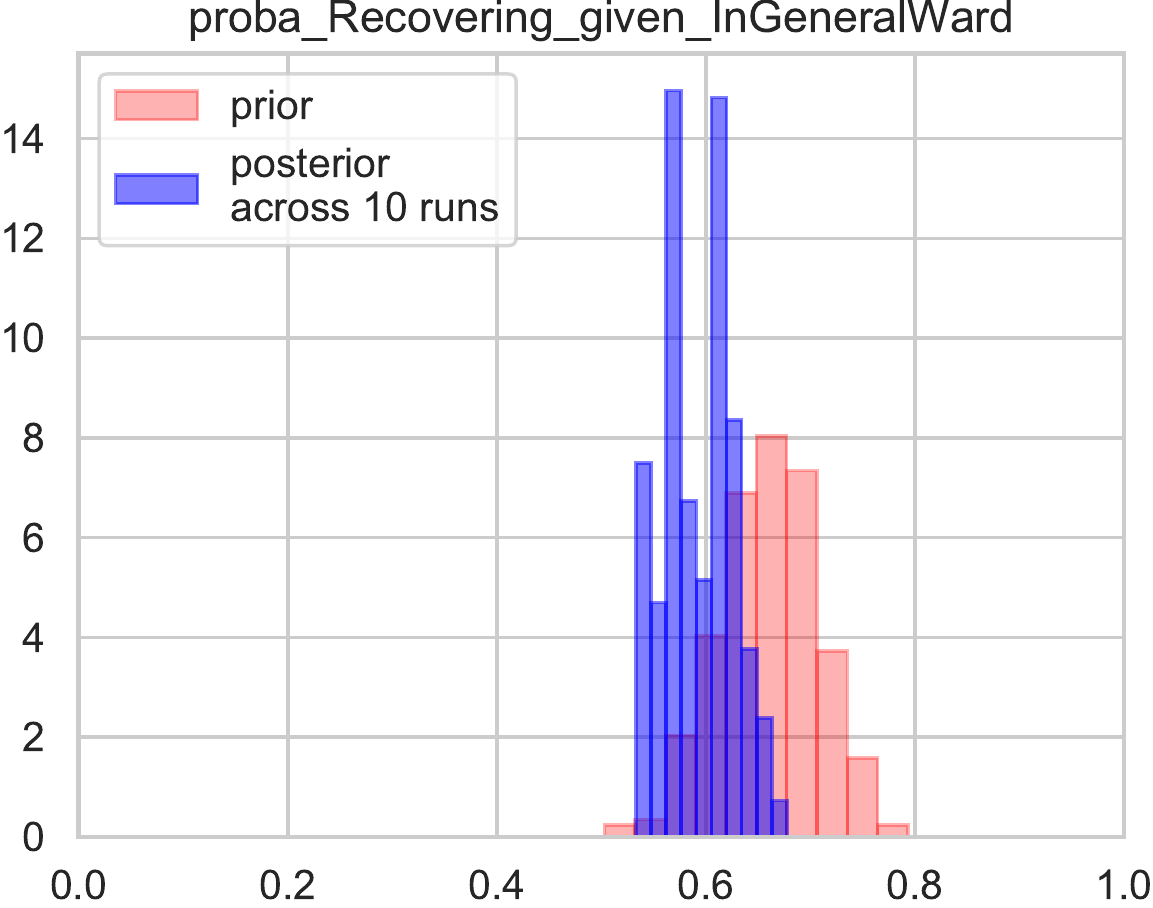}
	& 
	\includegraphics[width=\PWW\textwidth]{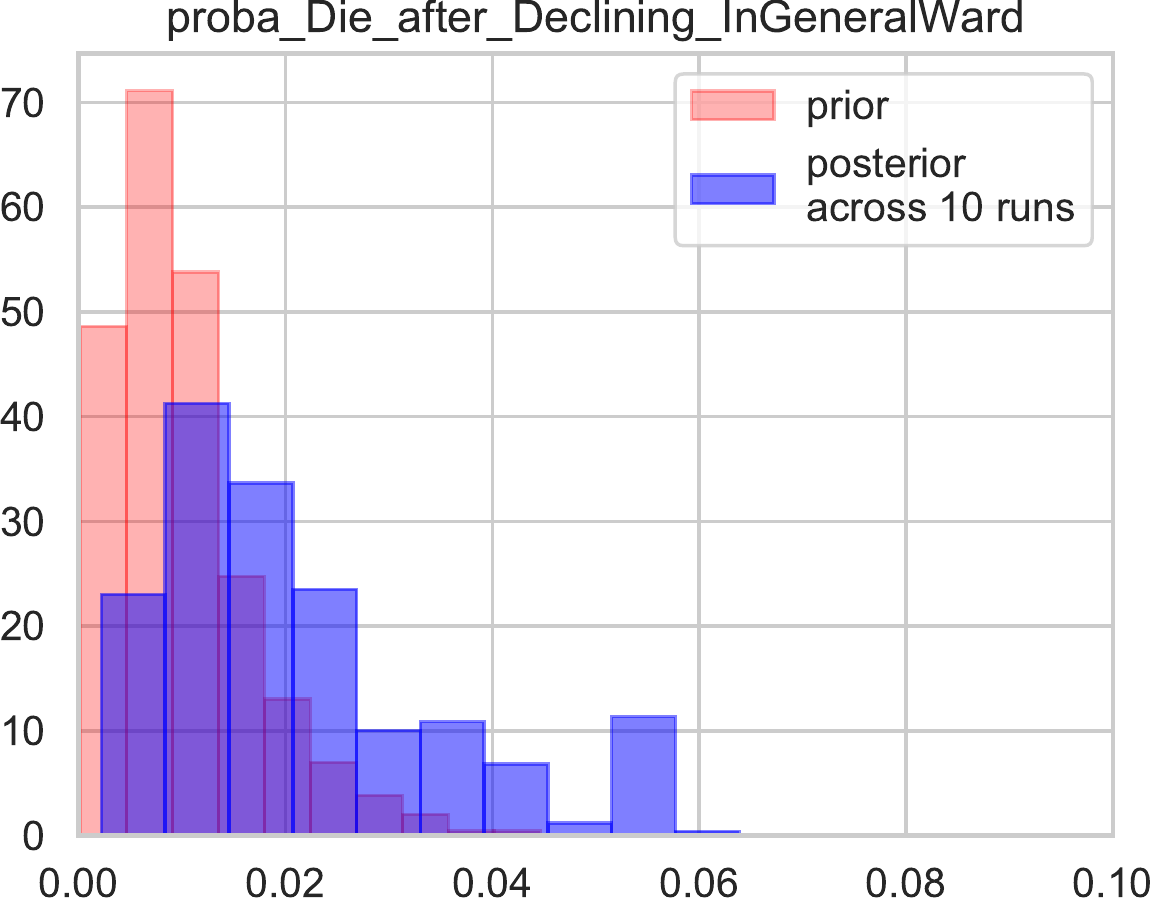}
	&
	\includegraphics[width=\PWW\textwidth]{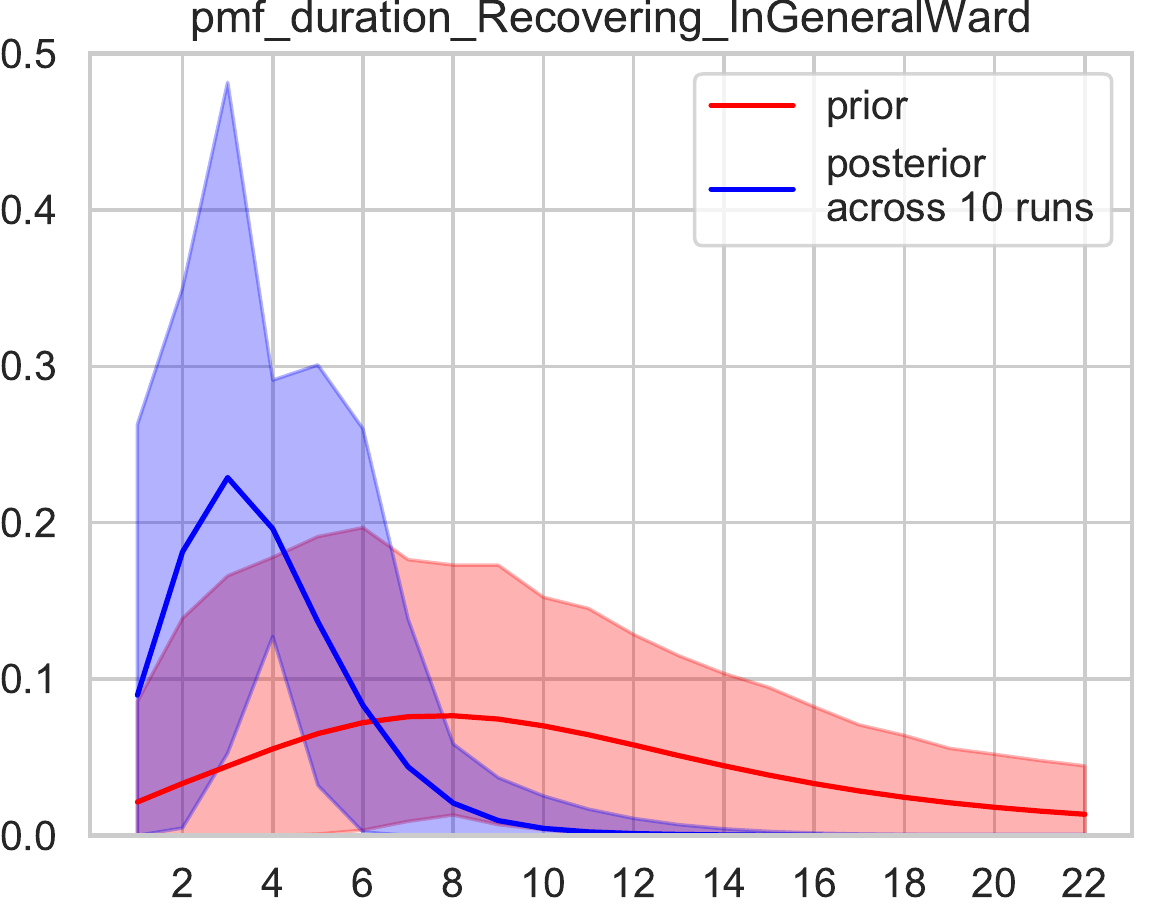}
	&
		\includegraphics[width=\PWW\textwidth]{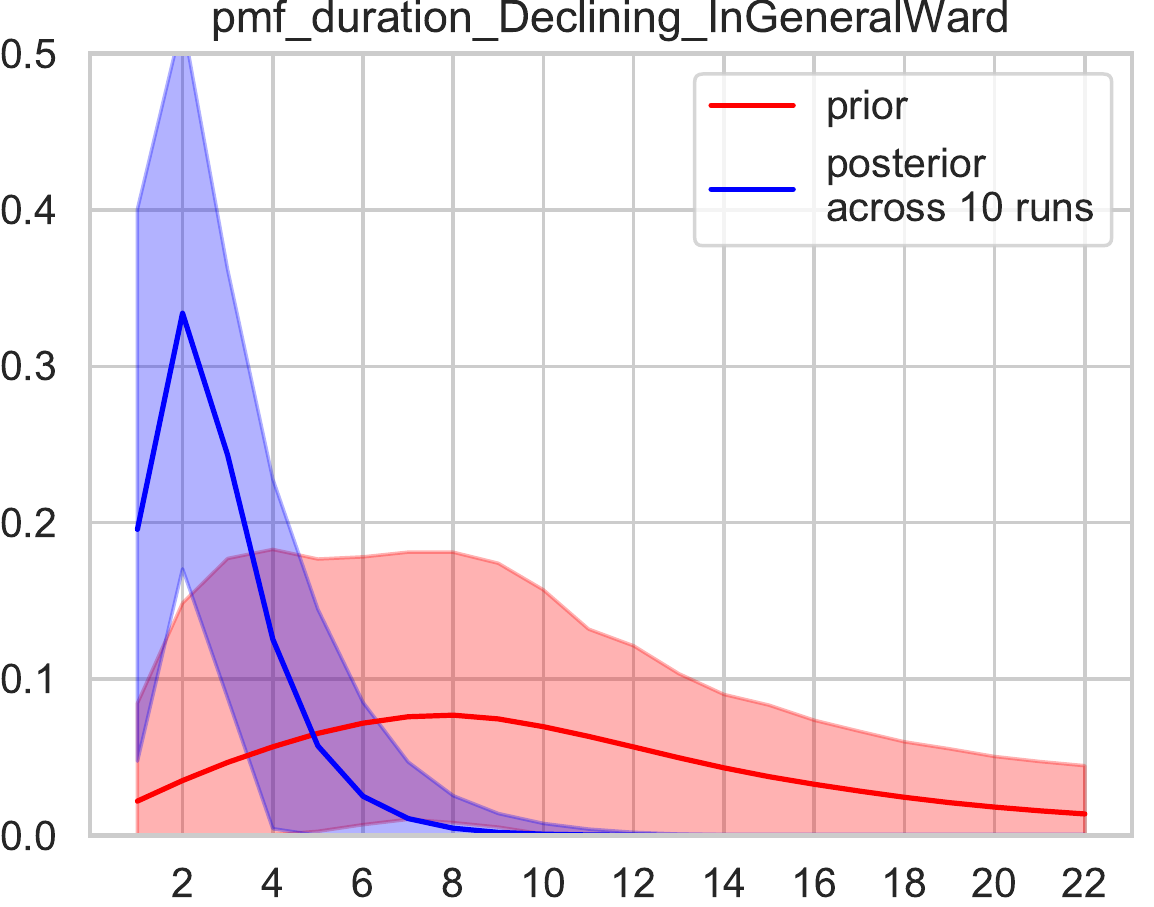}
\\
	\includegraphics[width=\PWW\textwidth]{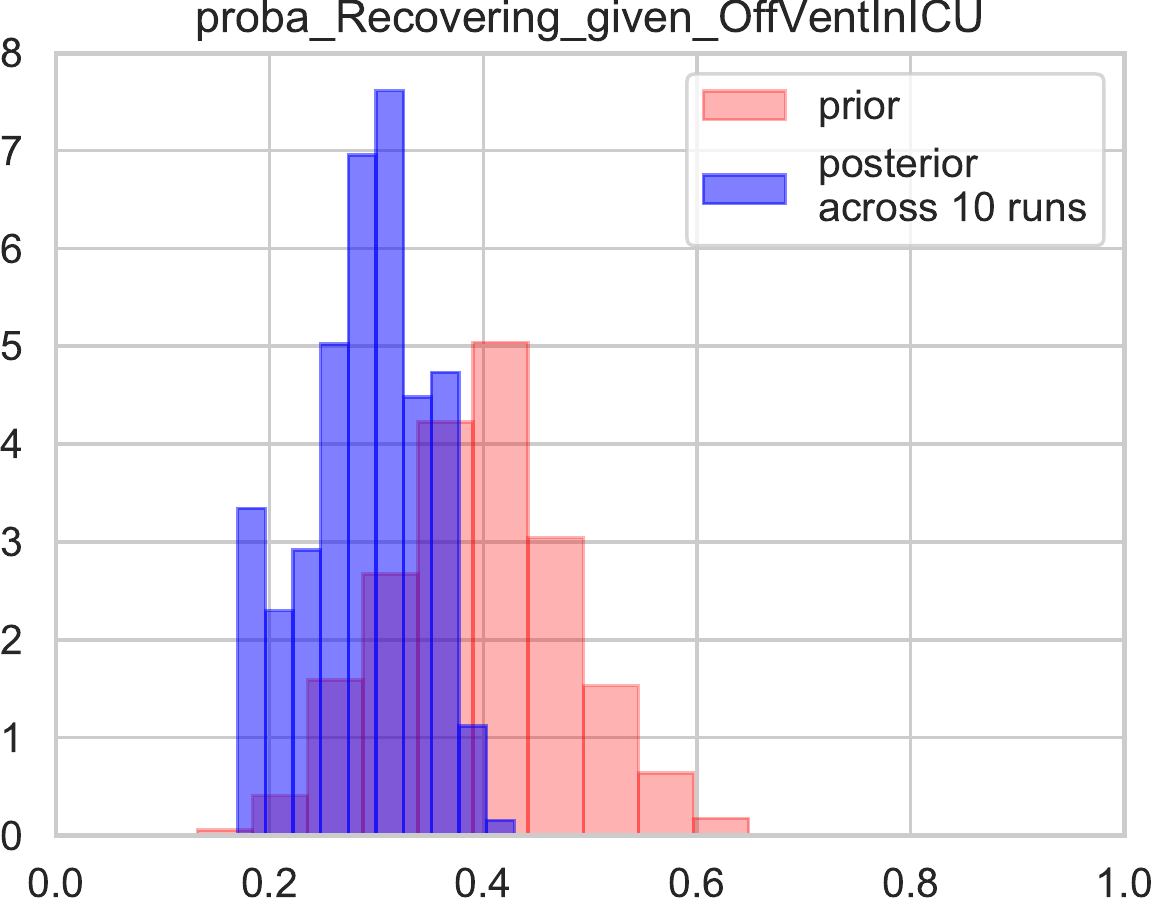}
	& 
	\includegraphics[width=\PWW\textwidth]{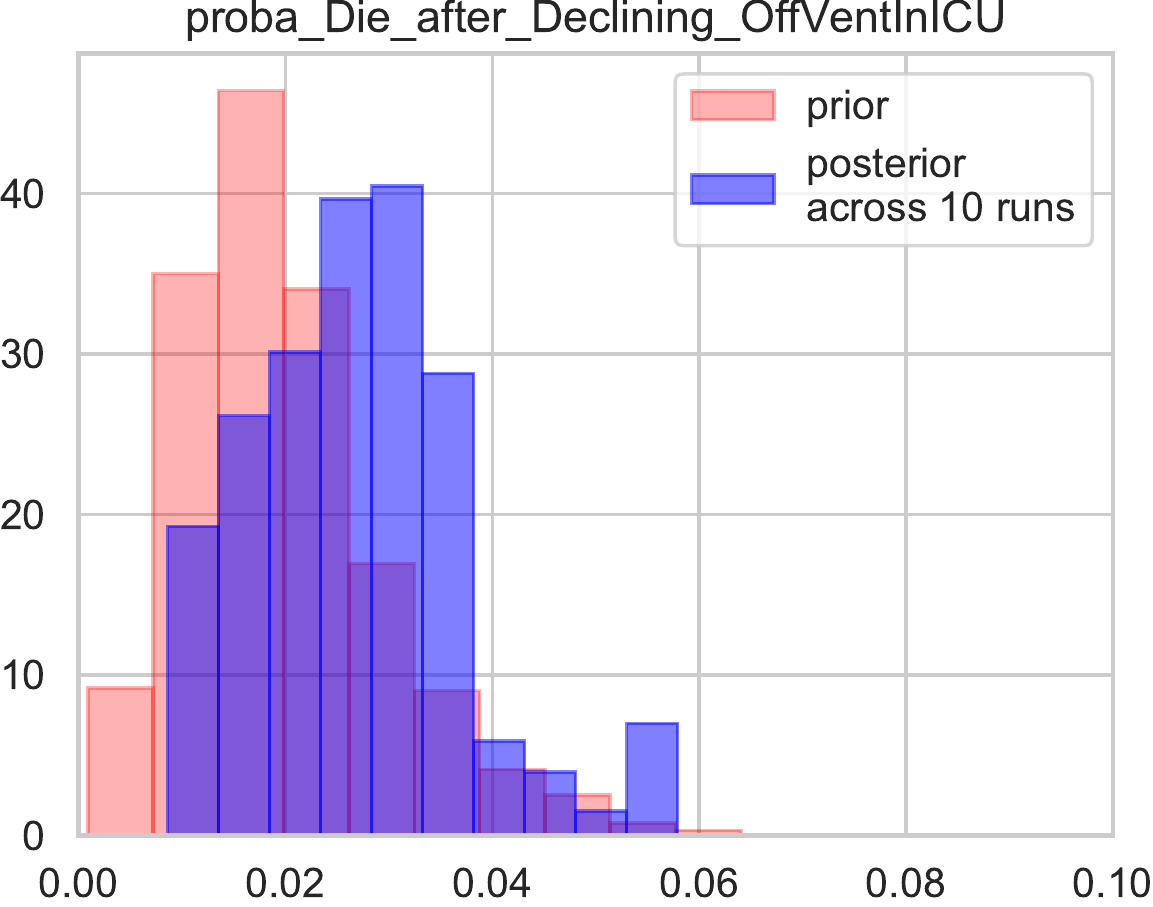}
	&
	\includegraphics[width=\PWW\textwidth]{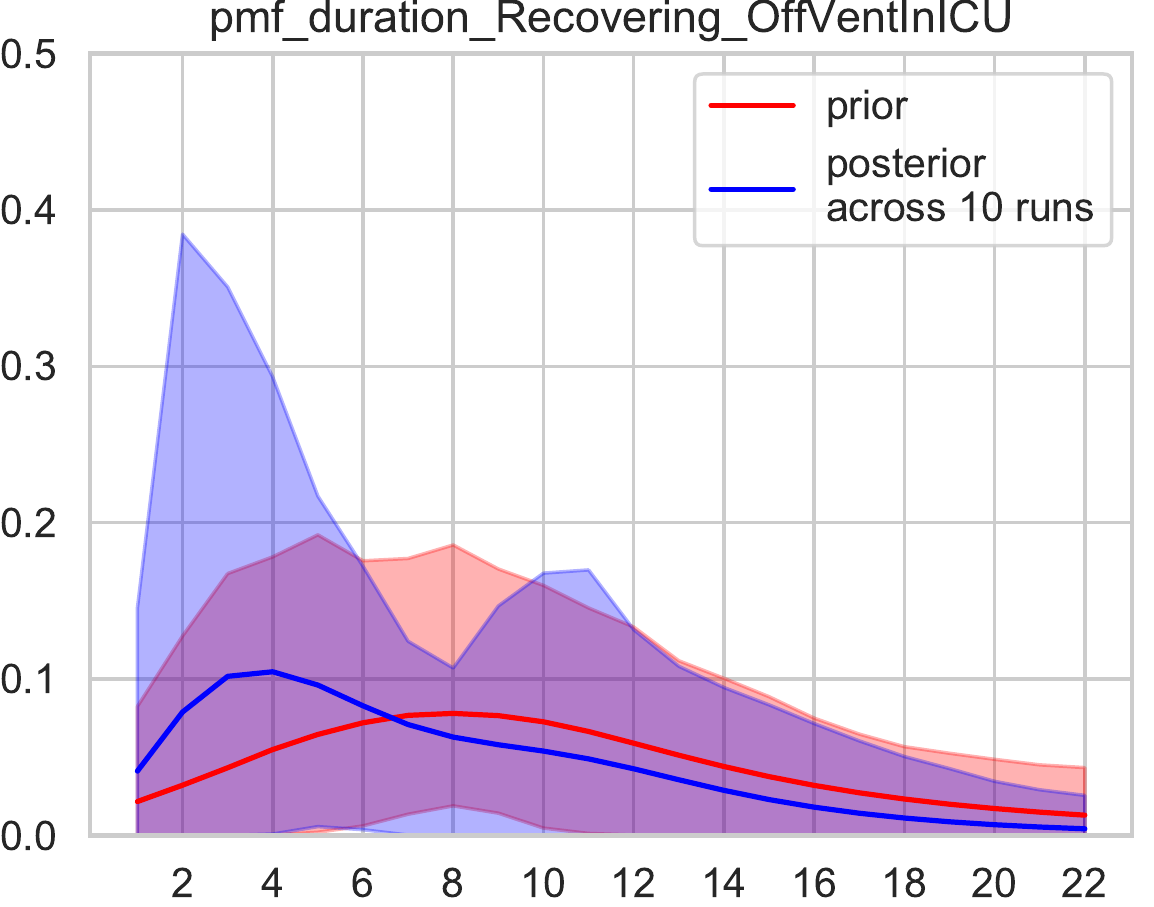}
	&
	\includegraphics[width=\PWW\textwidth]{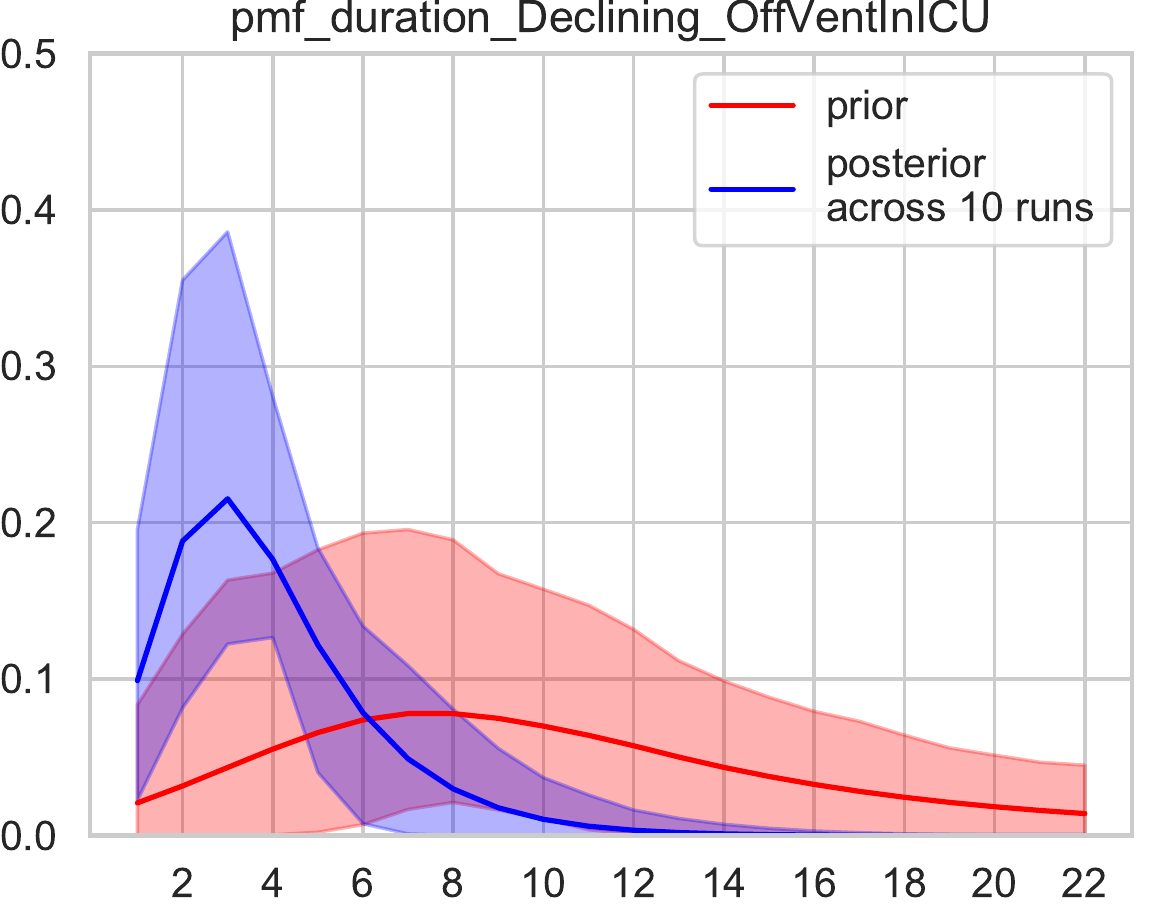}
\\
	\includegraphics[width=\PWW\textwidth]{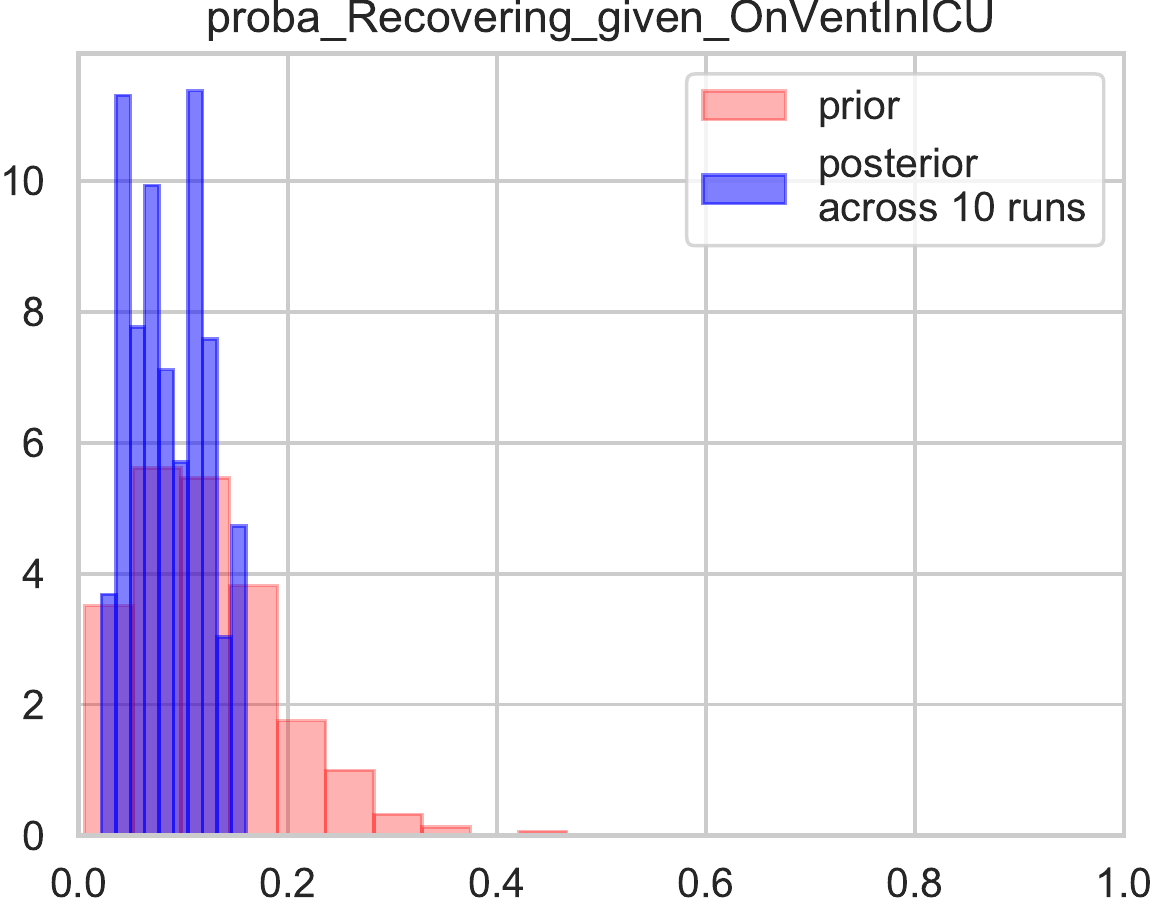}
	& 
	&
	\includegraphics[width=\PWW\textwidth]{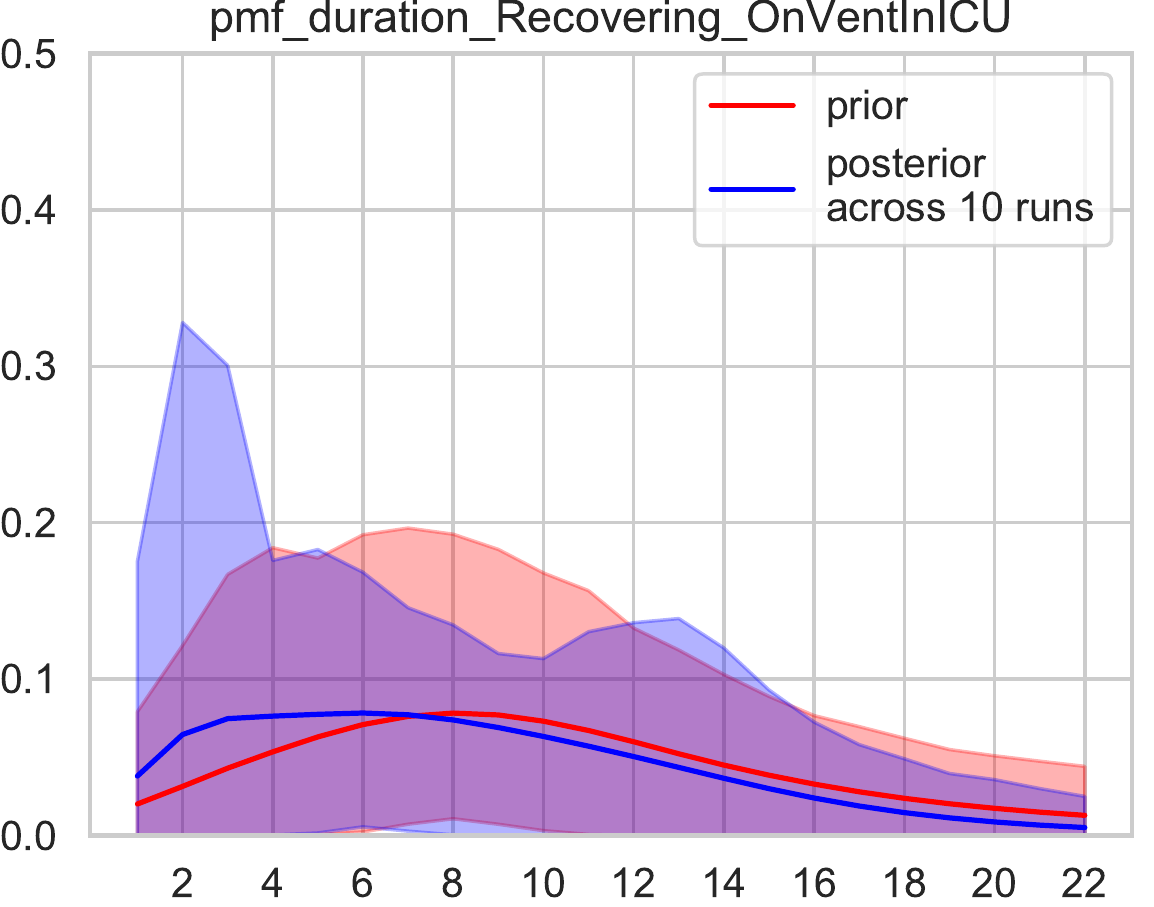}
	&
	\includegraphics[width=\PWW\textwidth]{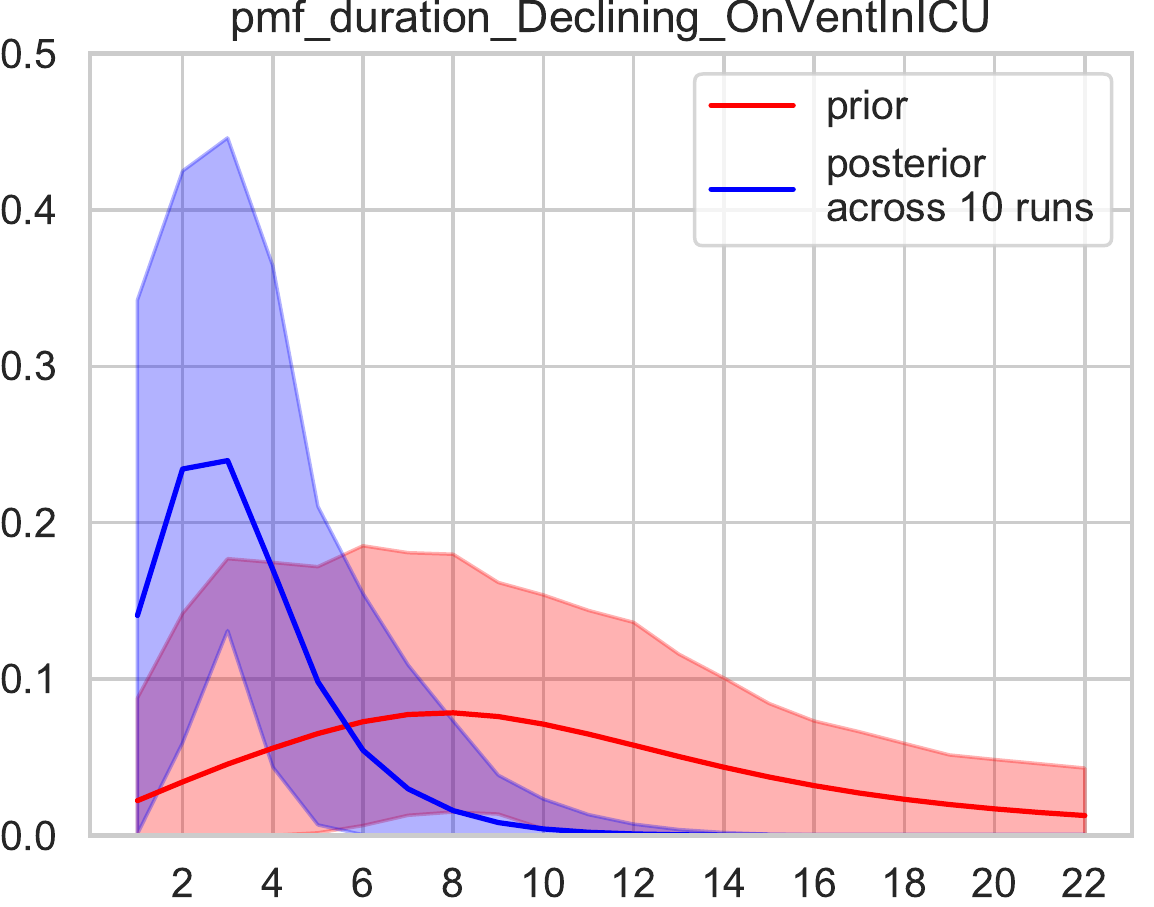}
\end{tabular}
    \caption{
    \textbf{Posterior distributions over parameters for MA obtained from our data granularity ablation study.}
    Here, the ACED-HMM was trained to match the total number of hospital beds ($\GG+\II+\VV$) as well as terminal counts ($\XX$) for MA state-level data. (Compare to Fig.~\ref{fig:posterior_visualization_MA}, which fit on counts for individual stages $\GG$, $\II$, $\VV$, and $\XX$).
    The colored interval of duration plots shows the 2.5 - 97.5th percentile intervals of 2000 samples (10 runs, each with 200 samples).
}
\label{fig:posterior_visualization_MA_ablation}
\end{figure}

\newpage

\subsection{Scalability Assessment}
\label{sec:app_scalability}

\begin{figure}[!b]
\begin{tabular}{c c}
	Exact & 5x Scaling
	\\
	\includegraphics[width=0.47\textwidth]{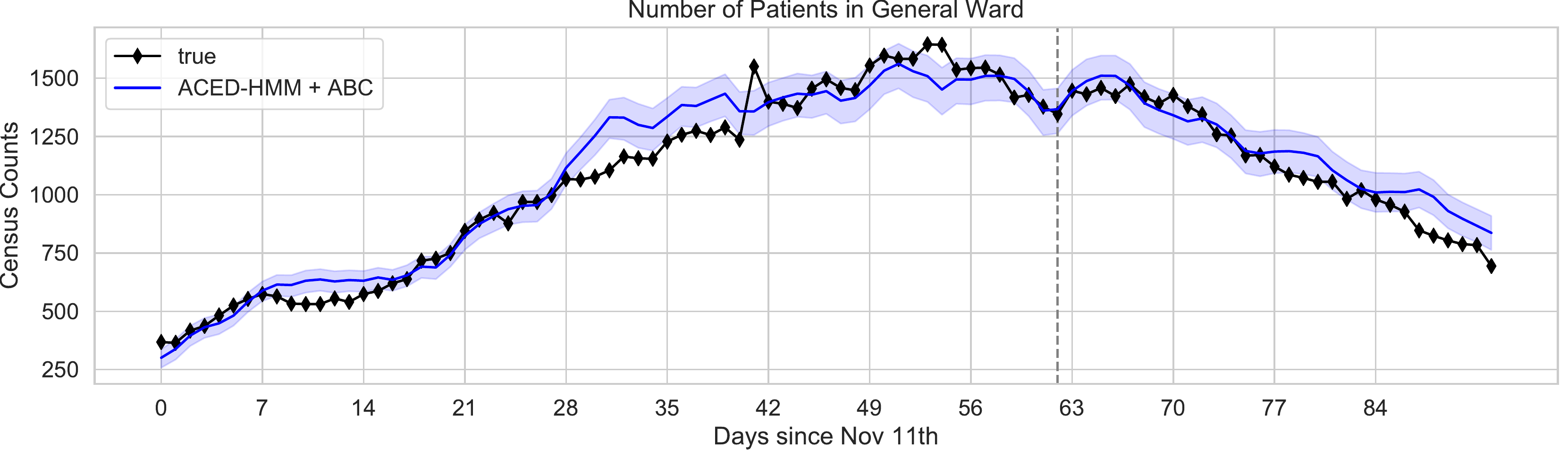}
	&	
	\includegraphics[width=0.47\textwidth]{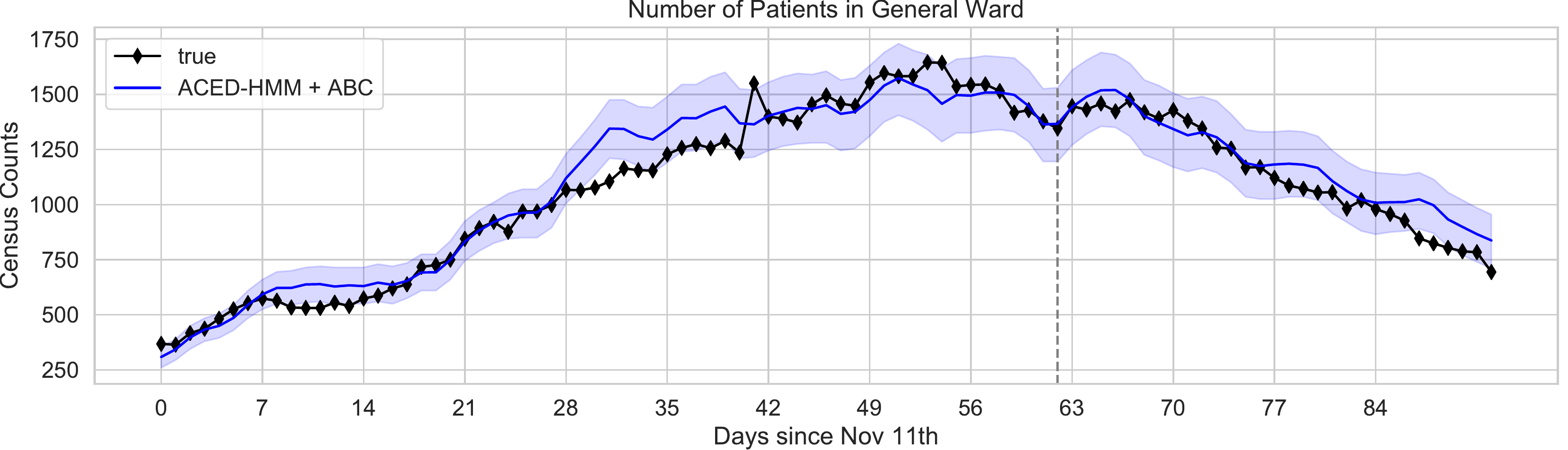}
	\\
	\includegraphics[width=0.47\textwidth]{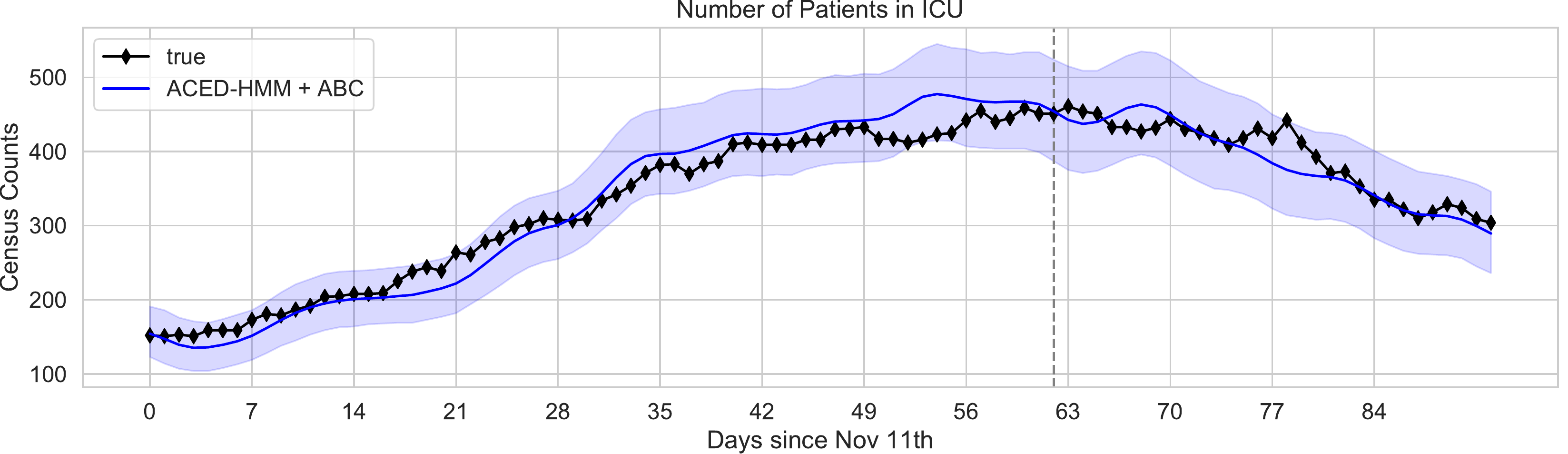}
	&	
	\includegraphics[width=0.47\textwidth]{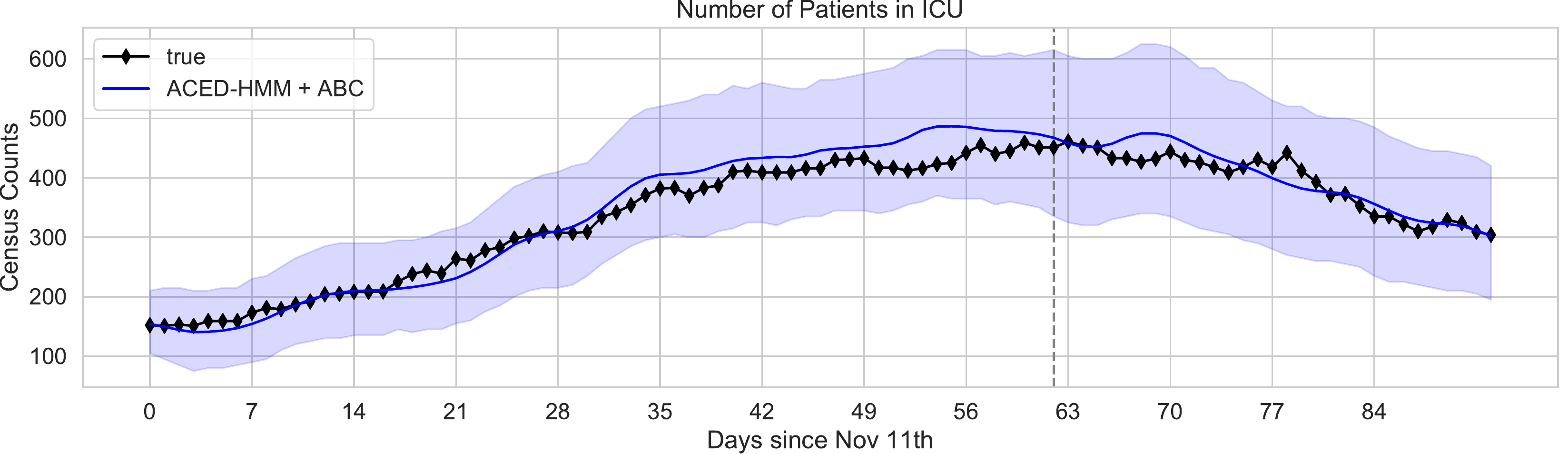}
\end{tabular}
\caption{\textbf{Scalability assessment for MA data.} We show the predictions for our ABC method using all admissions recorded in the training period (left), as well as our proposed scaling procedure with a 5x scale factor (right).
The scaling simulates only 20\% of observed admissions but then scales all counts by 5.
The scaling procedure allows simulations to run 5 times faster with the same qualitative trend recovery and only modest increases in uncertainty intervals.
}
\label{fig:exact_vs_5x_scaling_MA}
\end{figure}

We visually assess the performance of our procedure for scaling up simulations described in Sec.~\ref{sec:scalability} on data from Massachusetts in Fig.~\ref{fig:exact_vs_5x_scaling_MA}.
Despite simulating 5x fewer patient trajectories (and thus being substantially cheaper), our scaling leads to reasonably similar overall trends in the forecasts with only modestly increased uncertainty intervals.

\subsection{Synthetic Data Assessment: Can we recover ``ideal'' model parameters?}
\label{sec:synthetic_data}
\begin{figure}[t!]
\centering
\begin{tabular}{c c | c c}
	\includegraphics[width=\PWW\textwidth]{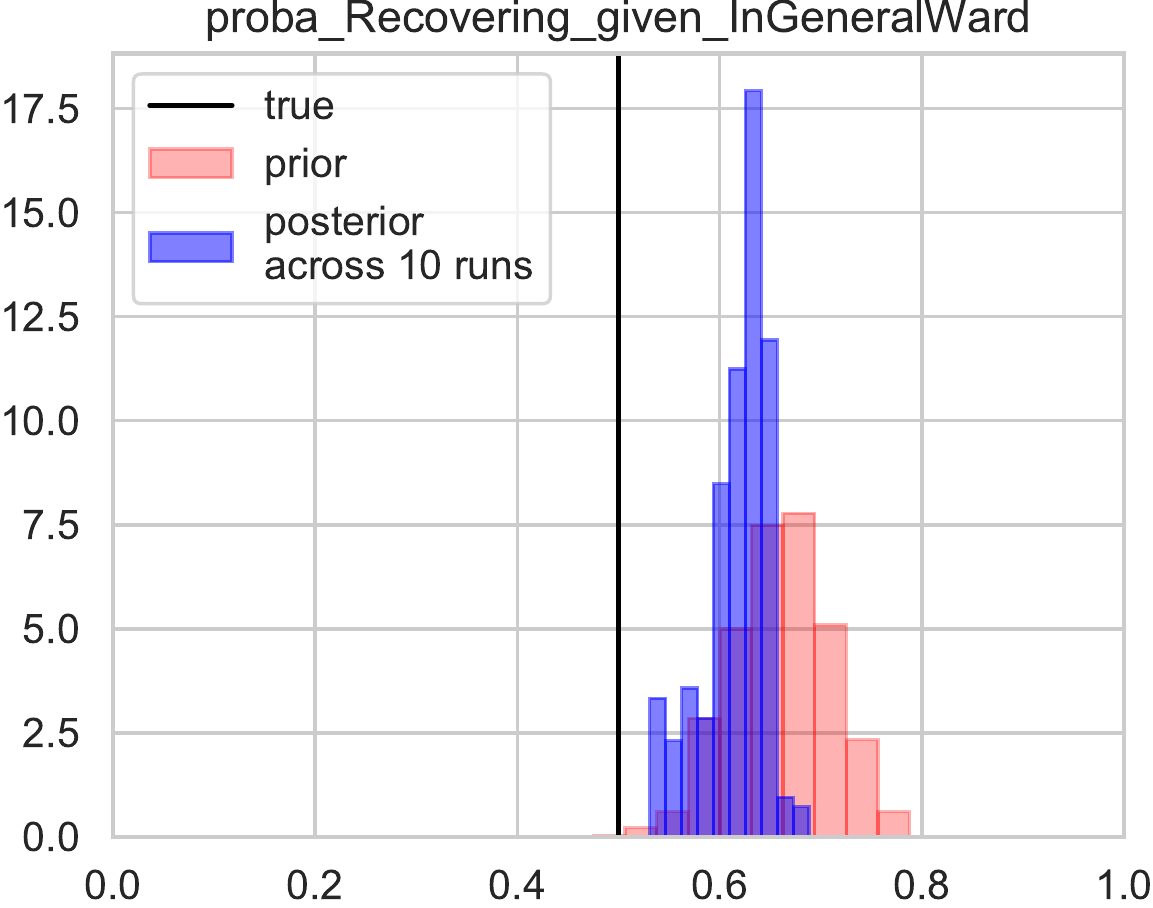}
	& 
	\includegraphics[width=\PWW\textwidth]{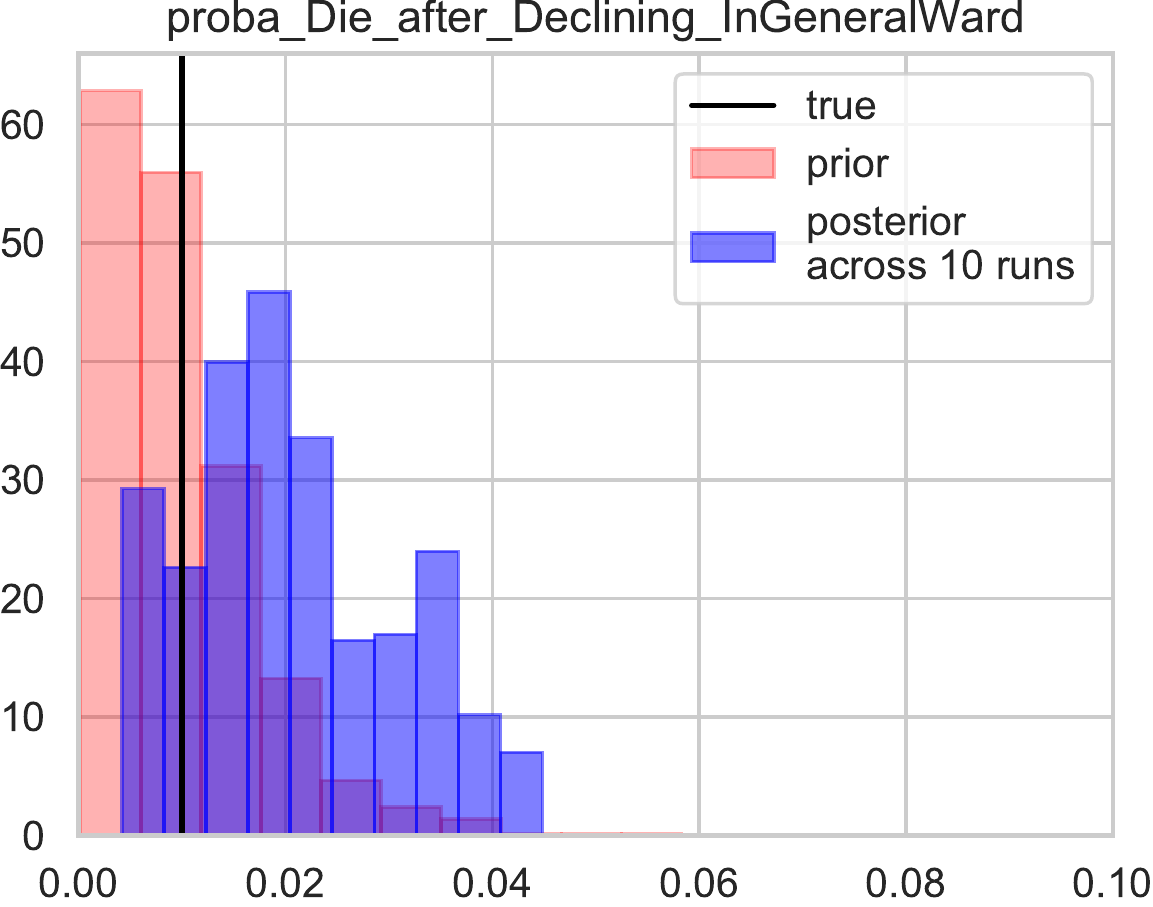}
	&
	\includegraphics[width=\PWW\textwidth]{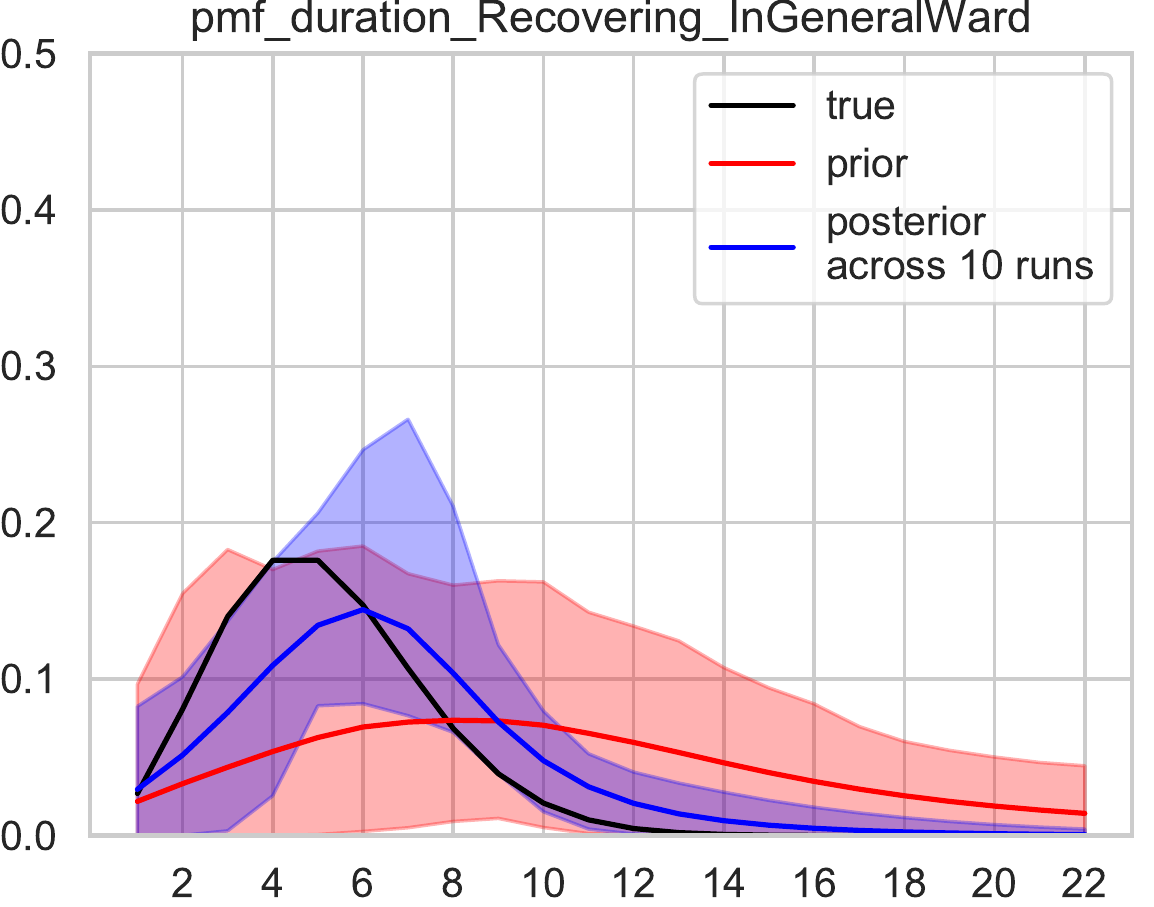}
	&
		\includegraphics[width=\PWW\textwidth]{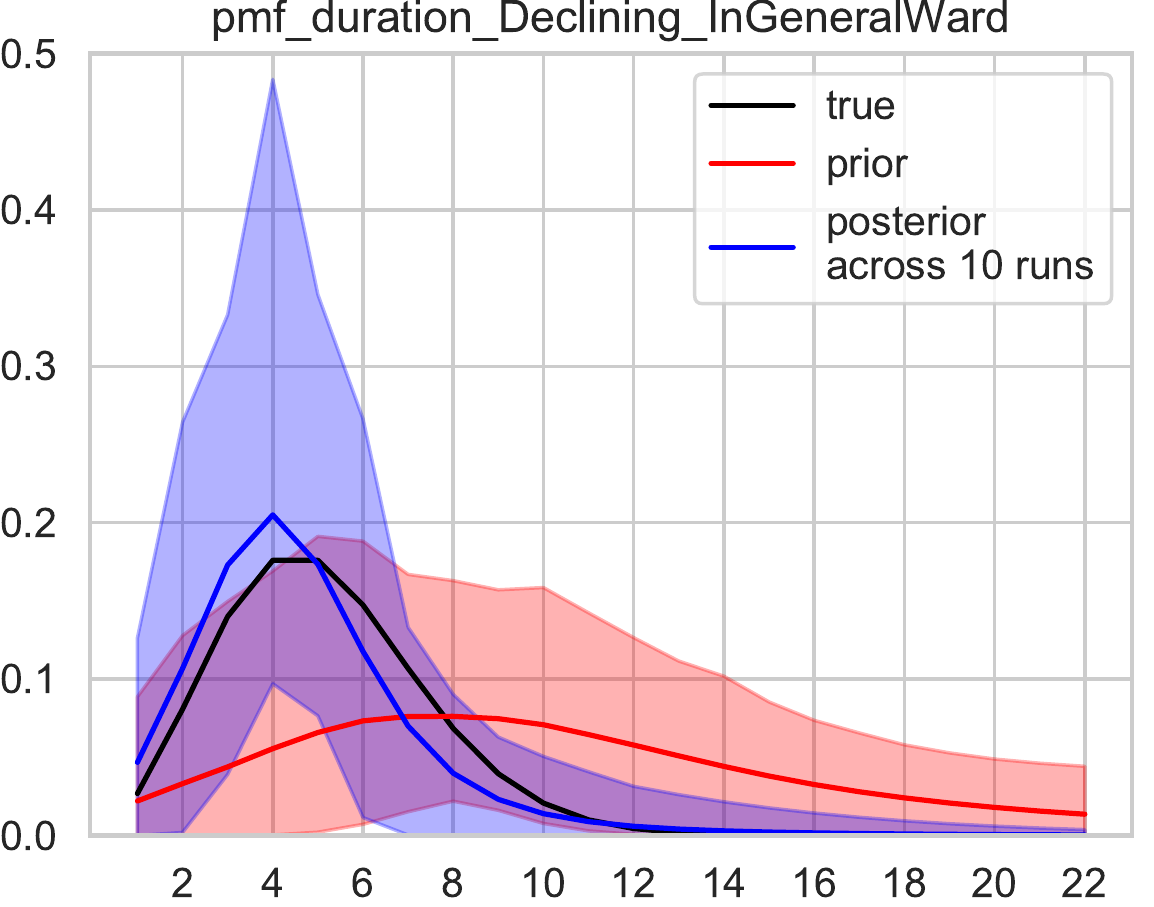}
\\
	\includegraphics[width=\PWW\textwidth]{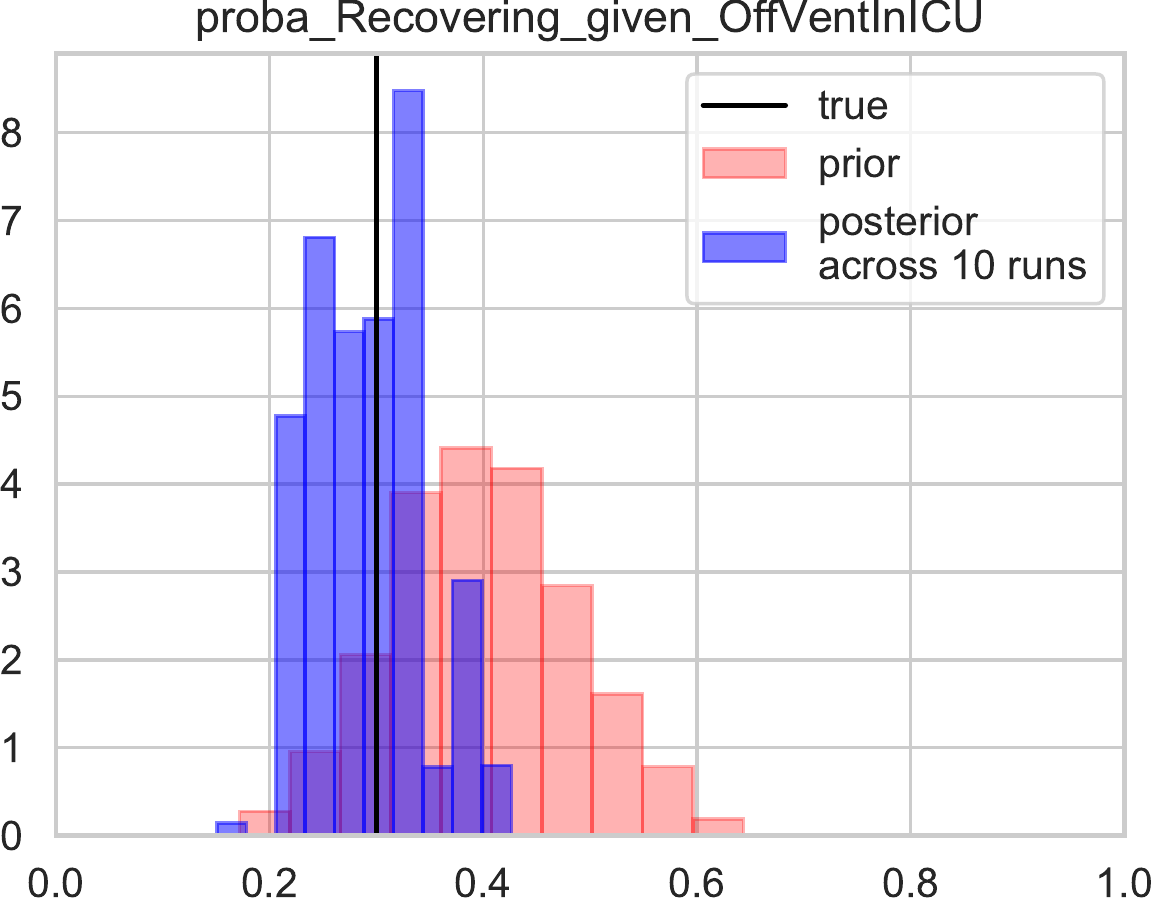}
	& 
	\includegraphics[width=\PWW\textwidth]{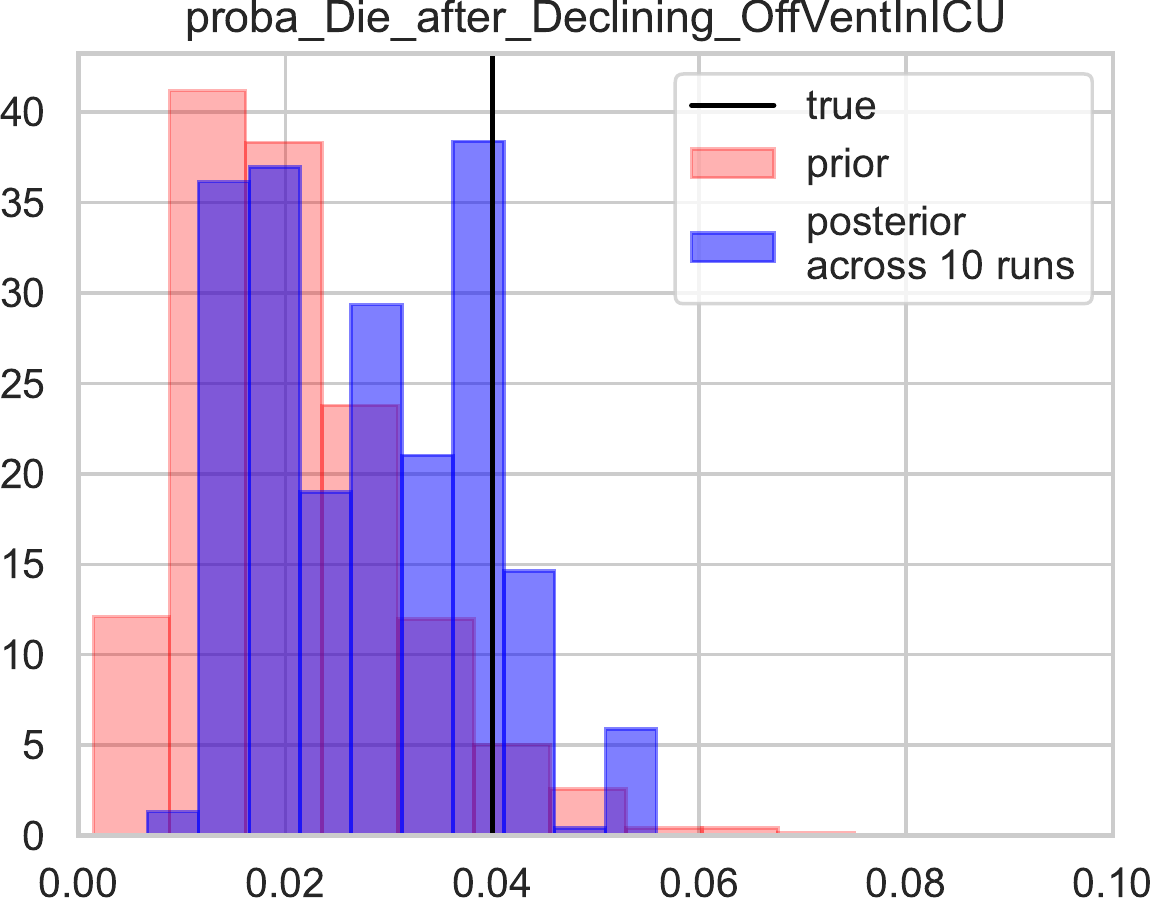}
	&
	\includegraphics[width=\PWW\textwidth]{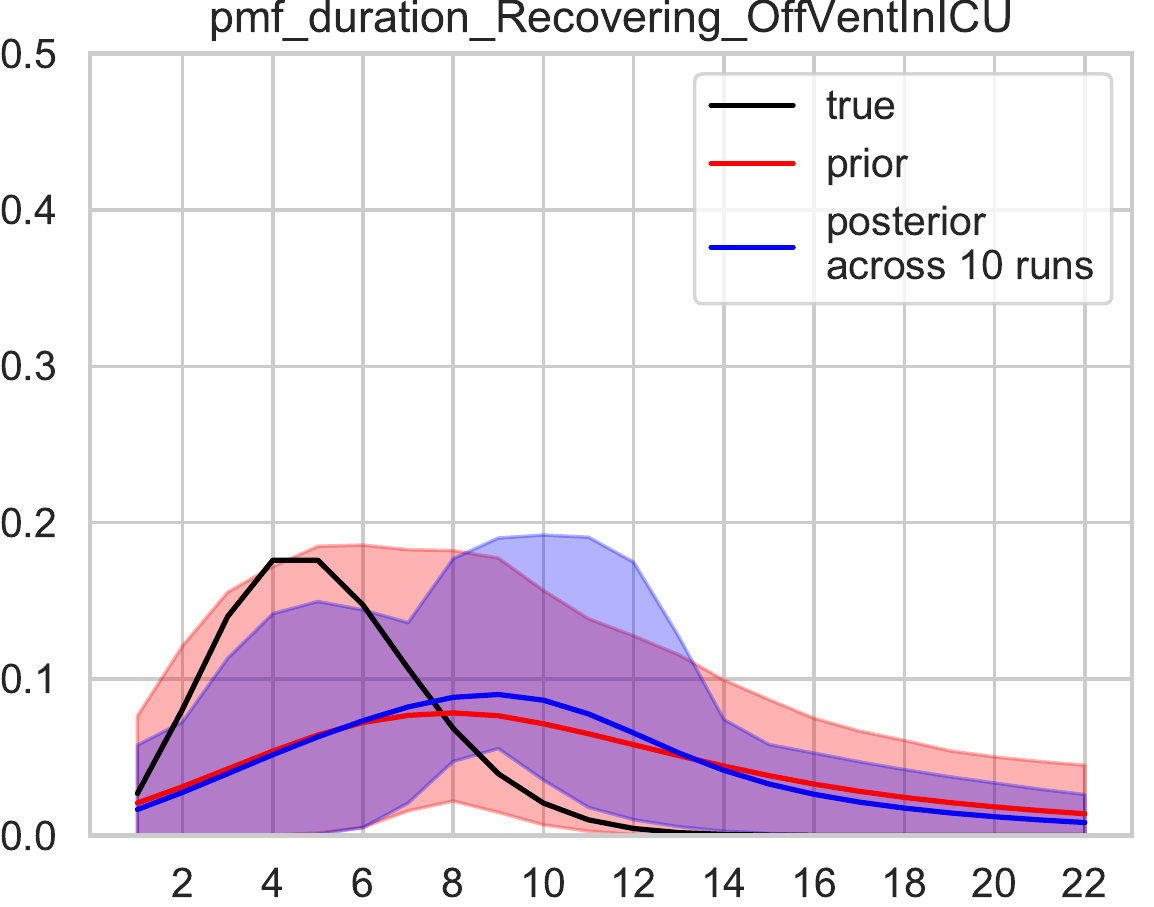}
	&
	\includegraphics[width=\PWW\textwidth]{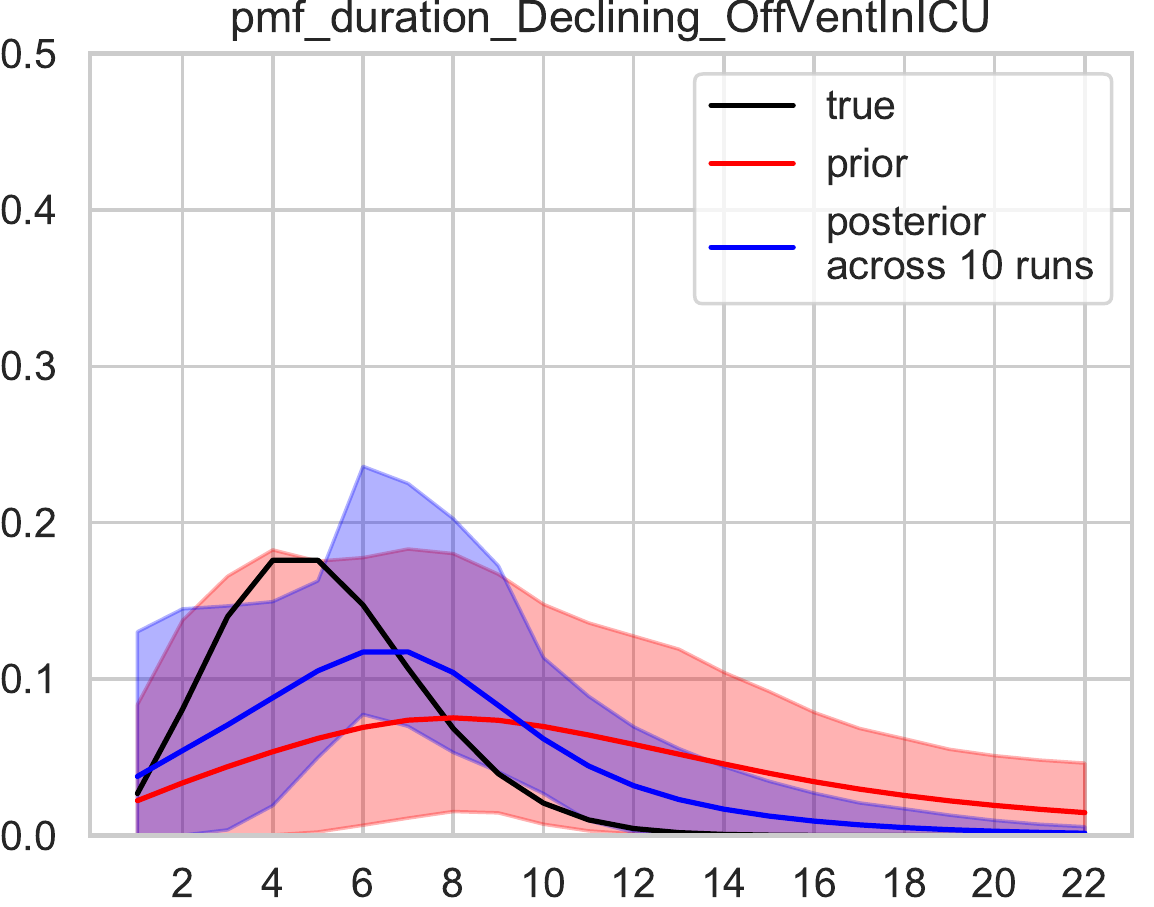}
\\
	\includegraphics[width=\PWW\textwidth]{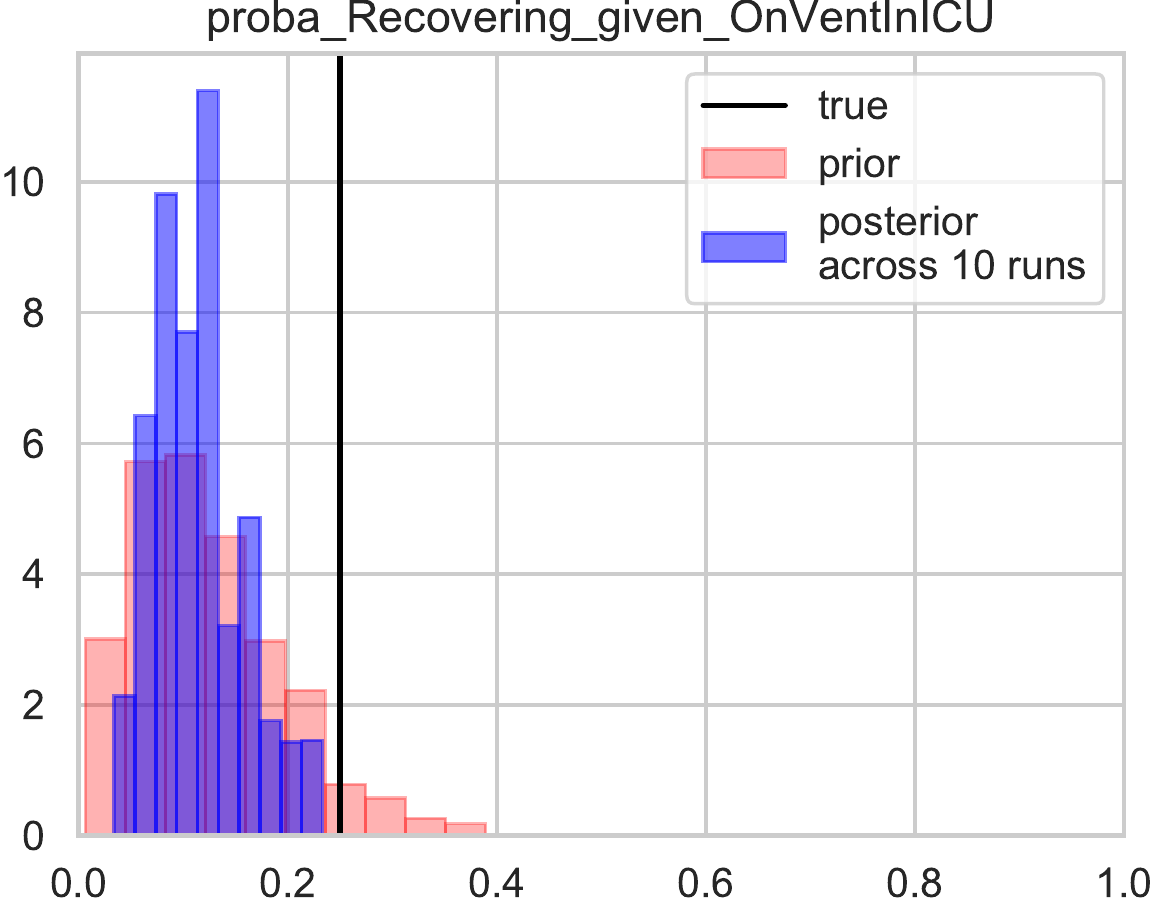}
	& 
	&
	\includegraphics[width=\PWW\textwidth]{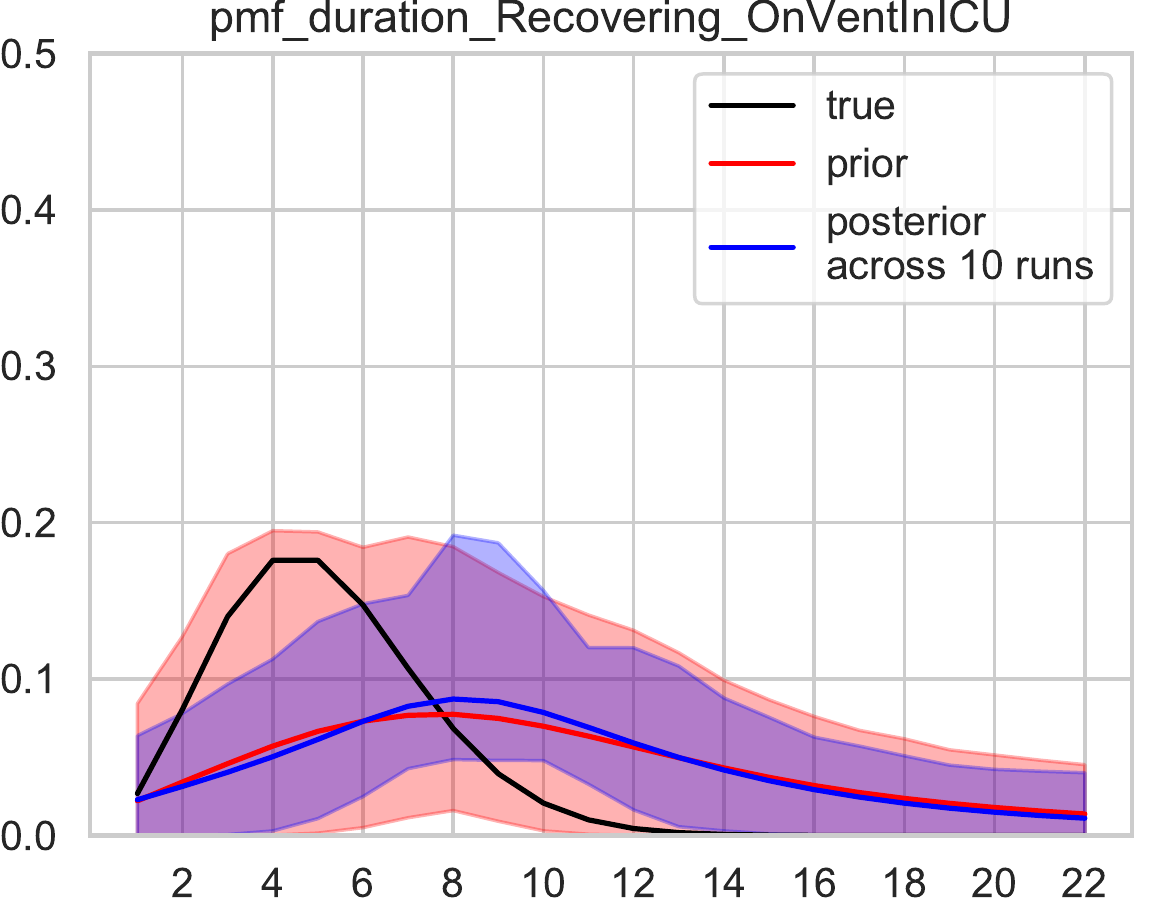}
	&
	\includegraphics[width=\PWW\textwidth]{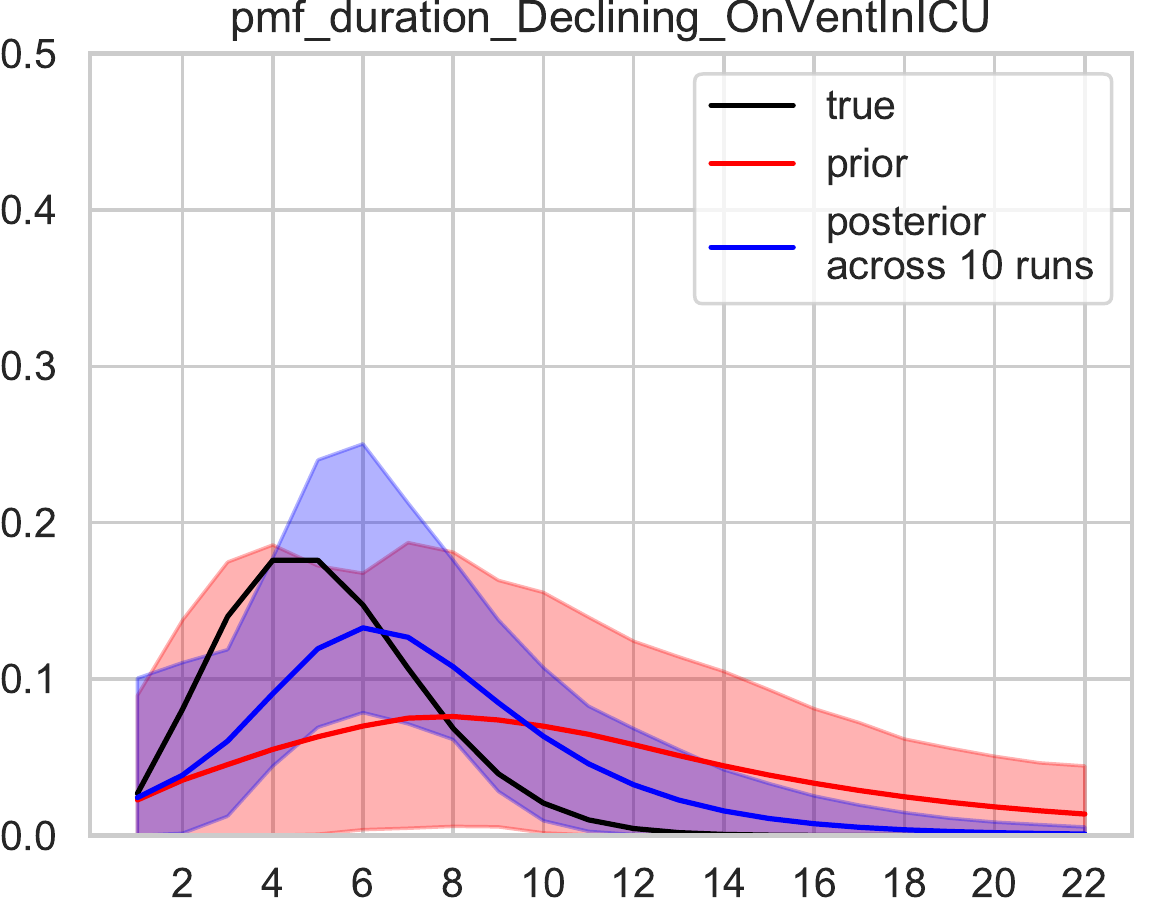}
\end{tabular}
    \caption{
    \textbf{Posterior distributions over parameters for synthetic data with hospital-level admissions.}
    We show transition parameters (left) and duration parameters (right) after fitting on 61 days of counts, where each day we used available simulated census counts for $\RR, \GG, \II, \VV$, and $\XX$.
The colored interval of duration plots shows the 2.5 - 97.5th percentile intervals of 2000 samples (10 runs, each with 200 samples).
The prior is also shown for comparison.
}
\label{fig:posterior_visualization_toy_data_1x}
\end{figure}

\begin{figure}[b!]
\centering
\begin{tabular}{c c | c c}
	\includegraphics[width=\PWW\textwidth]{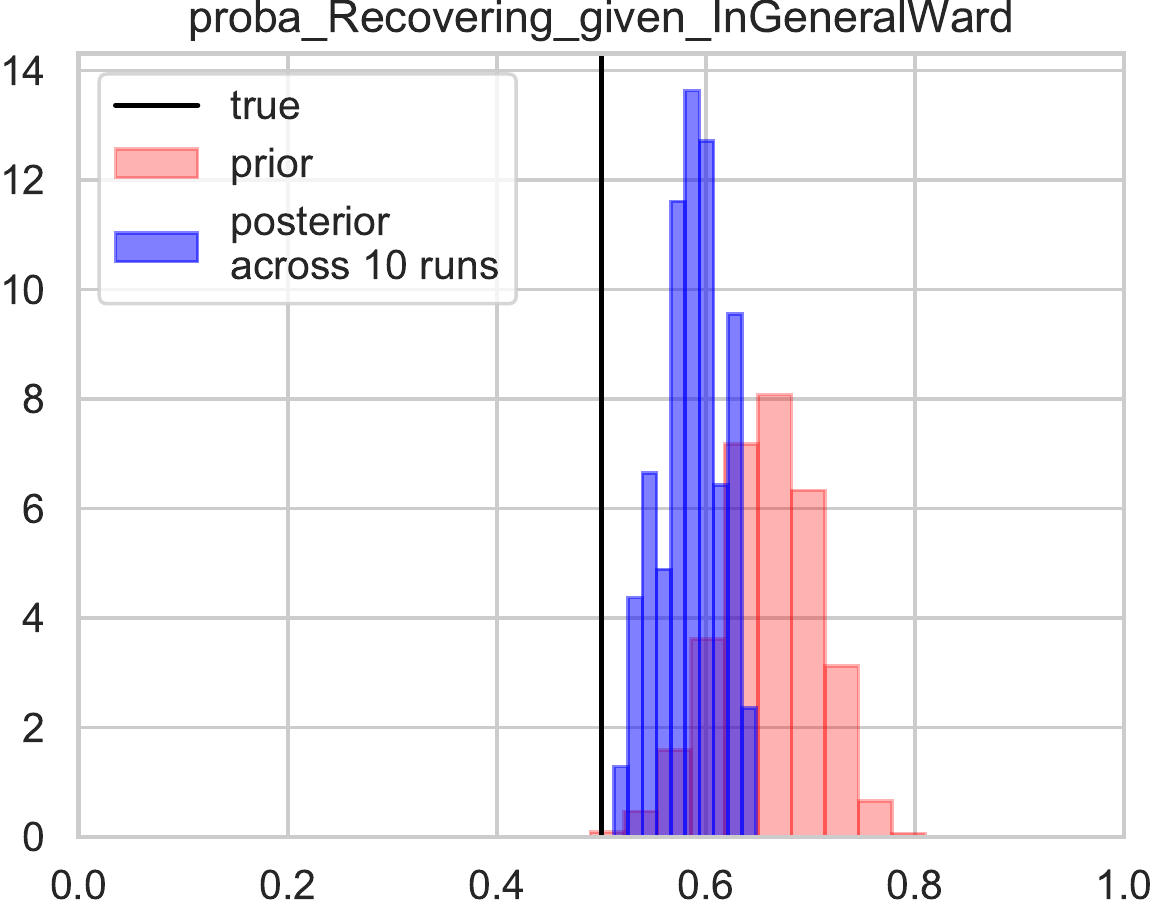}
	& 
	\includegraphics[width=\PWW\textwidth]{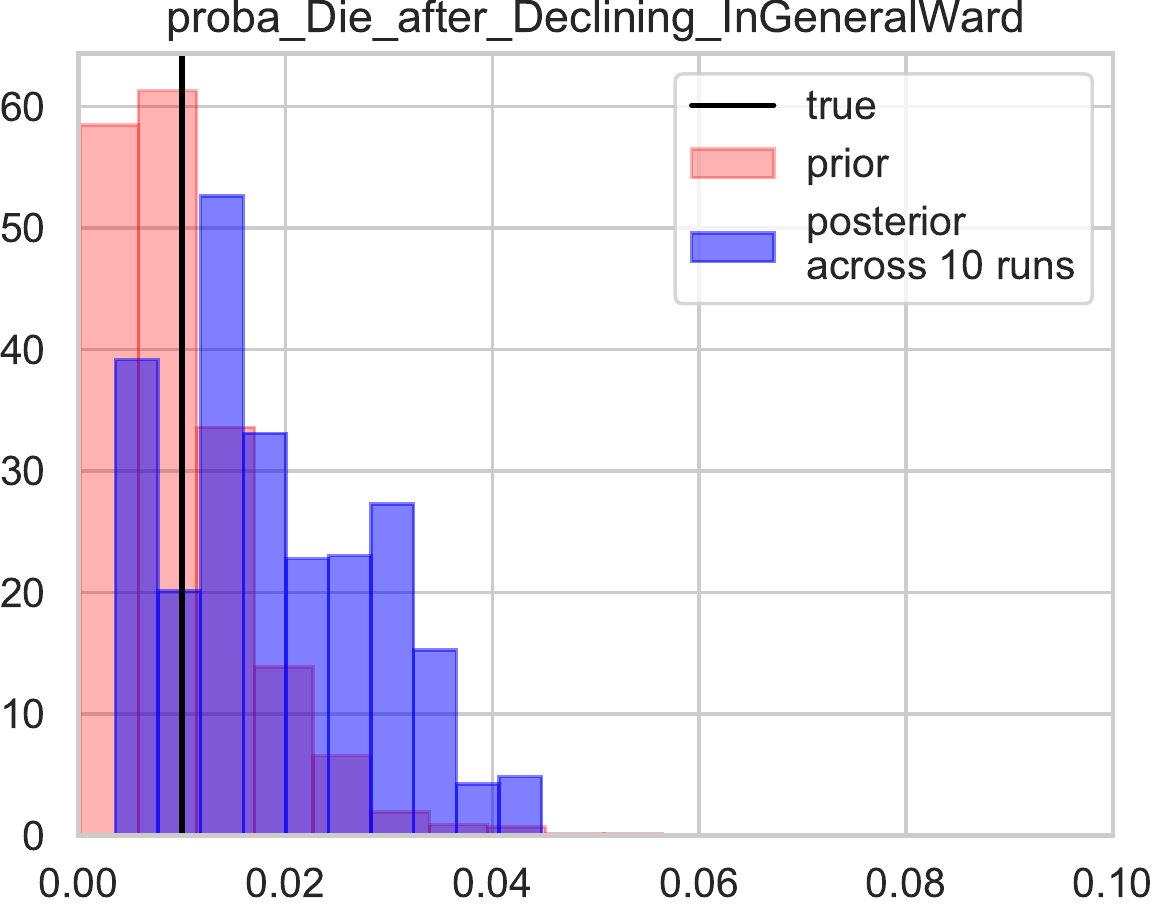}
	&
	\includegraphics[width=\PWW\textwidth]{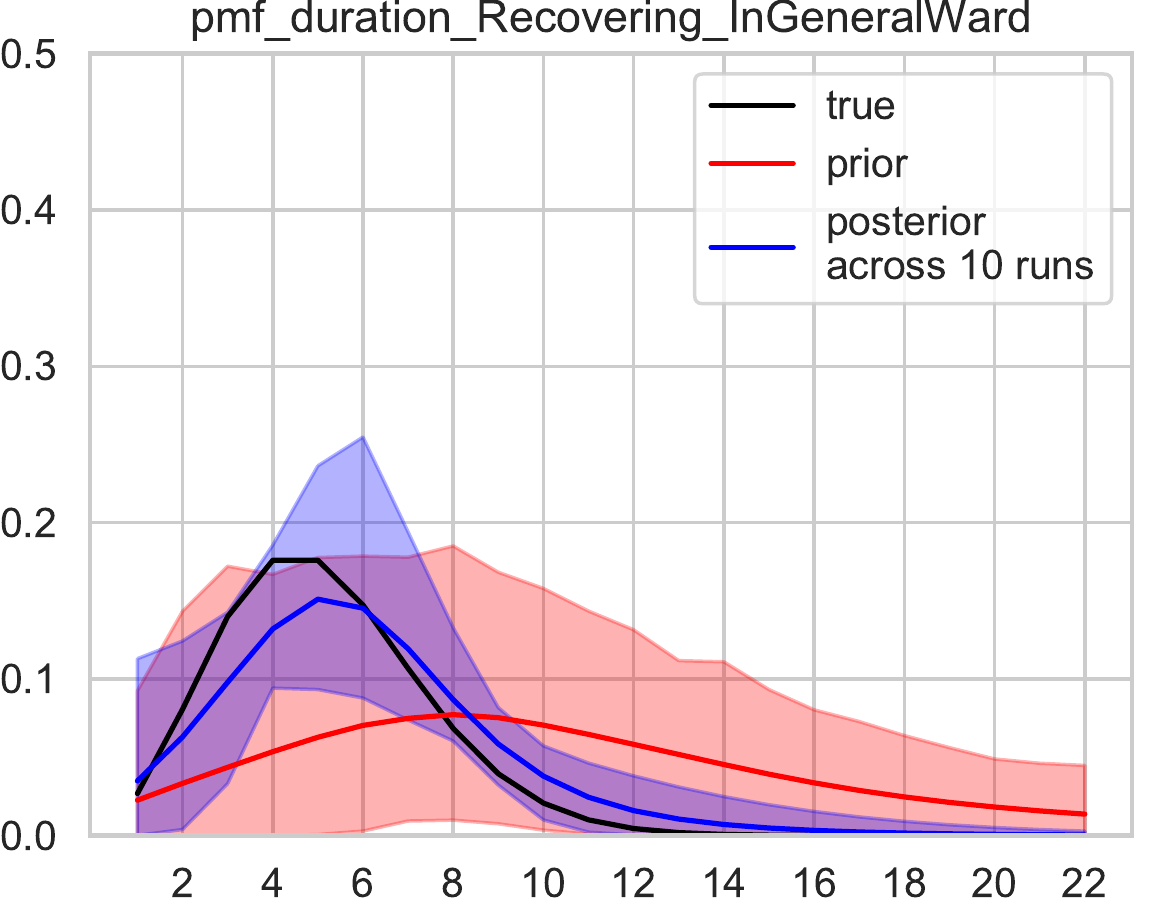}
	&
		\includegraphics[width=\PWW\textwidth]{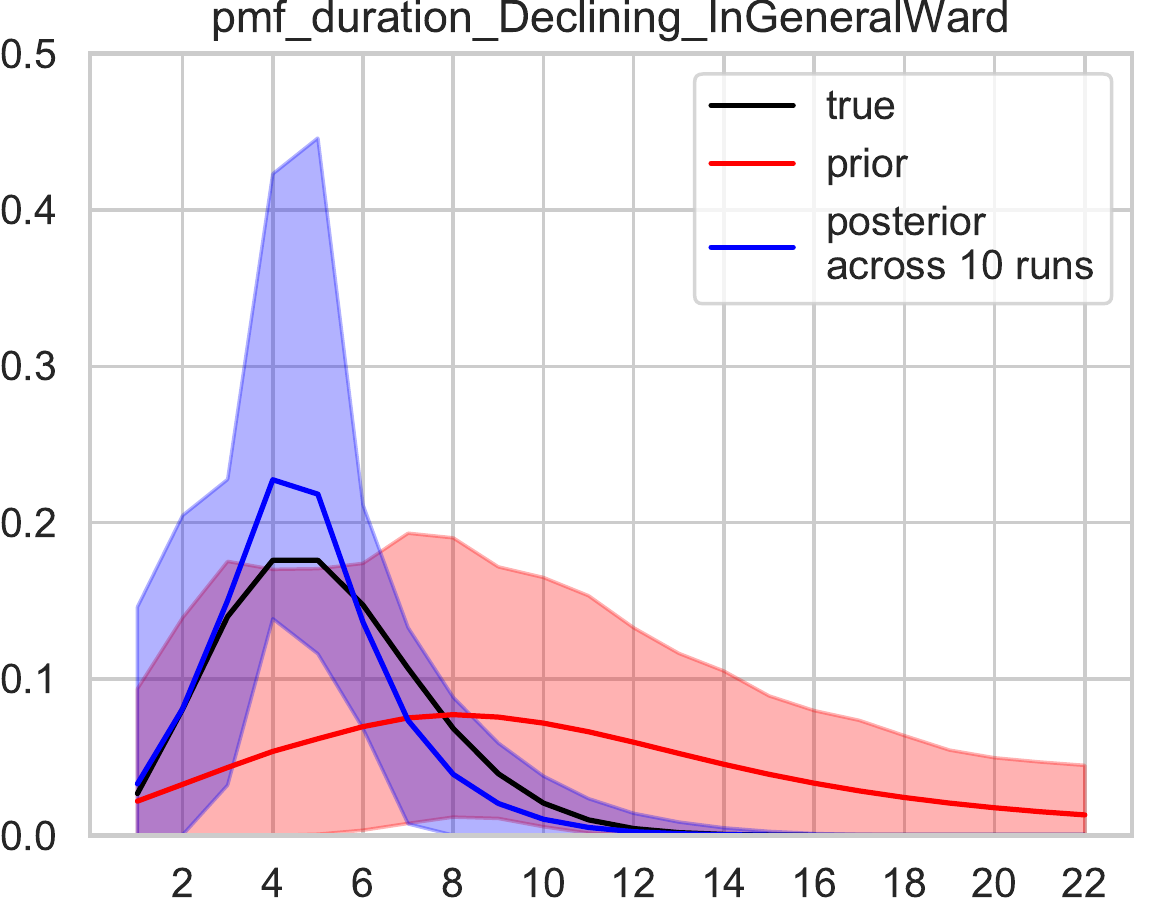}
\\
	\includegraphics[width=\PWW\textwidth]{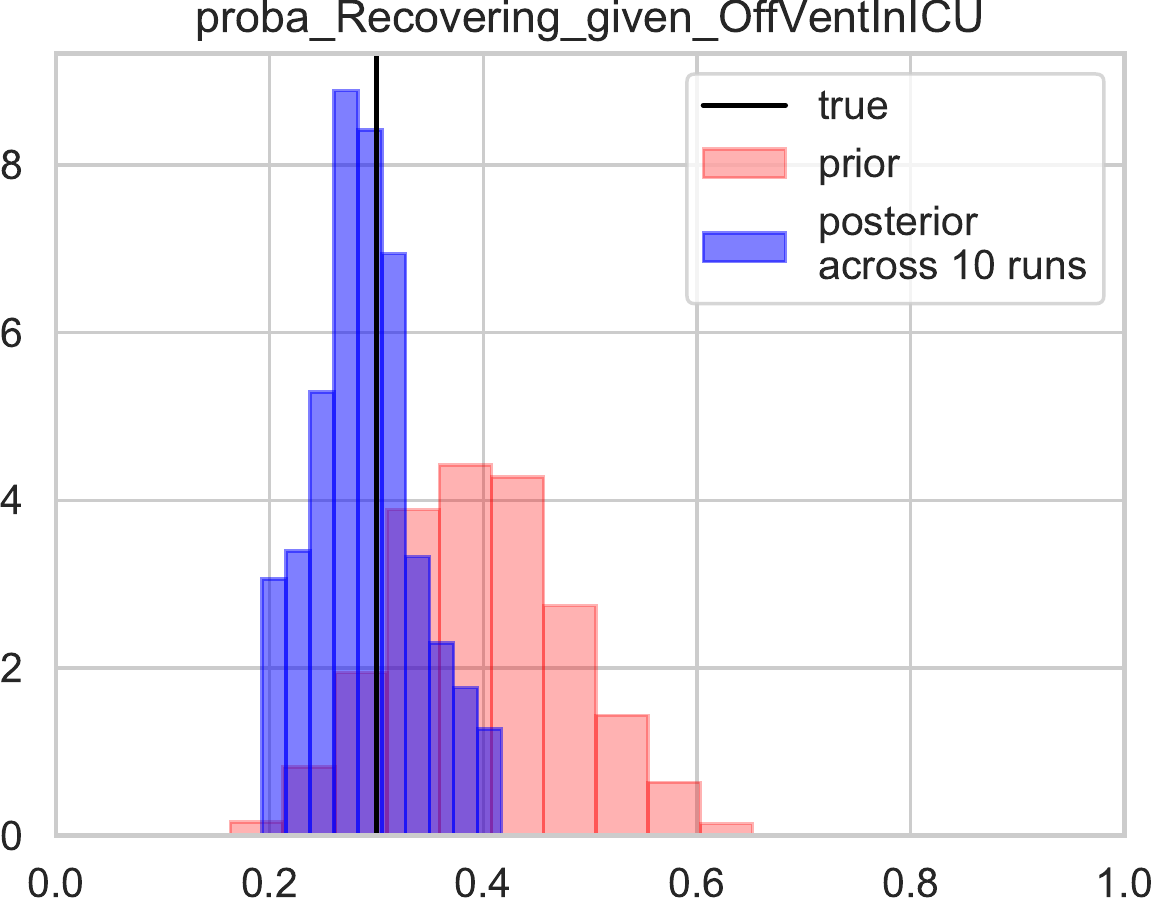}
	& 
	\includegraphics[width=\PWW\textwidth]{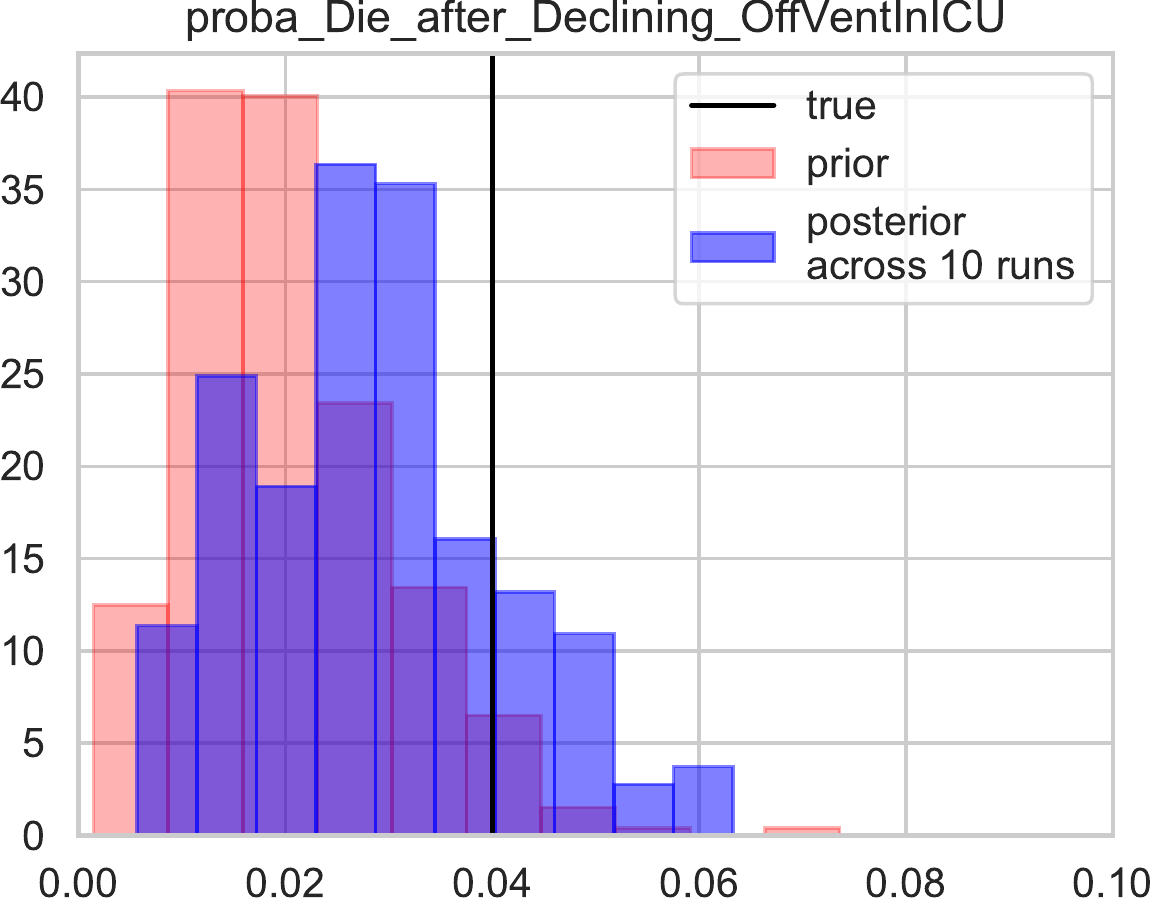}
	&
	\includegraphics[width=\PWW\textwidth]{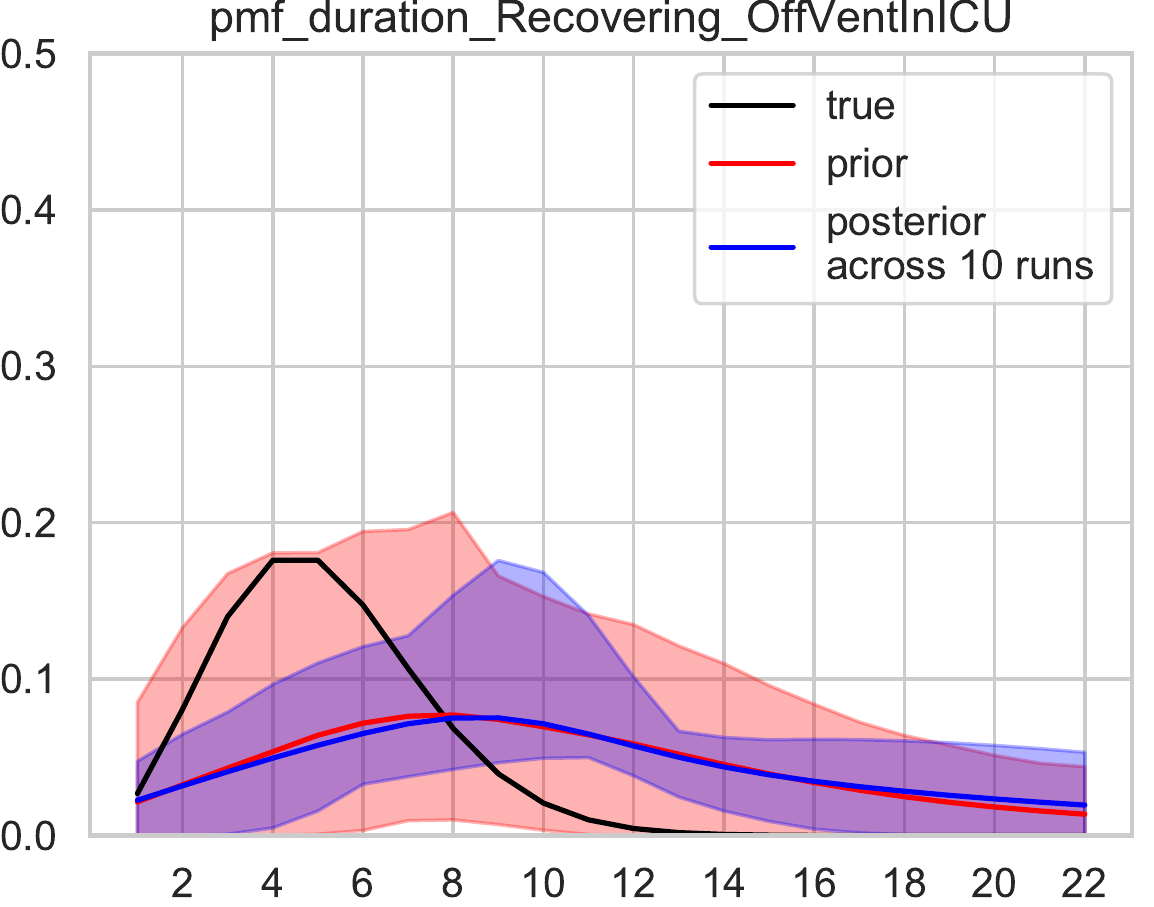}
	&
	\includegraphics[width=\PWW\textwidth]{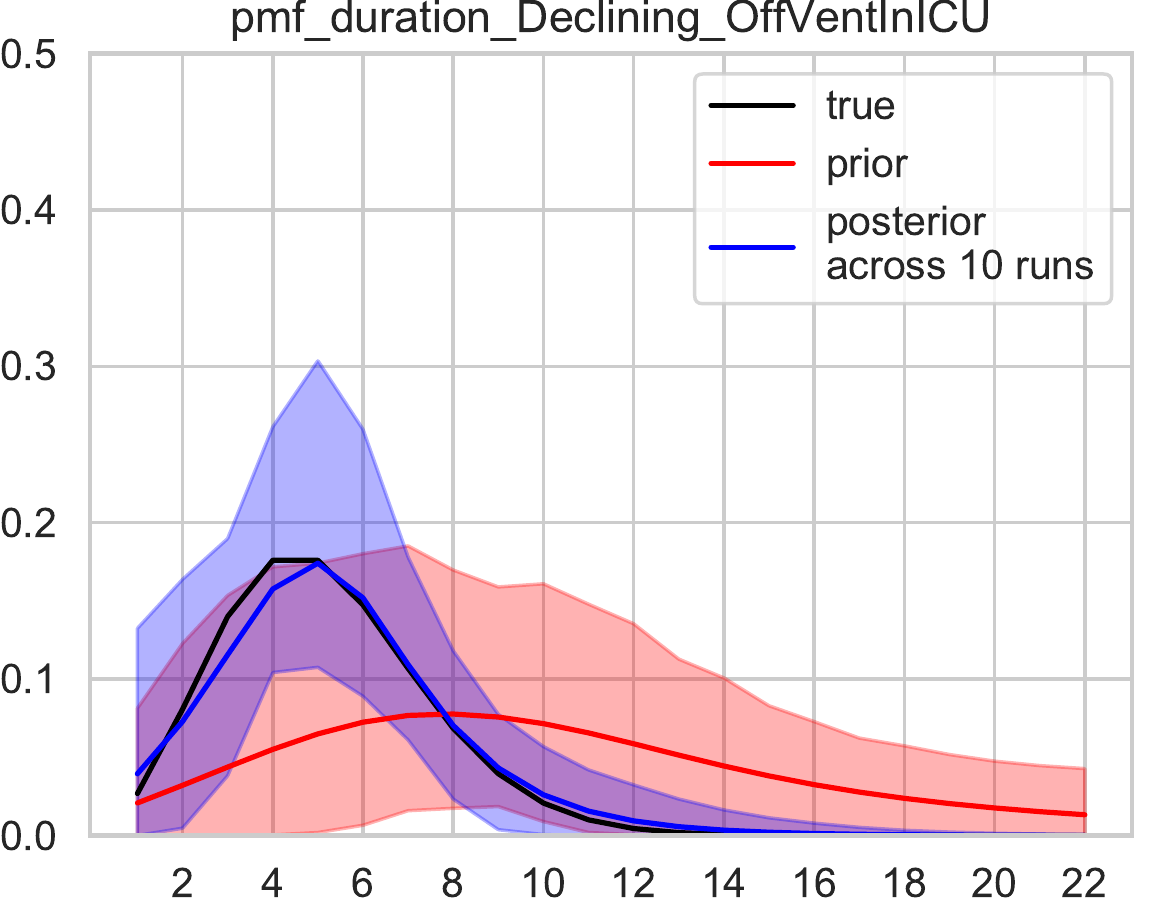}
\\
	\includegraphics[width=\PWW\textwidth]{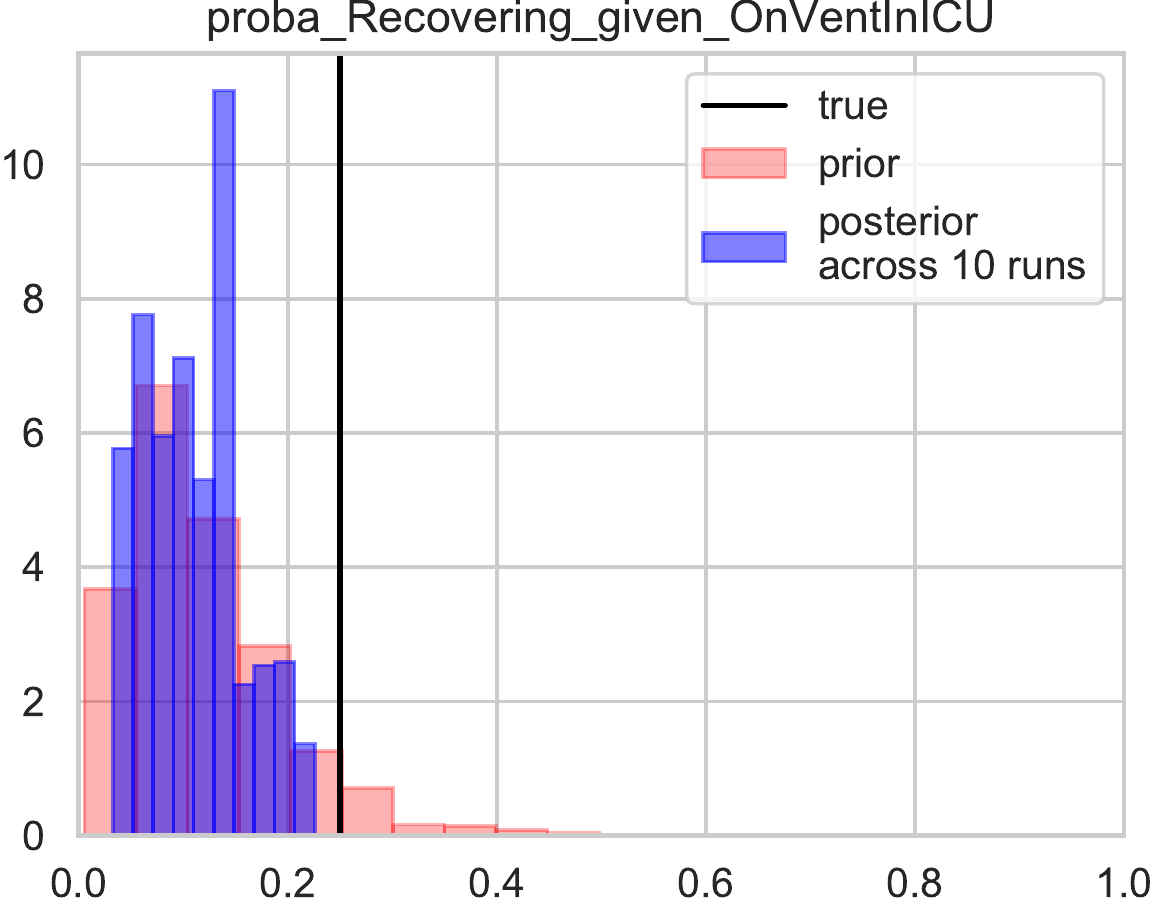}
	& 
	&
	\includegraphics[width=\PWW\textwidth]{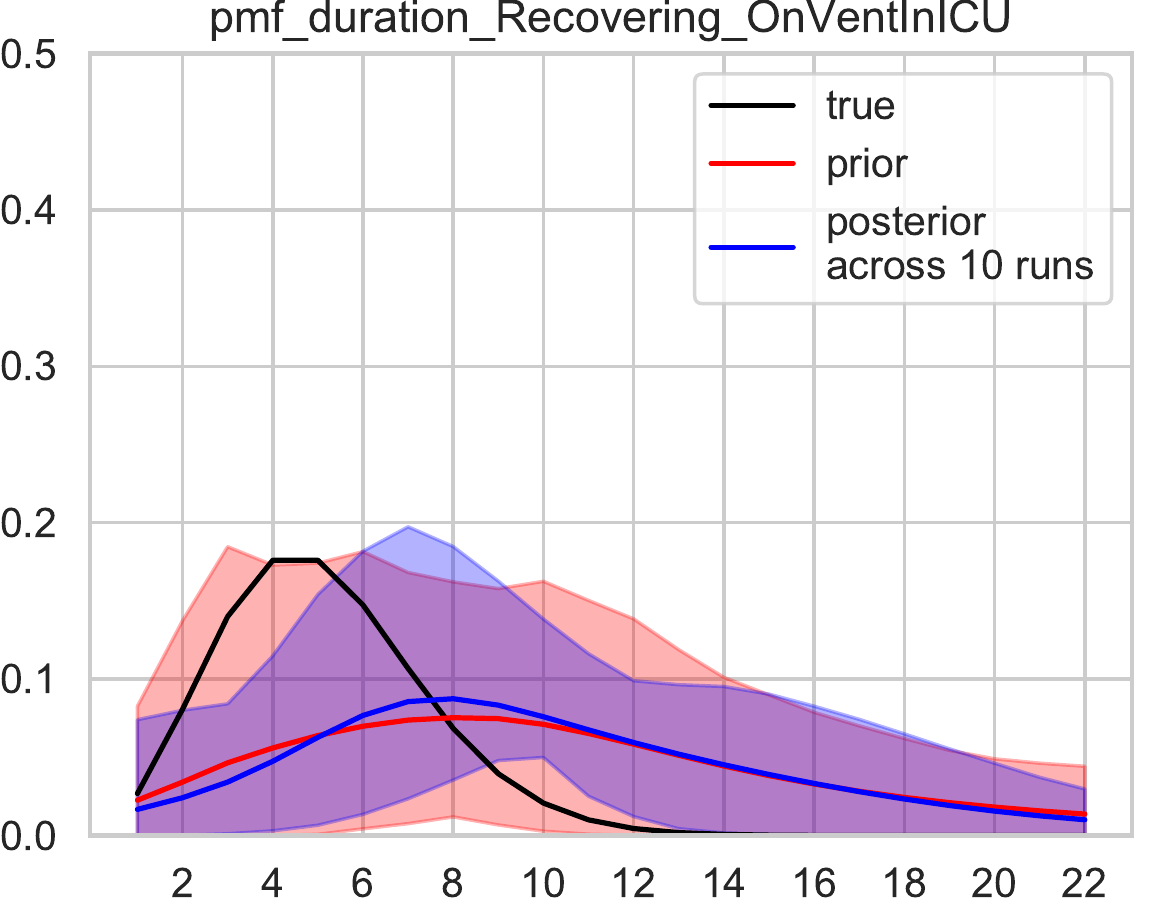}
	&
	\includegraphics[width=\PWW\textwidth]{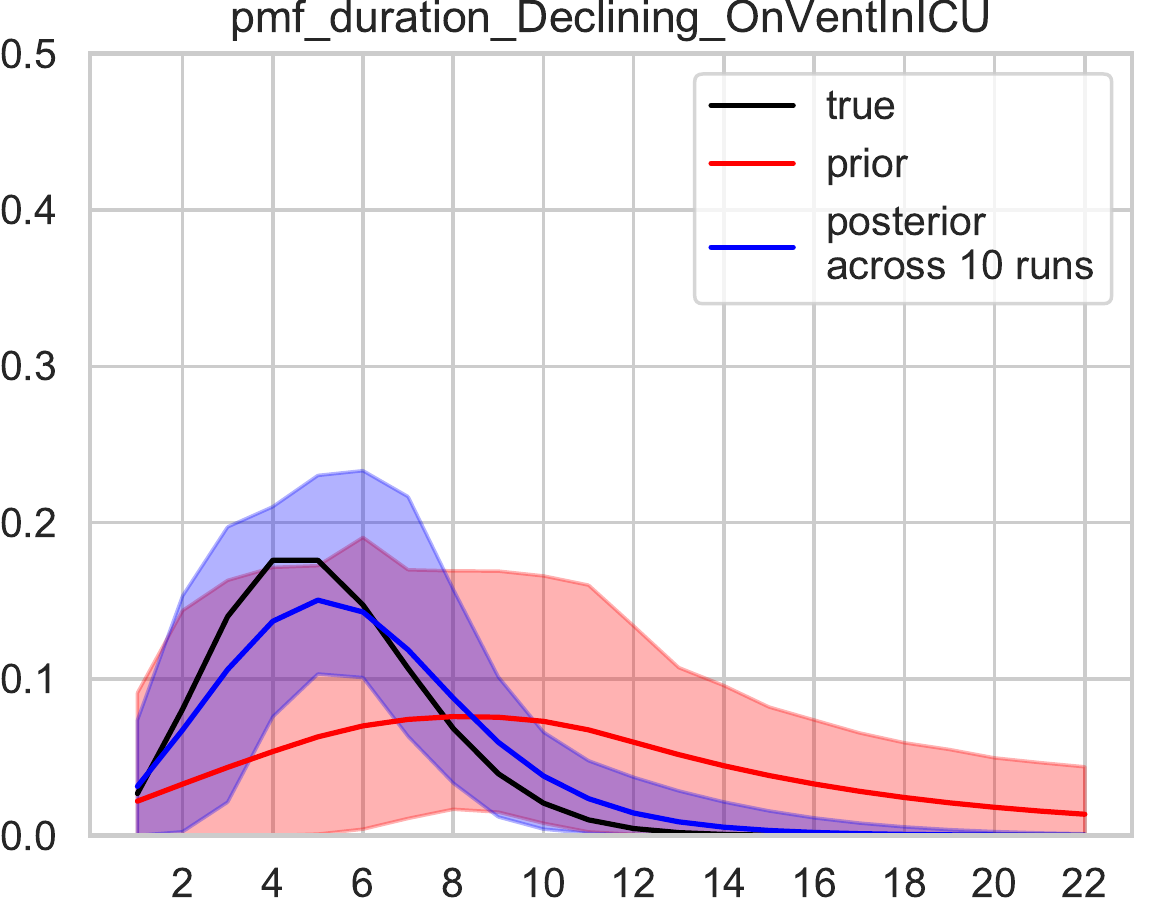}
\end{tabular}
    \caption{
    \textbf{Posterior distributions over parameters for synthetic data with regional-level admissions.}
    We show transition parameters (left) and duration parameters (right) after fitting on 61 days of counts, where each day we used available simulated census counts for $\RR, \GG, \II, \VV$, and $\XX$.
The colored interval of duration plots shows the 2.5 - 97.5th percentile intervals of 2000 samples (10 runs, each with 200 samples).
The prior is also shown for comparison.
}
\label{fig:posterior_visualization_toy_data_9x}
\end{figure}

We assessed the ability of our procedure to recover ''ideal'' model parameters under different admissions regimes. We selected a set of parameters that deviated significantly, though not excessively from the prior we derived from CDC data. Then, we simulated sets of counts using these parameters and a fixed random seed under 4 admissions regimes, respectively with 1-x, 3-x, 6-x and 9-x the amounts of true admissions at South Tees Hospitals between November 3rd and February 3rd. We note that the 9-x admissions regime is comparable to that of Massachusetts in terms of total admitted patients, and it thus can be considered a regional-level regime. We then used ABC with the same prior to learn 5 sets of model parameters to fit each of the 5 simulated counts, respectively. We used all possible counts for training: $\RR$, $\GG$, $\II$, $\VV$ and $\XX$.\\
ABC was able to identify parameters that provided a close-to-optimal fit to the training counts as well as high-quality forecasts on test counts. While the parameters learned were optimal in their ability to explain the training data, in no case were they ''ideal'', i.e. the true parameters. In particular, we observed that, the higher the number of total admitted patients, the closer the learned parameters are to the true ones, as can be seen by comparing Fig.~\ref{fig:posterior_visualization_toy_data_1x} and Fig.~\ref{fig:posterior_visualization_toy_data_9x}, which show the posterior distributions for the 1-x (hospital-level) and 9-x (regional-level) regimes, respectively. Under the regional-level regime, 5 parameters are fully recovered (i.e. mean of posterior matches true parameters), 2 parameters are partially recovered (i.e. posterior shifts from prior towards true parameter), 3 parameters are not recovered (i.e. no shift from prior towards true parameters).\\
Under our modeling assumptions, higher admissions regimes generate counts which are reproducible by fewer sets of parameters. In other words, the space of ''optimal'' parameters shrinks. Thus, on higher admissions regimes, our procedure converges closer to the true parameters.\\
Furthermore, we observed that certain parameters are easier to recover than others. For instance, parameters that have a direct effect on $\RR$ and $\XX$ counts seem to be among the first to be recovered, this includes the duration probabilities recovering in $\GG$ and declining in $\VV$. By contrast, the parameters at the interface between hospital stages seem harder to recover.\\
We believe another useful experiment using synthetic data to be one that answers the question \emph{How informative does the prior need to be for the posterior to recover all the ''ideal'' parameters?}. We leave the answering of this question for future work.

\newpage 


\subsection{Importance of weighting on distance computation}
\label{sec:appendix_distance_ablation}

Here, we assess the impact of weighting counts within our distance computation (via the weight parameter $w_{tk}$ in Eq.~\ref{eq:weighted_distance_abc}) on forecast quality. To do so, we fit the parameters of ACED-HMM to the Massachusetts and Utah datasets used in our experiments using the same procedure described in Sec.~\ref{sec:learning_abc}, with the only difference that $w_{tk}$ is set to 1.0 for each timestep $t$ and stage $k$, thus effectively removing any timestep-specific and stage-specific biasing in the distance computation. 


Table~\ref{tab:error_distance_ablation_US} reports the test-set MAE scores of posteriors trained with this unweighted distance, and compares them to the test MAE scores of posteriors trained with our weighted distance in Eq.~\ref{eq:weighted_distance_abc}.
On Utah data, the use of weighted distance does not appear to have a meaningful impact on forecast quality over using the simpler distance: across all stages the estimated MAE interval shows substantial overlap. On Massachusetts data, on the other hand, the use of weights results in slightly better forecasts of $\II + \VV$ and $\VV$ counts, with slightly worse forecasts on $\GG$ counts. This effect was sought after in our weight calibration, as, for the purposes of evaluation societal impact, we care more to forecast ICU counts accurately than General Ward counts.
Overall, as the differences between weighted and unweighted versions of our ACED-HMM seem to be minor, we do not deem the timestep-specific and stage-specific weighting as leading factors in the success of our method.

\begin{table}[!h]
\centering
\resizebox{\textwidth}{!}{%
\texttt{
\begin{tabular}{c|r |r r|r r|r r|r r| }
  & 
    & $\GG$ & \textnormal{\bf InGeneralWard} 
    & $\II + \VV$ & \textnormal{\bf InICU }
    & $\VV$ & \textnormal{\bf OnVentInICU }
    & $\XX$ \textbf{sm.} & \textnormal{\bf Death}  \\ 
  & \textnormal{\bf Method}
    & MAE~ & lower - upper~
    & MAE~ & lower - upper
    & MAE~ & lower - upper
    & MAE~ & lower - upper
\\  
 \hline
 \hline
 \multirow{3}{*}{MA}
  & ACED-HMM + ABC ~~weighted dist.
                &  65.0~ & ~61.5 - ~~68.9  
                &  15.7~~& ~14.5 - ~16.9 
                &  34.0~~& ~32.6 - ~35.7
                & ~8.0~ &   7.8 - ~8.3~~\\
  & ACED-HMM + ABC unweighted dist.
                & 60.1~ & ~58.0 - ~~62.7  
                & 16.4~~& ~15.0 - ~17.6  
                & 37.8~~& 35.2 - ~40.4
                &  7.8~ &  7.3 - ~8.2~~\\                
  \cline{2-10}
  & Mean Test $y$
  				& 1141.6~ &                
  				& 392.5~~&               
  				&  249.1~ &              
  				& 66.5~ & \\ 
 \hline
 \hline
 \multirow{3}{*}{UT}
  & ACED-HMM + ABC ~~weighted dist.
                &  20.8~ & ~20.2 - ~~21.3  
                &  18.2~~& ~17.4 - ~19.2 
                &  NA~   &
                &  2.4~ &  2.3 - ~2.5~~\\ 
  & ACED-HMM + ABC unweighted dist.
                & 20.8~ & ~19.9 - ~~21.7  
                & 18.8~~& ~16.9 - ~20.7
                &  NA~  &
                &  2.5~ &  2.3 - ~2.7~~\\                
  \cline{2-10}
  & Mean Test $y$ 
  				& 272.9~ &               
  				& 164.3~ &               
  				& NA~    &             
  				& 11.7~ & \\ 
\end{tabular}}}
\caption{\textbf{Quantitative error assessment comparing two versions of the distance function used in ABC: unweighted and time-specific and stage-specific weighting}.  
We report results for both MA and UT during the testing period (Jan. 12 - Feb. 11, 2021).
The method for computing the MAE is analogous to that of Table~\ref{tab:error_metrics_US}.}
\label{tab:error_distance_ablation_US}
\end{table}

\newpage

\subsection{Exploring different duration parameterizations}
\label{sec:appendix_durations_ablation}

We perform two sets of experiments exploring different parameterizations of the duration distributions within ACED-HMM, described below.

\begin{figure}[!b]
\centering
\begin{tabular}{c c | c c}
	22 days & & 44 days
	\\
	\includegraphics[width=\PWW\textwidth]{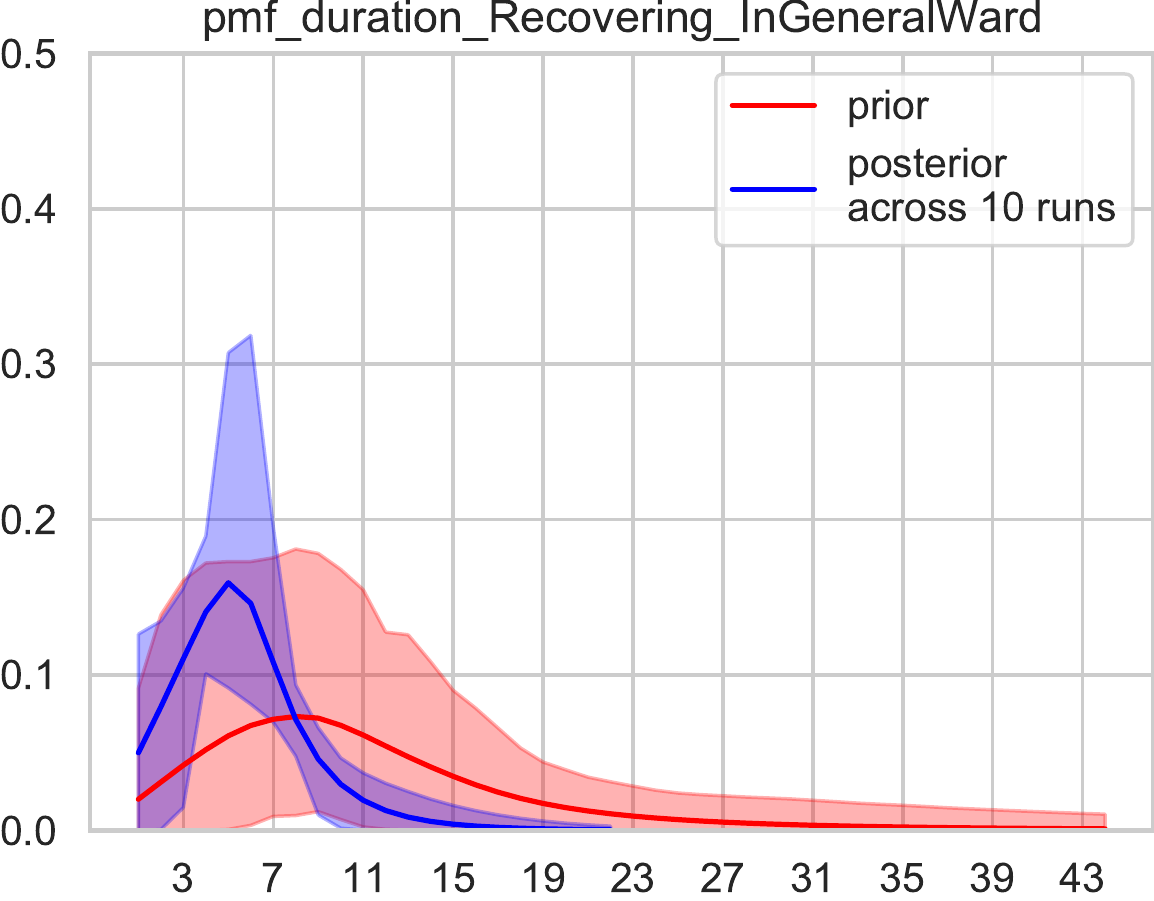}
	& 
	\includegraphics[width=\PWW\textwidth]{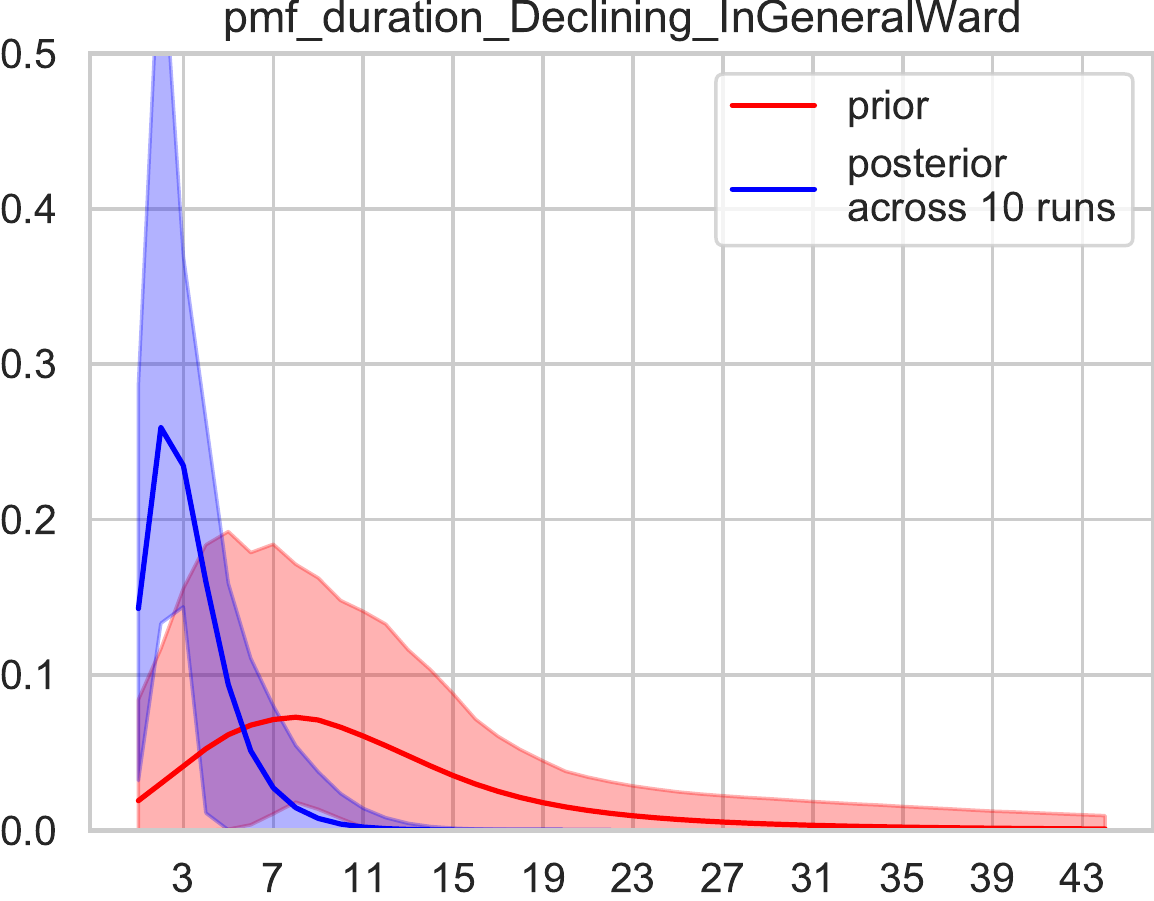}
	&
	\includegraphics[width=\PWW\textwidth]{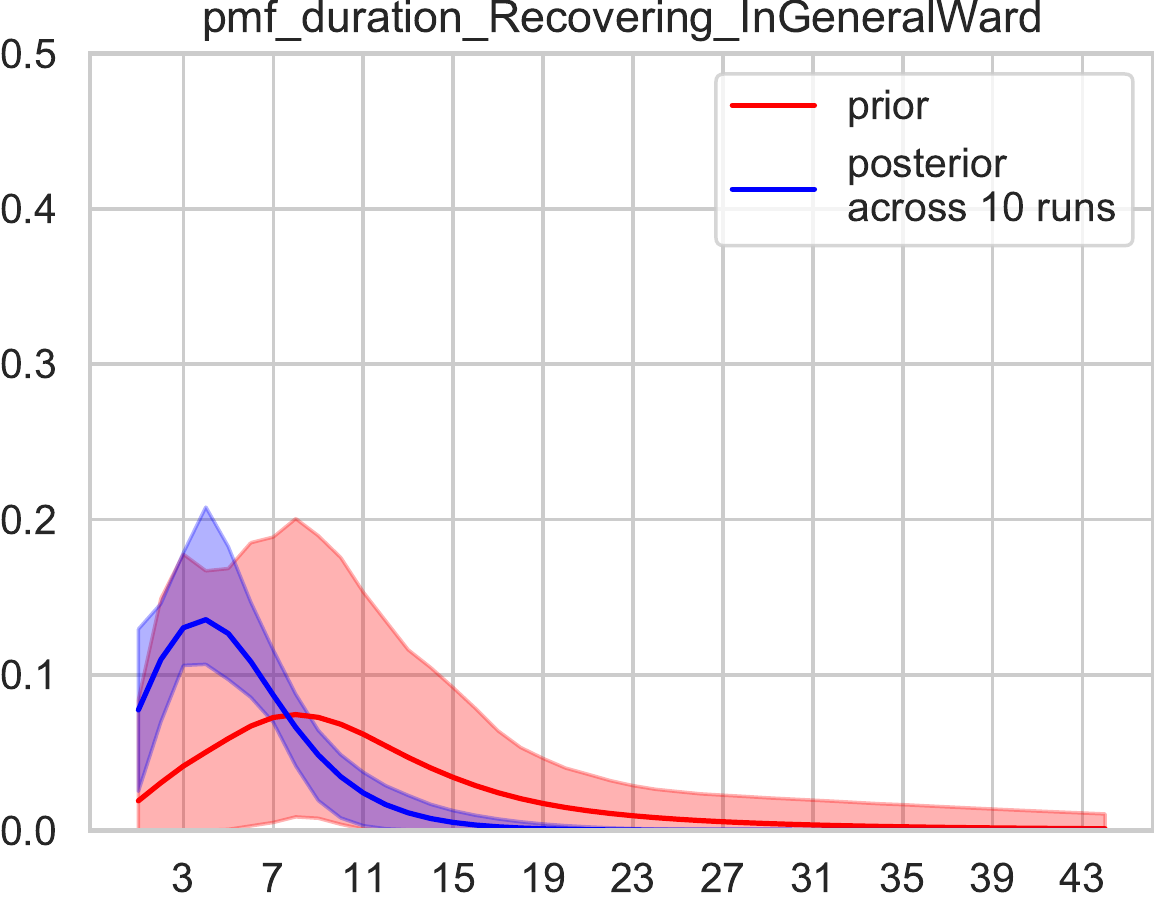}
	&
	\includegraphics[width=\PWW\textwidth]{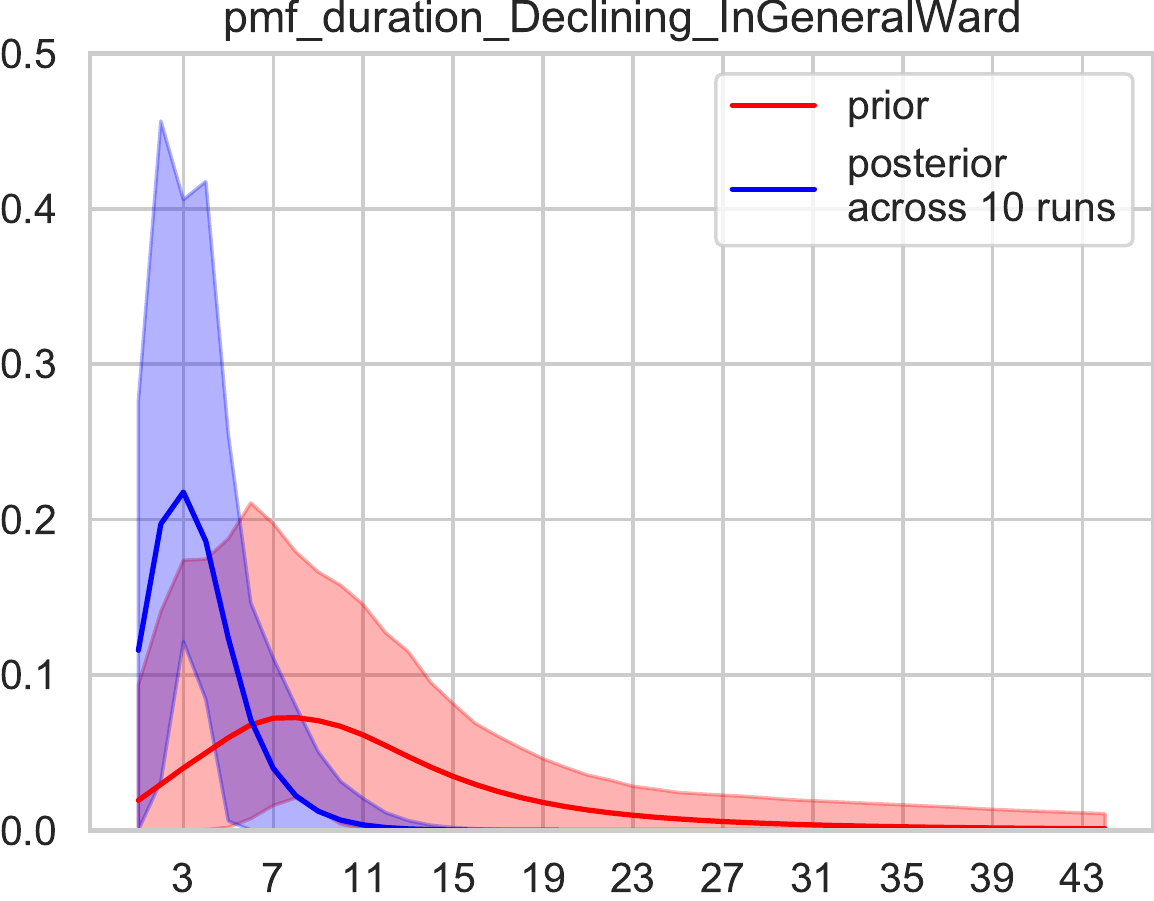}
\\
	\includegraphics[width=\PWW\textwidth]{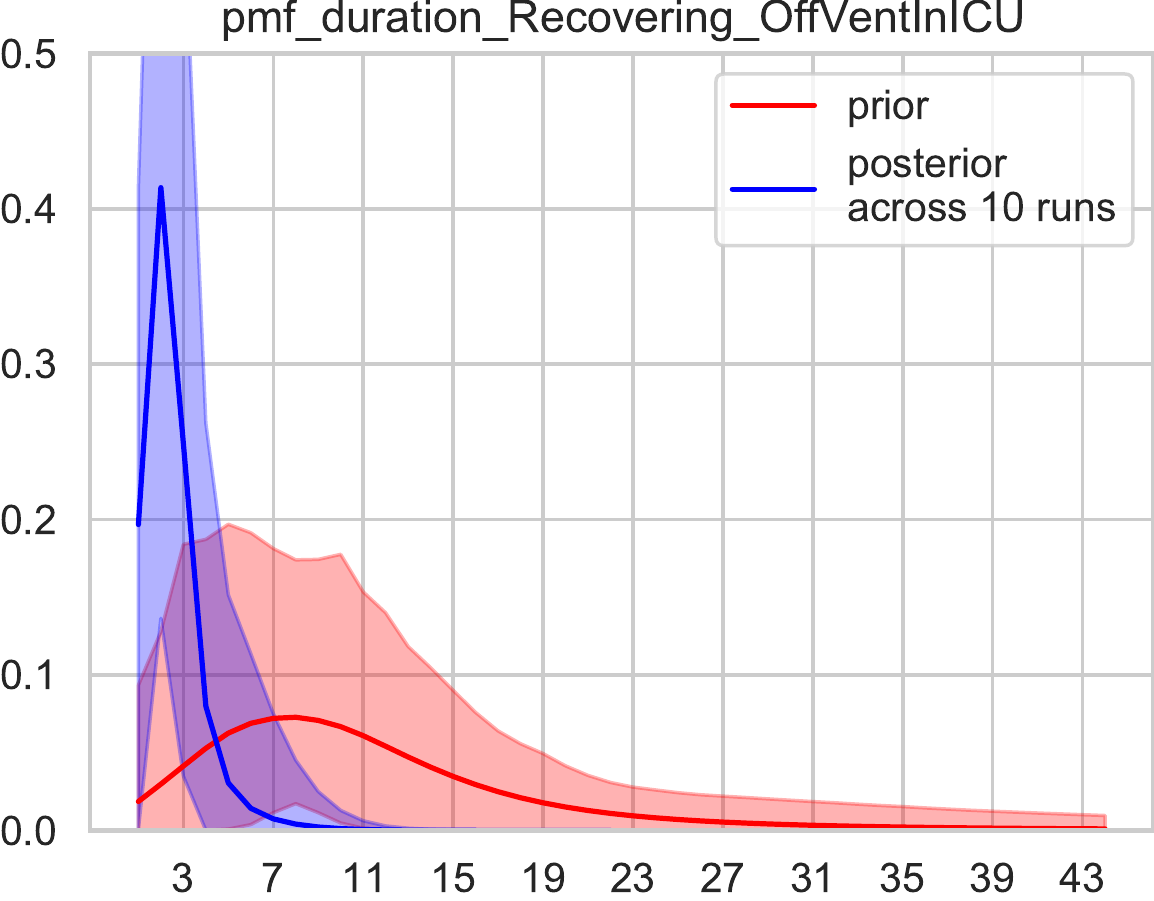}
	& 
	\includegraphics[width=\PWW\textwidth]{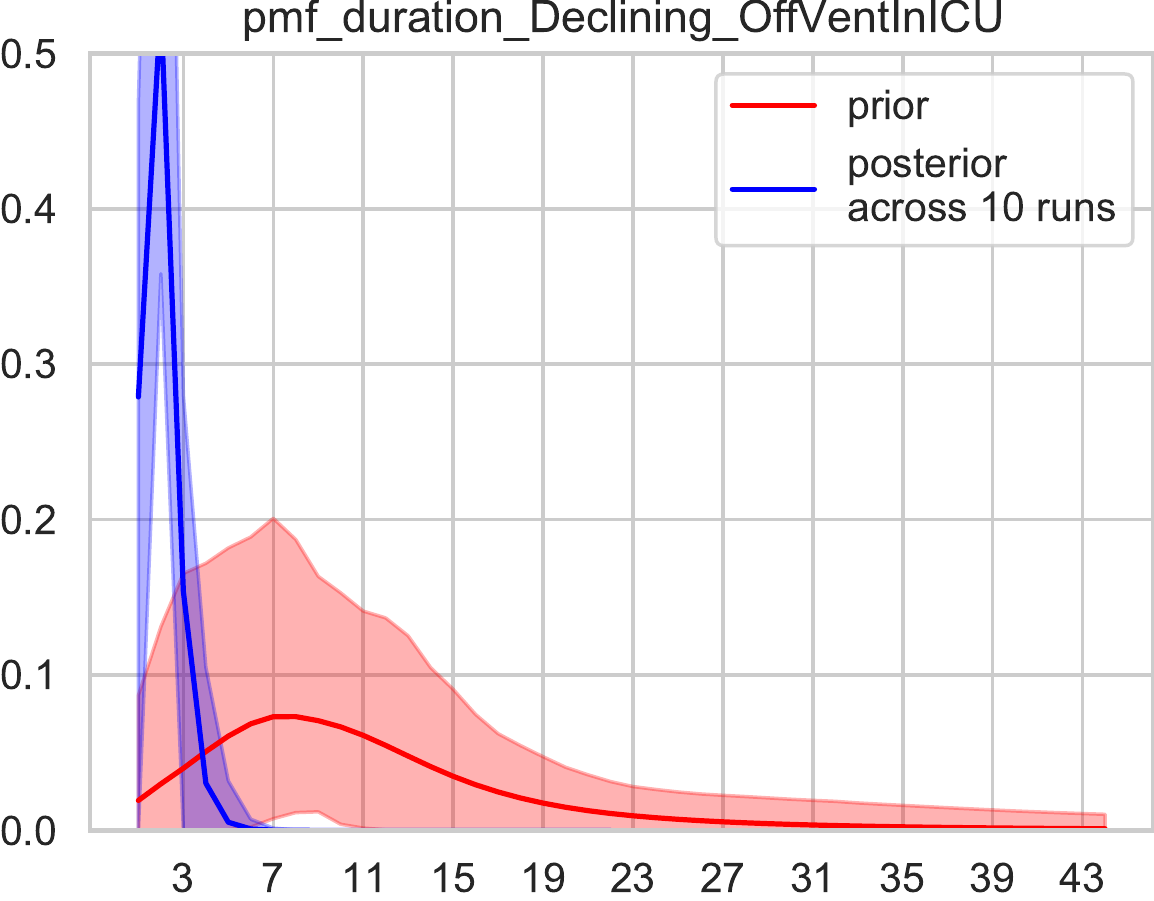}
	&
	\includegraphics[width=\PWW\textwidth]{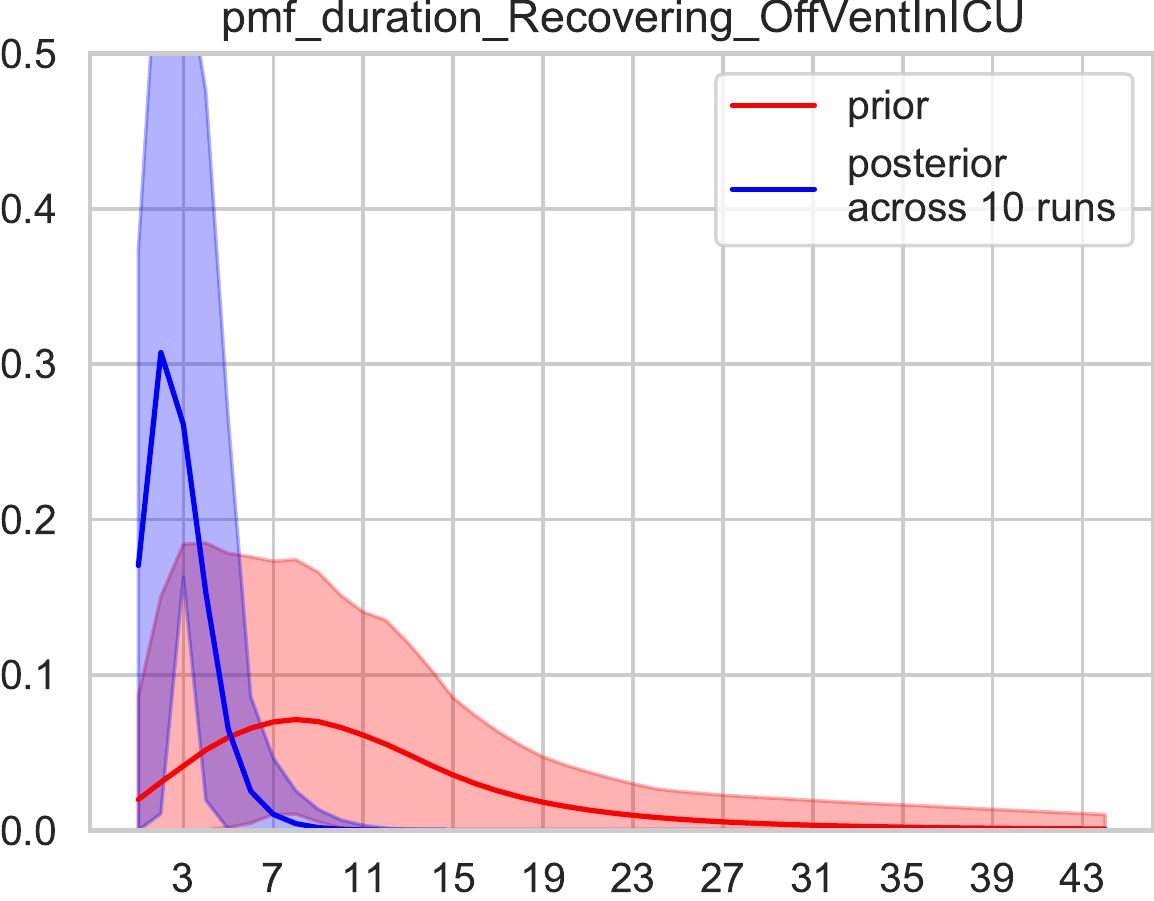}
	&
	\includegraphics[width=\PWW\textwidth]{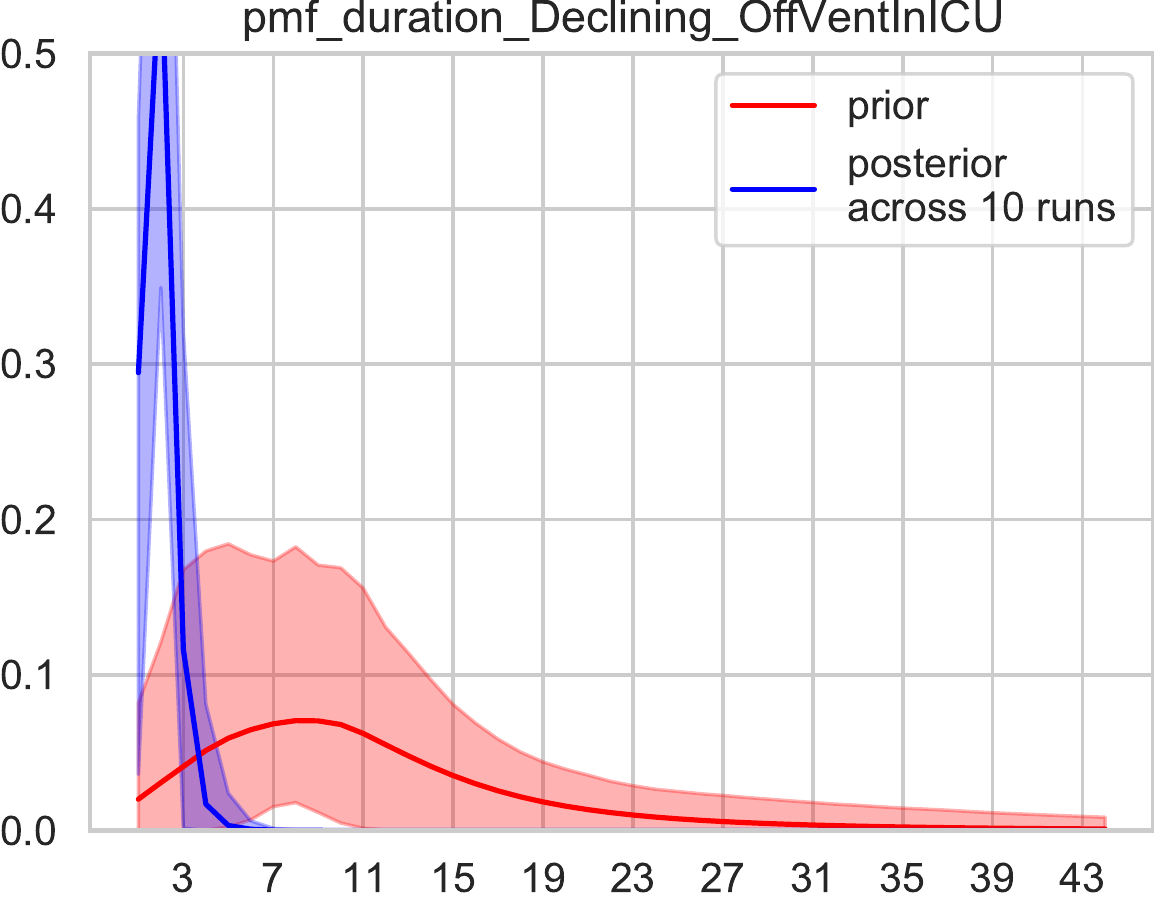}
\\
	\includegraphics[width=\PWW\textwidth]{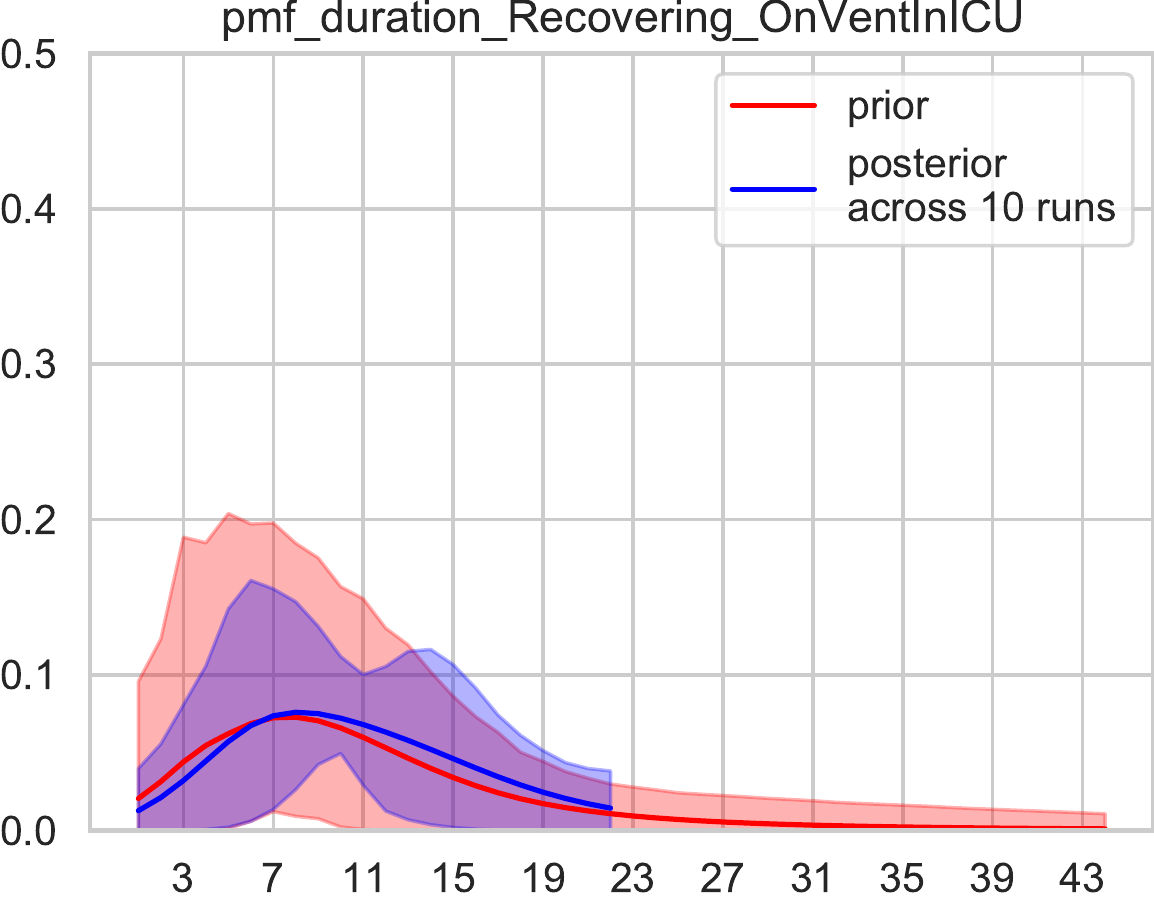}
	& 
	\includegraphics[width=\PWW\textwidth]{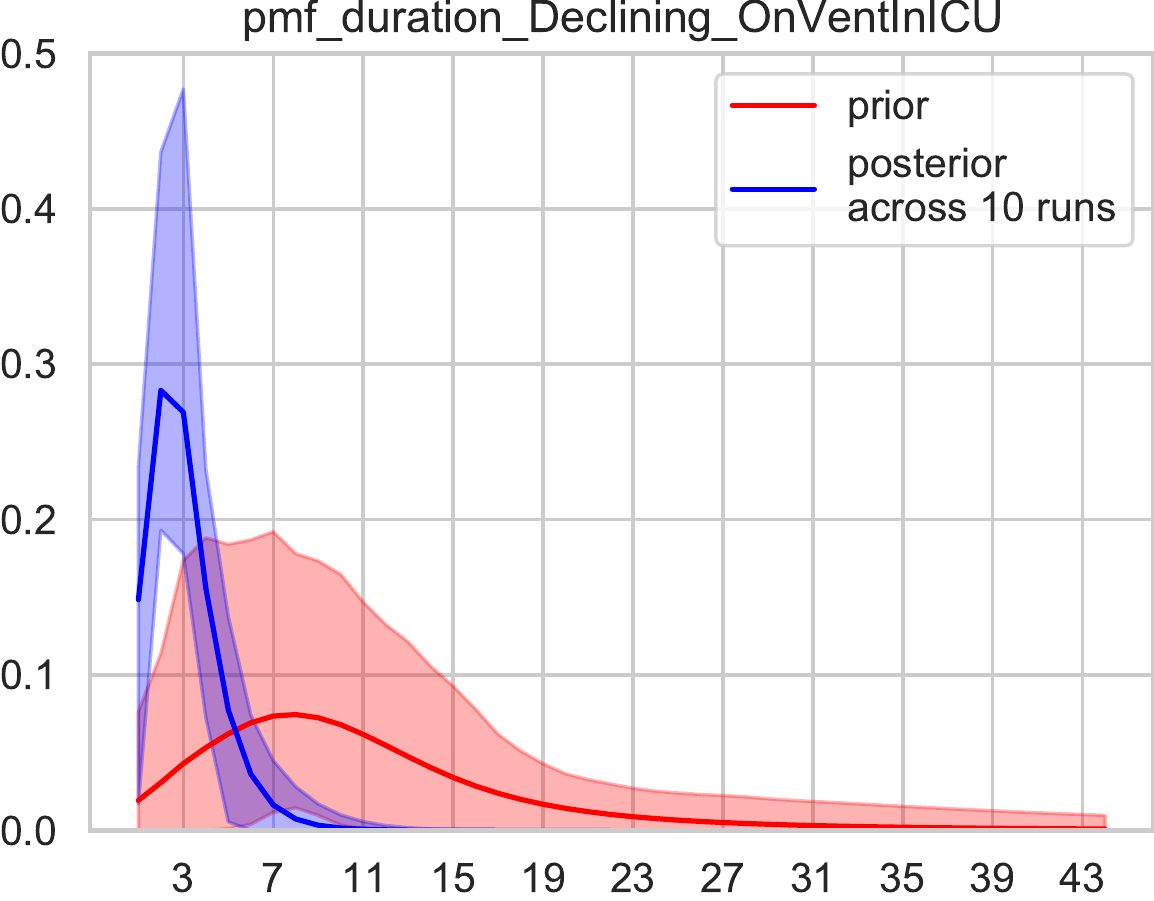}
	&
	\includegraphics[width=\PWW\textwidth]{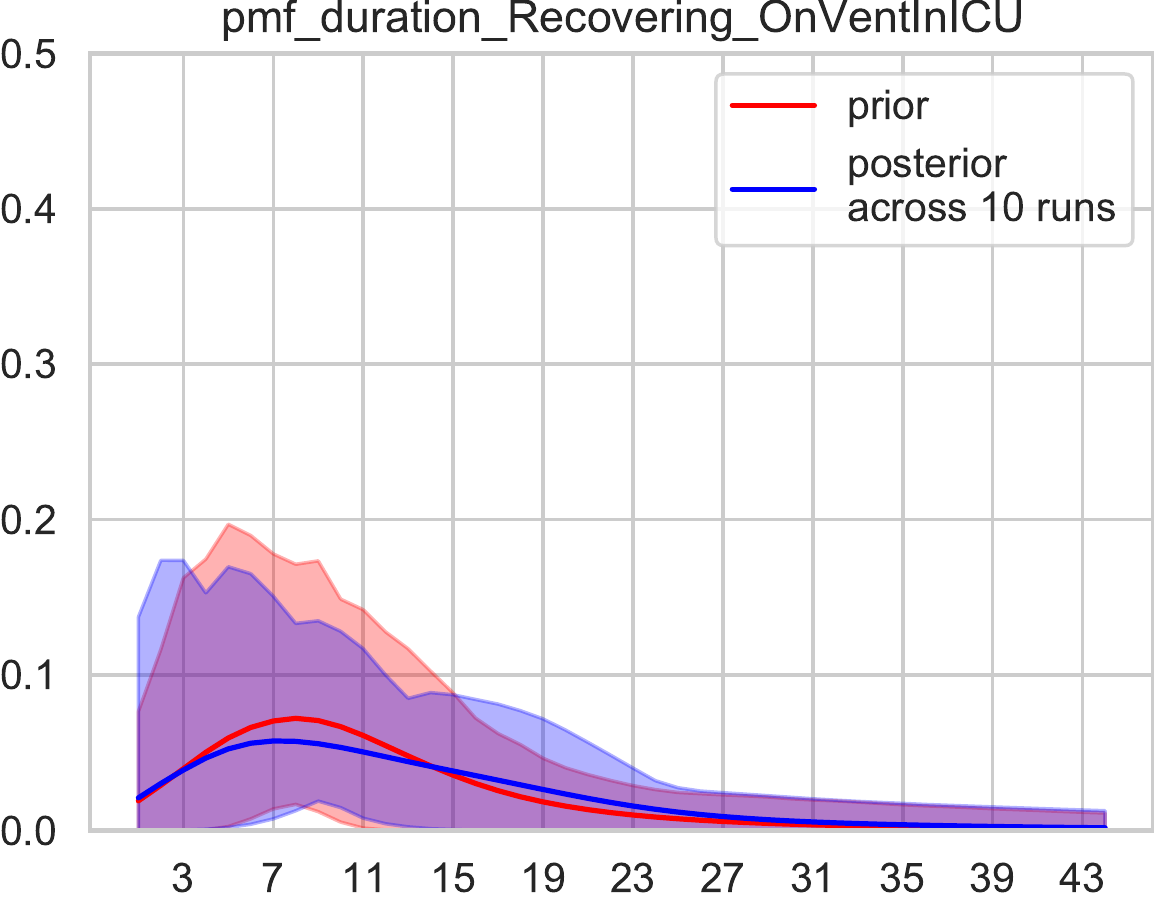}
	&
	\includegraphics[width=\PWW\textwidth]{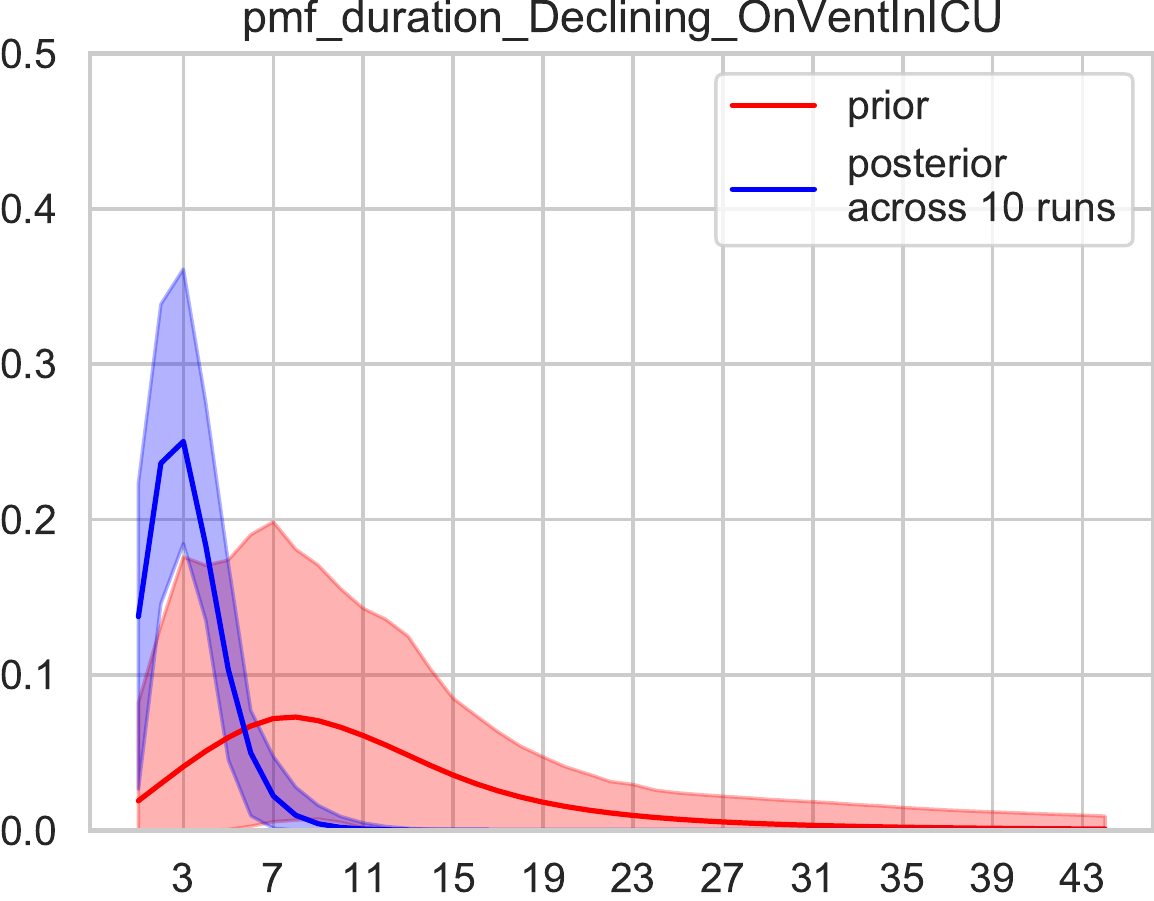}
\end{tabular}
\caption{\textbf{Comparison of posterior distributions of durations for ACED-HMM with different duration upper bounds.} The plots on the left are for ACED-HMM with durations capped at 22 days, whereas the plots on the right are for ACED-HMM with durations capped at 44 days.}
\label{fig:durations_length_ablation_visualization_MA}
\end{figure}

\paragraph{Alternative Duration Model Experiment 1: Larger truncation limit $D=44$.}
In the first experiment, we increase the upper bound of the per-segment duration distributions from 22 to 44, and train the modified ACED-HMM using our procedure on Massachusetts data. Figure~\ref{fig:durations_length_ablation_visualization_MA} shows the learned posterior duration distributions for the original upper bound ($D=22$) and less restrictive upper bound ($D=44$). The respective posterior distributions of both models look similar, indicating that the algorithm has converged to similar parameter values. Furthermore, almost all distributions have no weight after day 22. The only exception to these two observations is the duration distribution while \emph{recovering on the ventilator}, which does change noticeably between $D=22$ and $D=44$. 

We note two things about the duration distribution for recovering in stage $\VV$, which  changes between $D=22$ and $D=44$. 
First, the estimated posterior here is very similar to the prior even for $D=44$. 
Because the prior does put some weight after day 22 when $D=44$, perhaps this the reason for the visible change in the posterior.
Second, this distribution is the hardest to recover, since very few patients go on the ventilator, and, of those, only about 10\% recover, according to our learned posterior in Figure~\ref{fig:posterior_visualization_MA}. In our earlier experiments with synthetic data,  we showed it is difficult to recover the true parameters related to recovering on the ventilator (Figure~\ref{fig:posterior_visualization_toy_data_9x}).


\setlength{\tabcolsep}{0.1cm}

Nevertheless, Table~\ref{tab:error_durations_ablation_US} shows that ACED-HMM with a 44-day-per-segment upper bound achieves better MAE on $\VV$ counts for Massachusetts data (MAE reduced from 34 to 27). We believe this improvement may be attributable to the duration distribution while recovering on the ventilator placing weight past day 22, even if, as we discussed above, this is difficult to learn from limited data and may be strongly influenced by the choice of prior. Given this improved MAE result, we recommend setting $D=44$ or at least higher than 22 days.

\begin{table}[!t]
\centering
\resizebox{\textwidth}{!}{%
\texttt{
\begin{tabular}{c|r |r r|r r|r r|r r| }
  & 
    & $\GG$ & \textnormal{\bf InGeneralWard} 
    & $\II + \VV$ & \textnormal{\bf InICU }
    & $\VV$ & \textnormal{\bf OnVentInICU }
    & $\XX$ \textbf{sm.} & \textnormal{\bf Death}  \\ 
  & \textnormal{\bf Method}
    & MAE~ & lower - upper~
    & MAE~ & lower - upper
    & MAE~ & lower - upper
    & MAE~ & lower - upper
\\  
 \hline
 \hline
 \multirow{4}{*}{MA}
  & ACED-HMM + ABC
                &  65.0~ & ~61.5 - ~~68.9  
                &  15.7~~& ~14.5 - ~16.9 
                &  34.0~~& ~32.6 - ~35.7
                & ~8.0~ &   7.8 - ~8.3~~\\
  & ACED-HMM + ABC 44 days
                & 68.1~ & ~63.2 - ~~75.7  
                & 15.8~~& ~14.2 - ~17.6  
                & 27.3~~& 24.2 - ~30.6
                & 7.7~ &  7.3 - ~8.1~~\\ 
  & ACED-HMM + ABC poisson
                & 56.4~ & ~54.5 - ~~58.5  
                & 16.3~~& ~15.0 - ~17.8  
                & 32.0~~& 29.1 - ~34.4
                &  8.1~ &  7.7 - ~8.5~~\\  
  \cline{2-10}
  & Mean Test $y$
  				& 1141.6~ &                
  				& 392.5~~&               
  				&  249.1~ &              
  				& 66.5~ & \\ 
 \hline
 \hline
 \multirow{4}{*}{UT}
  & ACED-HMM + ABC
                &  20.8~ & ~20.2 - ~~21.3  
                &  18.2~~& ~17.4 - ~19.2 
                &  NA~   &
                &  2.4~ &  2.3 - ~2.5~~\\ 
  & ACED-HMM + ABC 44 days
                & 20.7~ & ~19.8 - ~~21.4  
                & 19.6~~& ~18.1 - ~21.4  
                & NA~   &
                & 2.5~ &  2.3 - ~2.7~~\\ 
  & ACED-HMM + ABC poisson
                & 21.0~ & ~20.2 - ~~21.8  
                & 18.4~~& ~16.8 - ~19.7  
                & NA~   &
                &  2.5~ &  2.3 - ~2.7~~\\                
  \cline{2-10}
  & Mean Test $y$ 
  				& 272.9~ &               
  				& 164.3~ &               
  				& NA~    &             
  				& 11.7~ & \\ 
\end{tabular}}}
\caption{Quantitative error assessment for predictions for two US states during the testing period (Jan. 11 - Feb. 11, 2021). We compare the performance of three versions of ACED-HMM: 1) using our duration parameterization with each state's maximum length truncated to 22 days; 2) using our duration parameterization with each state's maximum length truncated to 44 days; 3) using a poisson parameterization for the durations with each state's maximum length truncated to 22 days. The method for computing the MAE is analogous to that of Table~\ref{tab:error_metrics_US}.}
\label{tab:error_durations_ablation_US}
\end{table}


\paragraph{Alternative Duration Model Experiment 2: Truncated Poisson.}
In the second experiment investigating modeling assumptions about durations, we parameterize each duration distribution as a \emph{truncated Poisson}.
This is equivalent to fixing the temperature parameter $\nu^{k,h}$ to 1.0 in Eq.~\eqref{eq:tractable_duration_two_parameter_family}. This restrictive choice results in a less flexible parameterization that is not able to generate distributions with varying entropy. 
However, this parameterization also reduces the number of trainable parameters from 17 to 11, thus reducing training time as well as potentially reducing the difficulty of posterior estimation, as there are less possible combinations of parameters with equally good fits to the data.

We fit each version of our model -- fixing $\nu^{s,h} = 1$ and estimating its posterior -- to the Massachusetts and Utah datasets and performed both a quantitative and a qualitative evaluation.

\begin{figure}[h!]
\centering
\begin{tabular}{c c}
MA & UT
\\
\includegraphics[width=0.46\textwidth]{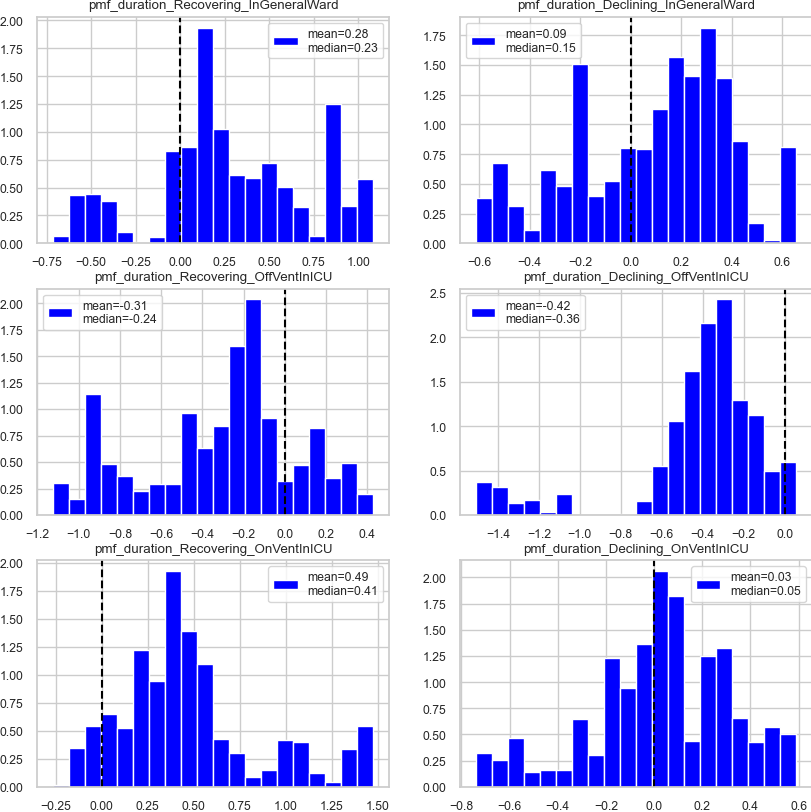}
&
\includegraphics[width=0.46\textwidth]{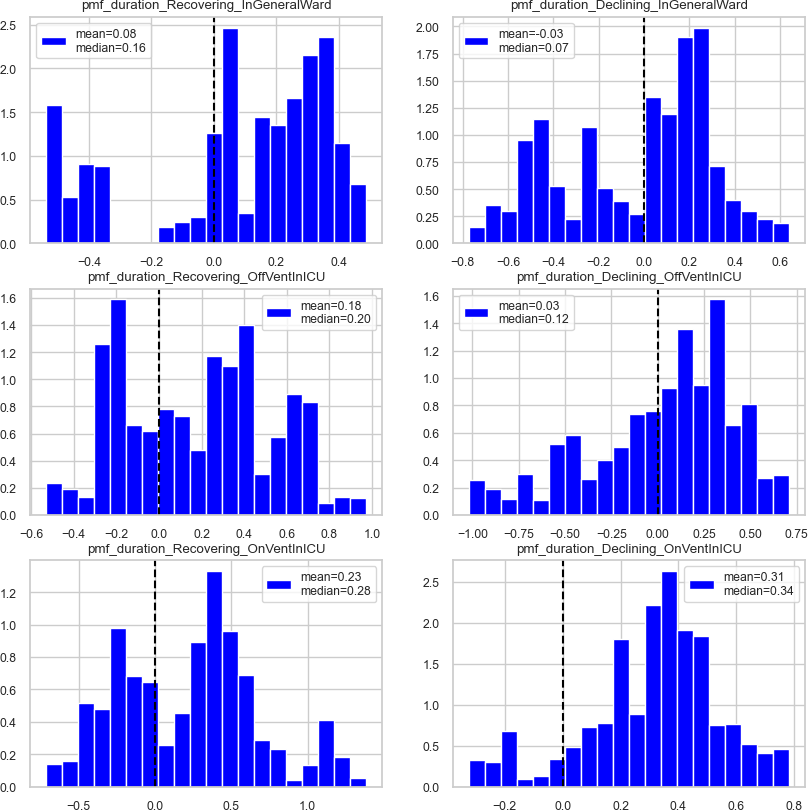}	
\end{tabular}
\caption{Posterior distribution of the log temperature for durations in MA (left) and UT (right). For each state, we show samples from the posterior of the log temperature $\log_{10} \nu^{k,h}$ for all stages $k$ and health states $h$.}
\label{fig:MA_log_temperature_histogram}
\label{fig:UT_log_temperature_histogram}
\end{figure}

As a qualitative evaluation, we plot the learned posterior temperature distributions (in log scale), for both MA and UT in Figure~\ref{fig:UT_log_temperature_histogram}. For UT, we see that the learned posteriors over the log temperature parameter often place non-trivial mass outside a narrow interval around zero.
Similar findings can be seen in the posteriors for MA.
If mass concentrated near $\log_{10} \nu^{k,h} = 0.0$, then we could conclude that $\nu^{k,h} \approx 1.0$ and suggest the posterior was well-approximated by a truncated Poisson).
The non-trivial variation around 0.0 in almost all stages suggests that we should continue using our 2-parameter flexible parameterization: at worst, it can match the truncated Poisson as a special case.

As a quantitative evaluation, Table~\ref{tab:error_durations_ablation_US} shows a comparison in test mean absolute error (MAE) between ACED-HMM with our suggested parameterization (learned $\nu^{s,h}$) and with the simpler truncated Poisson parameterization ($\nu^{s,h} = 1.0 \quad \forall s, h$). The truncated Poisson parameterization yields slightly better MAE scores for $\GG$ and $\VV$ counts, and slightly worse ones for $\II + \VV$ counts. 
Our ABC procedure is able to find suitable sets of parameters that fit the data well, even with a less flexible parameterization like the truncated Poisson.
We pointed out this issue of parameter identifiability (i.e. multiple distinct sets of parameters having similarly good fits to the data) in our experiment on synthetic data. 
While we do not have strong empirical evidence in favor of our suggested 2-parameter formulation, there is also little reason to avoid it given that it is competitive.

Overall, as our purpose is not simply to fit to past data and forecast on future data, but also to learn meaningful posterior distributions, the decision between using our flexible parameterization vs. a poisson one cannot be reduced to a comparison over MAE. Instead, one should use clinical knowledge to determine whether the simplification provided by the Poisson assumption is close enough to the truth for all possible regions and sites. As we lack such knowledge, we recommend using our more flexible parameterization, which may provide better fits and more realistic posterior distributions for regions or sites in which the Poisson assumption is far from the truth.

\paragraph{Must Durations Be Truncated?}
Conceptually, there is little need to truncate each segment's duration distribution. However, in practice truncation may improve efficiency (e.g. by constraining the memory required to represent each segment's duration distribution). 
Our public code implementation does not easily allow for an unspecified upper bound, but it should be possible to alter the implementation to achieve this if desired.
\newpage 
\section{Details of our proposed model: ACED-HMM}

\subsection{View of the ACED-HMM as a Hidden Markov Model}

In the main paper, we discuss how our proposed model can be seen as an explicit duration hidden Markov model with 9 states: $\SS, \GG0, \GG1, \II0, \II1, \VV0, \VV1, \XX, \RR$. We provide the transition matrix for this model in Table~\ref{tab:transition_probas}.

\begin{table}[!h]
\centering{
\scriptsize
\begin{tabular}{l l r r r r r r r r}
& & \multicolumn{8}{c|}{next state}
\\
	& &  $\RR$  &  $\GG1$  &  $\II1$  &  $\VV1$  &  $\GG0$  &  $\II0$  &  $\VV0$  &  $\XX$
 \\
\parbox[t]{2mm}{\multirow{7}{*}{\rotatebox[origin=c]{90}{prev. state}}}
& $\SS$ &        &   $\rho_G$  &    &    &    $1-\rho_G$
\\
&$\GG1$ & 1
 \\
&$\II1$ & $r_I$  &  $1-r_I$
 \\
&$\VV1$ & $r_V$  &        & $1-r_V$
 \\
&$\GG0$ &   &     &  $\rho_I (1- d_G)$ &    &      & $(1 - \rho_I) (1-d_G)$ & & $d_G$
\\
&$\II0$ &   &     &       &  $\rho_V (1- d_I)$    &     &  & $(1 - \rho_V) (1-d_I)$ & $d_I$
\\
&$\VV0$ &   &     &       &     &      & & & 1
\end{tabular}
}
\caption{
State transition probabilities assumed by the explicit-duration HMM equivalent of our model: rows correspond to the previous state and columns indicate possible next states. Blank entries all have zero probability, but are kept blank for visual clarity. It is not possible to transition into the start state $\SS$ or out of the terminal states $\RR$ and $\XX$, so these rows/columns are omitted.
}
\label{tab:transition_probas}
\end{table}

\subsection{Prior over recovery probabilities}
\label{sec:app_prior}

We first reproduce the relevant numbers from the CDC's Table 2~\citep{u.s.cdcCOVID19PandemicPlanning2020} in Table~\ref{table:cdc_table_2}.

\begin{table}[!h]
	\begin{tabular}{c c c c | c}
	Age & 18-49 & 50 - 64 & 65+ & Avg.
	\\
	Percent transfered to ICU &
	23.8 & 36.1 & 35.3 & 34.3
	\\
	Percent who receive ventilation
	& 12.0 & 22.1 & 21.1 & 20.4
	\\
	Percent who die
	& 2.4 & 10.0 & 26.6 & 19.3
	\end{tabular}
	\caption{Reproduced Table 2 of \citet{u.s.cdcCOVID19PandemicPlanning2020}, accessed Mar. 19, 2021. We added the Avg. column.}
	\label{table:cdc_table_2}
\end{table}

We then describe how these estimates are turned into a prior over $\rho$. We first convert these estimates to single, age-independent estimates by computing the average across age groups weighted by the US country-level age distribution of hospital patient provided by the CDC~\footnote{\url{https://gis.cdc.gov/grasp/covidnet/COVID19\_3.html}} during the week of November 7th. We show the computed average on the right-most column of Table~\ref{table:cdc_table_2}. Then, assuming fixed values for the probabilities of dying $d$ and given these numbers, we derive the probability of recovering at each stage. Let $p(\II)$ be the percent of patients transferred to the ICU, $p(\VV)$ the percentage of patients who receive ventilation, and $p(\XX)$ the percent of patients who die (all these numbers can be found in the table above).
Then, we compute the mean of each parameter's prior (denoted $\rho_{\GG}$, $\rho_{\II}$ and $\rho_{\VV}$ for convenience) by solving for them in the following equations:
$$
    1 - p(\II) = \rho_{\GG} + ((1 - \rho_{\GG})d_{\GG})
$$
$$
    p(\VV) = p(\II)(1 - \rho_{\II})(1 - d_{\II})
$$
$$
    p(\XX) = p(\VV)(1 - \rho_{\VV}) + p(\II)(1 - \rho_{\II})d_{\II} + d_{\GG}
$$
The found values for $\rho_{\GG}$, $\rho_{\II}$ and $\rho_{\VV}$ denote the ratios of parameters for the Beta priors. To set the variance for the prior, we set the beta parameters for the prior over each $\rho$ by multiplying $\rho$ and $1 - \rho$ by a scalar value $r$. For $\rho_{\GG}$, we empirically set $r_{\GG} = 100$. Then, we set $r_{\II} = r_{\GG}(1 - \rho_{\GG})$ and $r_{\VV} = r_{\II}(1 - \rho_{\II})$. This increases the uncertainty in prior in proportion to the \emph{influx} of patients in the different stages expected under the regime specified by the prior mean probabilities.\\
For both probabilities of death, we set $r = 200$ to indicate strong confidence in the low value that we set and discourage exploration towards unrealistic values.

\section{Details of ABC Learning Procedure}
\label{sec:appendix_abc}

\paragraph{Biasing the distance computation.} For every dataset, we set weight $u_k$ for each count type $k$ such that counts that are generated at stages farther away from admissions are worth more, and such that their average is 1.0, so that the upper bound on the distance is preserved at 1.0. The rationale behind this is that counts generated farther away from admissions are generally harder to recover, as indicated by the greater uncertainty of ABC in predicting those counts that we see in our experiments.\\
For MA and SD, where $\GG, \II, \VV$ and $\XX$-smoothed counts are available, we set $u_{\GG} = 0.7$, $u_{\II} = 0.9$, $u_{\VV} = 1.1$ and $u_{\XX} = 1.3$.
For UT and CA, where $\GG, \II+\VV$ and $\XX$-smoothed counts are available, we set $u_{\GG} = 0.8$, $u_{\II+\VV} = 1.0$ and $u_{\XX} = 1.2$.
For the data ablation experiment on MA, we set $u_{\GG+\II+\VV} = 0.8$ and $u_{\XX} = 1.2$.
For both UK hospitals, we set $u_{\RR} = 0.8$, $u_{\GG+\II+\VV} = 1.0$ and $u_{\XX} = 1.2$.
For the experiments on synthetic data, we set $u_{\GG} = 0.8$, $u_{\RR} = 0.9$, $u_{\II} = 1.0$, $u_{\VV} = 1.1$ and $u_{\XX} = 1.2$.

\paragraph{Justification of the maximum-normalized distance computation.} 
At an individual timestep $t$, we assess distance via a mean absolute error normalized by the maximum between individual entries (which keeps this value on a unit scale).
A side-effect of this normalization is that parameters that overestimate the true counts are preferred over parameters that underestimate the true counts by the same margin.
For example, if true count $y^k_{t} = 20$ and simulated count $\tilde{y}^k_{t} = 22$, then the relevant distance will be $\frac{|20 - 22|}{22} = \frac{1}{11}$. Instead, if $\tilde{y}_t^k = 18$, the relevant unweighted portion of the distance will be $\frac{|20 - 18|}{20} = \frac{1}{10}$. 
This asymmetry enforces a pessimistic bias that in our experimental circumstances would be appropriate - e.g., when developing forecasts that will be used to determine what resources a hospital will need to meet future demand.
In any case, the scheme at most moderately favors over prediction of utilization, compared to under utilization.
Alternative distances could be easily considered.

\paragraph{Scheduling the decay.}
We draw samples from a \textit{burn-in} phase and a \textit{sampling} phase. During the \textit{burn-in} phase, we anneal $\varepsilon$ across iterations with the goals of 1) making it converge to an optimal value for sampling, and 2) dragging the parameters to the high-density region of the parameter space.
Then, during the \textit{sampling} phase, we stop the annealing, and draw samples in standard ABC MCMC as described in \citet{marjoramMarkovChainMonte2003}, where $\varepsilon$ is held fixed across iterations.\\
Crucially, in the sampling phase, we raise $\varepsilon$ by a small value compared to the best value found during the burn-in phase, so that we accept slightly more diverse samples than it would have otherwise.

In the \textit{burn-in} phase, we anneal $\varepsilon$ in the following way. The upper-bound on our distance $d$ allows us to initialize $\varepsilon$ to $1.0$. In practice, we find that $\varepsilon$ can often be safely initialized to a lower value, thus reducing the number of iterations needed for convergence. We anneal $\varepsilon$ exponentially after each parameter proposal using hyperparameter $\gamma$. Crucially, we never allow $\varepsilon$ to take a value below the last accepted distance $d_{best}$, thus allowing the algorithm to naturally converge to a value of $\varepsilon$ that is optimal for sampling, as it is low enough that most proposals get rejected for not being able to provide a better match to the training data. We apply one more change to the annealing schedule: at regular intervals during the \textit{burn-in} phase, we increase $\varepsilon$ by a fixed value $p$. This allows the algorithm to ``escape'' a local optima that might have been encountered along the way. We found this to be particularly useful for fitting parameters to datasets containing fewer patients, such as single-hospital datasets. Indeed, with fewer patients to model, the variance in the counts generated by a given set of parameters is higher, thus it is more likely that a given set of proposed parameters gets accepted due to a particularly "lucky" simulation, which had assigned an unusually low distance to the proposed parameters.\\
\\
\textbf{Specific tuning of the decay schedule.}
In all our experiments, we initialize $\varepsilon$ to 0.7, as we find such value to be always greater than the distance generated using samples from the prior. We fix the number of \emph{burn-in} iterations to 24,000 (where one iteration makes and evaluates a proposal for each of the 17 parameters in turn, for a total of 408,000 total single-parameter iterations), $p$ to 0.05, and $f$ to 0.15. To set the annealing parameter $\gamma$, we consider the following. The optimal, convergence value of $\varepsilon$ varies by dataset. In particular, due to the design of our distance function, datasets with higher counts have a lower convergence value. An approximate lower bound to the convergence value can be quickly found for any dataset by computing the average distance between multiple sets of counts generated with the dataset's admissions and a fixed set of parameters, and a single representative of such counts. Though our annealing schedule is designed to be likely to escape local optima, we find it desirable, especially for lower-counts datasets, to set $\gamma$ such that $\varepsilon$ does not excessively undershoot the approximate lower bound by the end of the training iterations. Concretely, this translates to lower-counts datasets having higher $\gamma$, and thus slower annealing, in our experiments.

\paragraph{Details of proposal distributions.}
For each transition probability $\tau_k$, we propose a new value $\tau_k^*$ by sampling from a Beta distribution whose mean is the old value and with a scaling parameter $r >0$ that determines the variance: $\tau_k^* \sim \text{Beta}(r\tau_k, r (1 - \tau_k))$. We set $r = 100$ for the recovering probabilities and to $r = 200$ (lower entropy) for death probabilities. 

For the mode of each duration probability $\lambda^{s, h}$, we propose a new value $\lambda^{*s, h}$ by sampling from a truncated normal distribution with mean at $\lambda^{s, h}$ and a variance of 0.25: $\lambda^{*,s, h} \sim \text{TruncNorm}(\lambda^{s, h}, 0.25, [1, D])$. For the \emph{log} of each temperature parameter $\nu^{s, h}$, we sample a new value from a normal distribution centered at the old value, with a variance of 0.01: $\nu^{*s, h} \sim \text{Norm}(\nu^{s, h}, 0.01)$.

\paragraph{Warm start.}
At the start of the simulated training period, it is unrealistic to assume that the initial patients in each hospital stage have only just been admitted to said stage (i.e. they are all at day zero of their stay). Thus, in an effort to more realistically model the progression through the hospital of the initial population, we apply a simple 'warm start' heuristic, in which we simulate the admission of the initial population in each state by uniformly distributing them during 5 days previous to day zero. We increase the admissions by 3\% in both G and V to account for any patients who exit the hospital before the official start of the training period.

\paragraph{Details of ensembling.} For all experiment, we collect 200 samples each from 10 different runs of the algorithm, resulting in 2000 total samples. We only use runs which we have verified to have converged to similar values of $\varepsilon$, and thus produce samples that explain the training counts equally well.

\begin{figure}[h]
\centering
\includegraphics[width=0.75\textwidth]{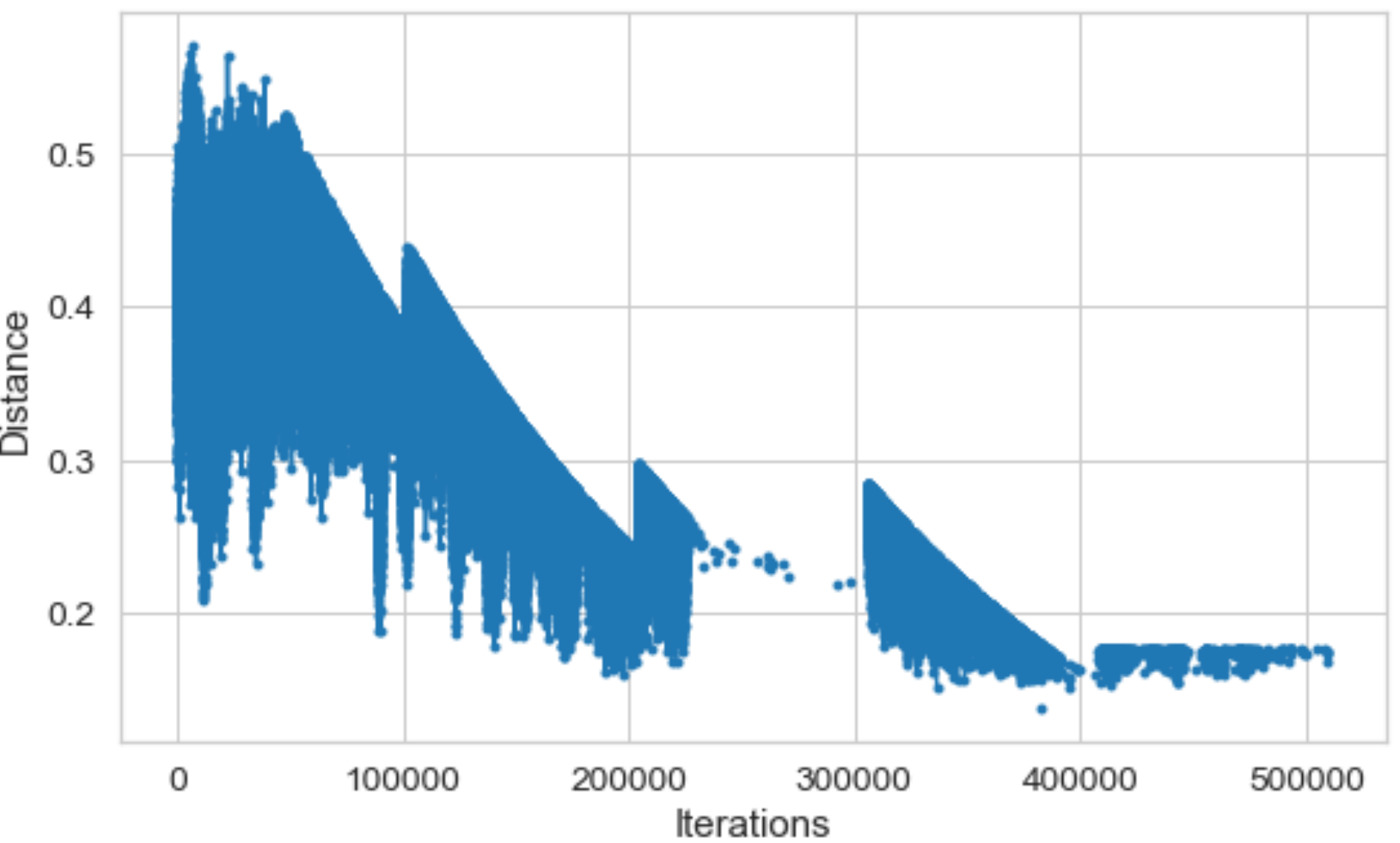}
\caption{\textbf{Trend of accepted distanced on one run of ABC for ACED-HMM on South Tees hospital data.} Each data point indicates a set of parametrs that has surpassed the first stage of acceptance (distance below $\varepsilon$). Around iteration 230,000 the algorithm got stuck in a local optimum, where few proposals were getting accepted. The resetting of $\varepsilon$ after iteration 300,000 helped the algorithm get unstuck from the local optimum.}
\label{fig:distances_plot_south_tees}
\end{figure}

\section{Details of Baseline AR-Poisson forecasting model}
\label{sec:appendix_gar}

We consider a simple probabilistic model for univariate counts, the \emph{latent autoregressive Poisson} or just AR-Poisson for short. This is inspired by previous work on single hospital site forecasting by ~\citet{leeForecastingCOVID19Counts2021}. The model can be fit separately to each univariate count time series of interest (e.g. separate fit to $\GG, \II,$ and $\VV$ counts).

Across $T$ days we observe a univariate time series $y_{1:T} = [y_1, y_2, \ldots y_T]$, where $y_t \in \{0, 1, 2, \ldots \}$ indicates the count of patients in a particular stage on day $t$.
Our goal is to develop \emph{forecasts} for the next $F$ days, given all previous observations, using a conditional probabilistic model $p( y_{(T+1):(T+F)} \mid y_{1:T} )$.

\paragraph{AR-Poisson Model.}
We consider an order-1 autoregressive process as our latent variable model, where each timestep has a latent real value $f_t \in \mathbb{R}$. We capture dependency across time in the latent series $f_{1:T}$, and model each count $y_t$ as conditionally independent given $f_t$:
\begin{align}
    p(f_{1:T}, y_{1:T} ) = \prod_{t=1}^T p( f_t \mid f_{t-1} ) \cdot \prod_{t=1}^T p(y_t \mid f_{t} ).
\end{align}
For most timesteps $t > 1$, we generate the latent value $f_t$ using the previous value $f_{t-1}$:
\begin{align}
	p(f_t \mid f_{t-1}) = \text{Normal}(\beta_0 + \beta_1 f_{t-1}, \sigma^2), ~ t \in 2, 3, \ldots
\end{align}
The very first value $f_1$ is generated with mean $\beta_0$:
\begin{align}
    p(f_1) = \text{Normal}(\beta_0, \sigma^2)
\end{align}
The latent-generating parameters are coefficients $\beta = [\beta_0, \beta_1]$ and standard deviation $\sigma > 0$.

We then generate each observed count $y_t$ using a Poisson likelihood, setting the mean parameter by transforming the latent $f_t$ to a positive value via the exponential:
\begin{align}
    p(y_t  \mid f_t) = \text{Poisson}(\exp(f_t)).
\end{align}

\paragraph{Forecasting Method.}
We wish to first sample from the posterior over parameters $\beta$, $\sigma$, and latent values $f$: $p(\beta, \sigma, f_{1:T} \mid y_{1:T})$.
We place the follow vague priors over the parameters:
\begin{gather}
    \beta_0 \sim \text{Normal}(0, 0.1), \\
    \beta_1 \sim \text{Normal}(1, 0.1), \\
	\sigma \sim \text{HalfNormal}(0.1).
\end{gather}
We then use the No-U-Turn sampler \citep{hoffmanNoUTurnSamplerAdaptively2014} to perform Markov chain Monte Carlo approximation of the posterior.
Our NUTS sampler is implemented using the PyMC3 toolbox~\citep{salvatierProbabilisticProgrammingPython2016}.

Given a single posterior sample indexed by $s$, with parameters $\beta^s, \sigma^s$ and latents $f^s_{1:T}$, we can then use the generative model to draw a \emph{forecast} of latents $f$ and counts $y$ for the next $F$ days:
\begin{align}
    f^s_{T+\tau} &\sim \text{Normal}( \beta^s_0 + \beta^s_1 f^s_{T+\tau-1}, \sigma^s ), &\tau \in 1, \ldots F
    \\
    y^s_{T+\tau} &\sim \text{Poisson}( \exp(f^s_{T+\tau}) ), &\tau \in 1, \ldots F
\end{align}

We typically draw a forecast of $S$ distinct samples, using $S=1000$ or more to be sure we're capturing the full distribution.

\begin{figure}[!h]
    \centering
    \begin{tabular}{c c}\includegraphics[width=0.5\textwidth]{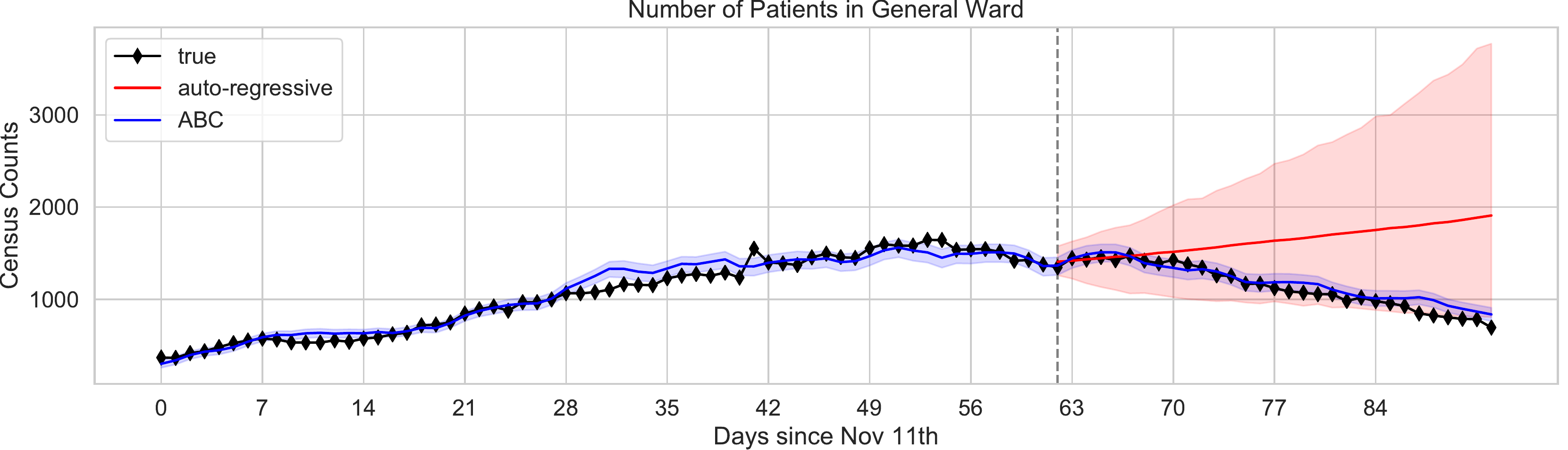}
    &\includegraphics[width=0.5\textwidth]{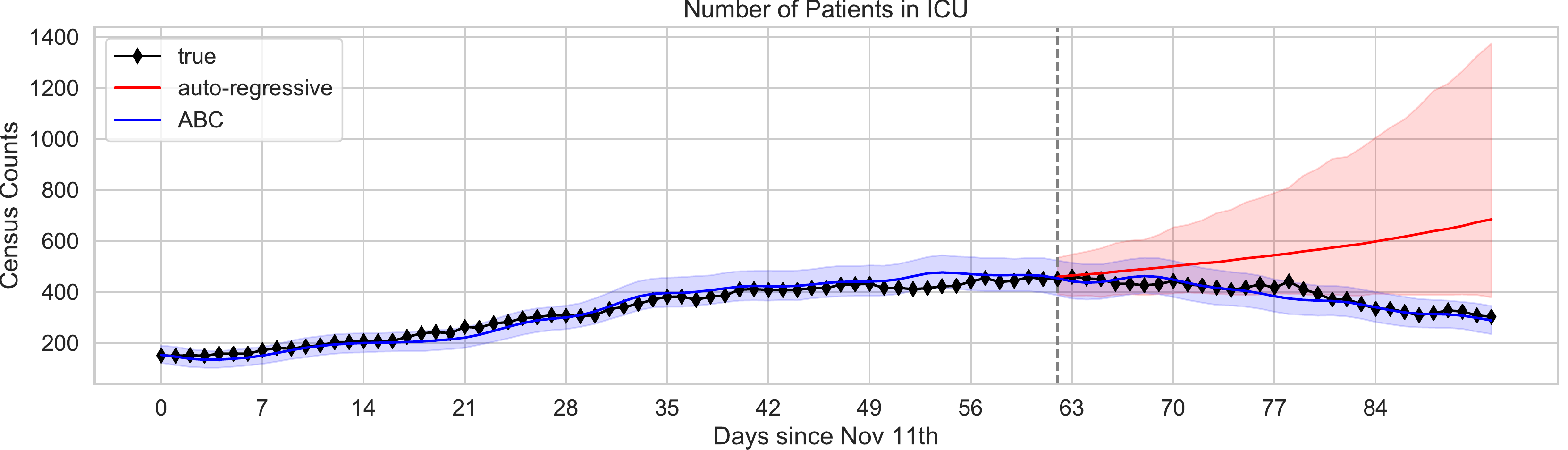}
    \end{tabular}
    \caption{\textbf{Fit and forecasts on General Ward ($\GG$) and Total ICU ($\II + \VV)$ counts for MA, comparing our ACED-HMM to the AR-Poisson baseline.} Compare to Fig.~\ref{fig:forecast_MA}
       }
\label{fig:forecast_MA_with_baseline}
\end{figure}

\paragraph{Results.}
In Fig.~\ref{fig:forecast_MA_with_baseline}, we compare our AR-Poisson with our mechanistic ACED-HMM. Given the series of counts from the last 28 days of training data, we forecast ahead $F=31$ days. We draw $S=1000$ forecast samples and depict the 2.5\textsuperscript{th}, 50\textsuperscript{th}, and 97.5\textsuperscript{th} percentiles.

We can see that our ACED-HMM model has clear advantages over the AR-Poisson, which is limited to capturing linear trends and makes the assumption that the future is like the past.

\section{Details of IHME Baseline forecasts for U.S. States}
\label{sec:appendix_ihme}

As a baseline, we consider the public forecasts produced by the IHME COVID-19 forecasting project~\citep{reinerModelingCOVID19Scenarios2021}. 
These forecasts are available online\footnote{\url{https://www.healthdata.org/covid/data-downloads}} and contain predicted daily occupancy count values for each U.S. state for several stages of hospital care: total beds, ICU beds, ventilators, and deaths.

The core IHME model is targeted at forecasting \emph{deaths} for each U. S. state, using an SEIR compartment model that accounts for a number of local factors.
Historical daily death counts represent the primary training data used, though other observed information such as testing and mobility are included.
Unlike our approach, the IHME model is not adapted to fit to the observed bed occupancy counts in hospitals in the region of interest.
The methodology behind IHME's forecasts for hospital resource use (total bed usage, ventilator usage, etc.) is described in Section 8 of the Supplementary Information~\citep{ihmecovid-19forecastingteamSupplementaryInformationModeling2020} released alongside their published research article on overall forecasting in the United States~\citep{reinerModelingCOVID19Scenarios2021}.

To forecast hospital resource utilization, the IHME team performs a ``microsimulation'' that starts at deaths and works backwards.
For each death, they assume the patient stayed the previous 6 days in the ICU.
They further simulate for each death, a predicted age bin using local age distribution data, and simulate a number of individuals in the same age group as having also entering the hospital on the same day but surviving.
Most of this ``age-hospital-cohort'' stay in the general ward (outside the ICU) for 8 days total.
A small fraction of this ``age-hospital-cohort'' (6.3\%) are assumed to be admitted to the ICU and stay for 20 days total (the middle 13 in the ICU).
85\% of individuals in the ICU are assumed to need the ventilator.
These parametric assumptions are reasonably informed by the literature from a specific location (New York state).

Compared to this approach, we argue that our approach of adapting a flexible hospital model with learnable transition probabilities and durations to a region of interest, rather than simply making fixed assumptions for all states, has substantial benefits.

\paragraph{Comparison of IHME forecasts to our methods.}
Several factors make it difficult to perform a fair head-to-head comparison between our model and the published IHME forecasts.
First, the data signals used to train each method differ: while our method uses counts from all available stages, the IHME model relies primarily on daily deaths.
Second, our forecasts require an admission count time series for the test period, while IHME does not.
Finally, while our model is open source and can be customized to any training period and any region of interest provided the requisite data, we could not find a way to easily get produce customized forecasts from IHME, so we rely on their published numbers.
Thus, any comparison between our methods and IHME should be more of a sanity check than a true comparison of equals. We expect that our method, because it is provided more detailed training data for the region of interest and especially given admissions for the testing period, should outperform the IHME baseline.

Acknowledging these challenges, we try to set up as fair a comparison as possible by taking the IHME forecasts released during our target testing period (Jan. 11 - Feb. 11, 2021), which are the ones dated 2021-01-15 on the IHME website. Manual inspection of the CSV files released on this date indicates that these models had access to death counts until Jan. 11 but not afterwards. We suggest this because before this date values of the ``deaths\_mean'' field in the released spreadsheet are whole numbers, after this date values are fractional indicating an expected value forecast rather than an actual observed count.

\paragraph{Raw forecast information from IHME website.}
We used the ``reference'' forecast released by the IHME for 2021-01-15, which provides daily count forecasts of hospital usage for each state of interest.
IHME also releases best-case or worst-case projections (imagining different levels of mask usage or movement restrictions), but we did not use these, taking the ``reference'' forecasts as a reasonable projection of the status quo.

Each field in the released forecast has a ``mean'', ``lower'', and ``upper'' value indicating the expected value as well as lower and upper values to suggest an interval of uncertainty.

We used the following fields of IHME ``reference'' forecasts.

\begin{itemize}
\item admis(mean/lower/upper) : hospital admissions by day
\item allbed(mean/lower/upper) : total covid beds needed by day 
\item ICUbed(mean/lower/upper) : ICU covid beds needed by day 
\item InvVen(mean/lower/upper) : invasive ventilation needed by day
\item deaths(mean/lower/upper) : daily covid deaths
\end{itemize}

We used the PDF file \texttt{IHME\_COVID\_19\_Data\_Release\_Information\_Sheet\.pdf} included in the ZIP file downloaded from IHME as a ``data dictionary'' to understand these field names and values.

\paragraph{Standardizing IHME forecasts our data format.}
We transformed the raw data daily counts to obtain values for our stages by performing the following element-wise subtractions/additions of the above raw data fields:
\begin{itemize}
\item General Ward $\GG$ : allbed - ICUbed
\item In ICU without ventilator $\II$: ICUbed - InvVen
\item In ICU on ventilator $\VV$: InvVen
\item In ICU total $\II + \VV$: ICUbed
\item Terminal $\XX$: deaths
\end{itemize}

\paragraph{MAE computation.}
To obtain an MAE estimate from IHME at a specific stage ($\GG, \II, \VV, \XX$) for a given state, we look in the testing period simply evaluate the MAE for the mean forecast of that stage compared to our ground-truth count values of that stage (as described in App.~\ref{sec:app_datasets}).

\paragraph{Intervals}
To obtain some estimates of uncertainty, we compute the MAE between the true counts and the lower and upper estimates.
This is all we can do given only mean/lower/upper values at each day (rather than a proper distribution).

We emphasize the IHME reported interval is not comparable to the intervals produced by our models.
Our model's intervals are more easily interpreted in the standard way: they provide a range of plausible values for what the MAE would be given another similar-sized sample from the posterior.

\end{document}